\documentclass[preprint]{elsarticle}
\usepackage{amssymb}
\usepackage{amsmath}
\usepackage{graphicx}
\usepackage{lipsum}
\usepackage{mwe}
\usepackage{lineno}
\usepackage[greek,english]{babel}
\usepackage{url}

\journal{Journal of Nuclear Instruments and Methods in Physics Research A}

\begin{document}
\begin{frontmatter}
\title{Timing performance of a multi-pad PICOSEC-Micromegas detector prototype}

\author[aff3]{S. Aune}
\author[aff2]{J. Bortfeldt}
\author[aff2]{F. Brunbauer}
\author[aff2]{C. David}
\author[aff3]{D. Desforge}
\author[aff5]{G. Fanourakis}
\author[aff7]{M. Gallinaro}
\author[aff11]{F. Garc\'{i}a}
\author[aff3]{I. Giomataris}
\author[aff9]{T. Gustavsson}
\author[aff3]{F.J. Iguaz}
\author[aff3]{M. Kebbiri}
\author[aff1,aff1n]{K. Kordas}
\author[aff1,aff1n]{C. Lampoudis}
\author[aff3]{P. Legou}
\author[aff2]{M. Lisowska}
\author[aff4]{J. Liu}
\author[aff2,fn2]{\mbox{M. Lupberger}}
\author[aff3]{O. Maillard}
\author[aff1,aff1n]{{I. Maniatis}}
\author[aff1,aff1n]{\mbox{I. Manthos}}
\author[aff2]{H. M\"{u}ller}
\author[aff2]{\mbox{E. Oliveri}}
\author[aff3]{T. Papaevangelou}
\author[aff1]{K. Paraschou}
\author[aff10]{M. Pomorski}
\author[aff4]{B. Qi}
\author[aff2]{\mbox{F. Resnati}}
\author[aff2]{L. Ropelewski}
\author[aff1,aff1n]{D. Sampsonidis}
\author[aff2]{L. Scharenberg}
\author[aff2]{T. Schneider}
\author[aff3]{L. Sohl}
\author[aff2]{M. van Stenis}
\author[aff1,aff1n]{A. Tsiamis}
\author[aff6]{Y. Tsipolitis}
\author[aff1,aff1n]{S.E. Tzamarias\corref{cor1}}
\ead{tzamarias@auth.gr}
\author[aff2]{A. Utrobicic}
\author[aff8,fn3]{\mbox{R. Veenhof}}
\author[aff4]{X. Wang}
\author[aff2]{\mbox{S. White}}
\author[aff4]{Z. Zhang}
\author[aff4]{Y. Zhou}

\address[aff3]{IRFU, CEA, Universit´e Paris-Saclay, F-91191 Gif-sur-Yvette, France}
\address[aff2]{European Organization for Nuclear Research (CERN), CH-1211 Geneve 23, Switzerland}
\address[aff4]{State Key Laboratory of Particle Detection and Electronics, University of Science and Technology of China, Hefei CN-230026, China}
\address[aff1]{Department of Physics, Aristotle University of Thessaloniki, University Campus, GR-54124, Thessaloniki, Greece.}
\address[aff1n]{Center for Interdisciplinary Research and Innovation (CIRI-AUTH), Thessaloniki 57001, Greece.}
\address[aff5]{Institute of Nuclear and Particle Physics, NCSR Demokritos, GR-15341 Agia Paraskevi, Attiki, Greece}
\address[aff6]{National Technical University of Athens, Athens, Greece}
\address[aff7]{Laborat\'{o}rio de Instrumentac\~{a}o e F\'{i}sica Experimental de Part\'{i}culas, Lisbon, Portugal}
\address[aff8]{RD51 collaboration, European Organization for Nuclear Research (CERN), CH-1211 Geneve 23, Switzerland}
\address[aff9]{LIDYL, CEA, CNRS, Universit Paris-Saclay, F-91191 Gif-sur-Yvette, France}
\address[aff10]{CEA-LIST, Diamond Sensors Laboratory, CEA Saclay, F-91191 Gif-sur-Yvette, France}
\address[aff11]{Helsinki Institute of Physics, University of Helsinki, FI-00014 Helsinki, Finland}

\cortext[cor1]{Corresponding author}
\fntext[fn2]{Now at University of Bonn, D-53115 Bonn, Germany.}
\fntext[fn3]{Also at National Research Nuclear University MEPhI, Kashirskoe Highway 31, Moscow, Russia; and Department of Physics, Uludağ University, 16059 Bursa,Turkey.}
\begin{abstract}

The multi-pad PICOSEC-Micromegas is an improved detector prototype with a segmented anode, consisting of 19 hexagonal pads. Detailed studies are performed with data collected in a muon beam over four representative pads. We demonstrate that such a device, scalable to a larger area, provides excellent time resolution and detection efficiency. As expected from earlier single-cell device studies, we measure a time resolution of approximately 25 picoseconds for charged particles hitting near the anode pad centers, and up to 30 picoseconds at the pad edges. Here, we study in detail the effect of drift gap thickness non-uniformity on the timing performance and evaluate  impact position based corrections to obtain a uniform timing response over the full detector coverage.
 
\end{abstract}


\end{frontmatter}

\section{Introduction} \label{intro}
%\begin{linenumbers}
The ability to precisely measure the production time of minimum ionizing particles (MIPs) enables new capabilities to the current experiments in High Energy and Nuclear Physics and extend their physics reach. As the scale and level of performance of the timing components of current experiments continues to advance, so do the roles for timing of physics objects and the technologies available for sensors proliferate.
Aside from the traditional role of time-of-flight for charged particle  identification \cite{crispin}, the LHC experiments are implementing charged particle timing to mitigate pileup-induced backgrounds for the coming higher luminosity era (HL-LHC). At the LHC, precision timing is also being implemented in components of the calorimetry for the same purpose \cite{hgc}. 

Aside from the LHC, timing capability is also considered at new facilities and in underground experiments - the latter principally for discrimination between scintillation and Cherenkov photons \cite{THEA}.
While the LHC upgrades have focused on silicon-based sensors with internal gain %(i.e. low gain -so called LGAD- for the forward regions of the ATLAS and CMS detectors, and very high gain Geiger-mode devices -Silicon PMs or SiPM- in the CMS Barrel Timing Layer) 
for the approved projects, R$\&$D continues on timing with silicon without internal gain \cite{LIPTON} and devices with intermediate gain
%, between unity and Geiger
\cite{HFS}. Nevertheless, it has become clear that there is a need to keep the detector capacitance small for noise considerations, which can be achieved by increasing the detector granularity. However, the consequent cost per unit area of such highly granular systems are likely to make silicon-based detectors unattractive for applications requiring more than a few square meters of coverage.

Our collaboration demonstrated that charged particle time-of-arrival resolution of better than 25 picoseconds and single-photon timing precision of roughly 50 picoseconds   can be achieved \cite{pico24, LukasSienna} with an implementation of  single cell Micromegas detectors  coupled to a few-mm thick UV-transparent Cherenkov radiator (PICOSEC-Micromegas). 
%We studied the underlying properties of this detector allowing to optimize the performance with a 2-stage detector and the choice of gas filling. 
%which demonstrated charged particle time-of-arrival resolution of better than 25 picoseconds and single photon timing of roughly 50 picoseconds in single cell devices \cite{pico24, LukasSienna}. 
%Earlier work \cite{pico24, LukasSienna}  focused on enhancing the robustness of this detector which is currently limited by the CsI photocathode material. 
%We study in detail the effect of drift gap thickness non-uniformity on the timing performance and evaluate  impact position based corrections to obtain a uniform timing response  over the full detector coverage.
%This study reveals that $\sim 10$  $\textrm{\selectlanguage{greek}m\selectlanguage{english}m}$ fabrication tolerances are necessary  to achieve a multi-anode detector design that could be scaled to provide robust and affordable precise timing determination over a large area, where a typical timing detector cell dimension of the order of 1~cm$^2$ would be appropriate.
Here, we report on a new PICOSEC-Micromegas device with 19 hexagonal anode pads (a multi-pad PICOSEC-Micromegas prototype) and we evaluate in detail the timing performance for muons impinging onto four representative pads.  In the multi-pad device discussed here, we study in detail the effect of drift gap thickness non-uniformity on the timing performance and devise 
%dedicated impact position 
corrections to obtain a uniform timing response over the full detector coverage.
Our work highlights the critical role of tight mechanical tolerances on the depth of the drift/preamplification stage.\\

After describing the multi-pad PICOSEC-Micromegas prototype (Section \ref{setup}), the experimental setup used for collecting  calibration data in a 150 GeV muon beam and the processing of  the detector signals (Section \ref{exptest}), we report  on the method used for a precise detector alignment and on the findings indicating the non-uniformity  of the drift gap thickness (Section \ref{alignment}). In Section \ref{spad} we  investigate the effect of the  above non-uniformity on the timing characteristics of each of the instrumented pads and we evaluate corrections that restore uniformity and improve the time resolution. In Section \ref{multicomb} we report on a method to combine timing and charge information from neighboring pads to estimate the arrival times of MIPs that impact the detector close to pad edges. Section \ref{concl} contains concluding remarks.

\section{The multi-pad PICOSEC-Micromegas detector} \label{setup}

\begin{figure}
\centering
\begin{minipage}{.55\textwidth}
\centering
\includegraphics[width=1.\textwidth]{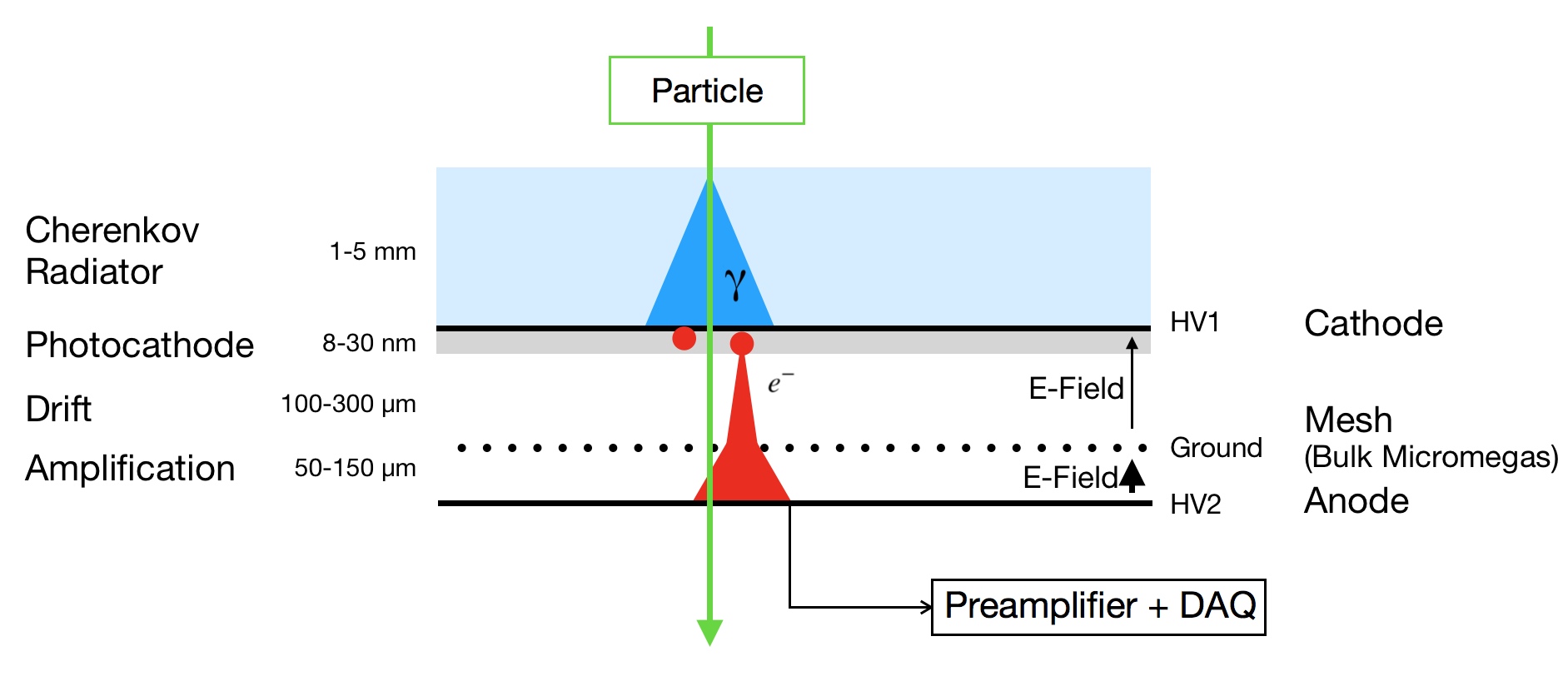}
\end{minipage}%
\begin{minipage}{.41\textwidth}
\centering
\includegraphics[width=1.\textwidth]{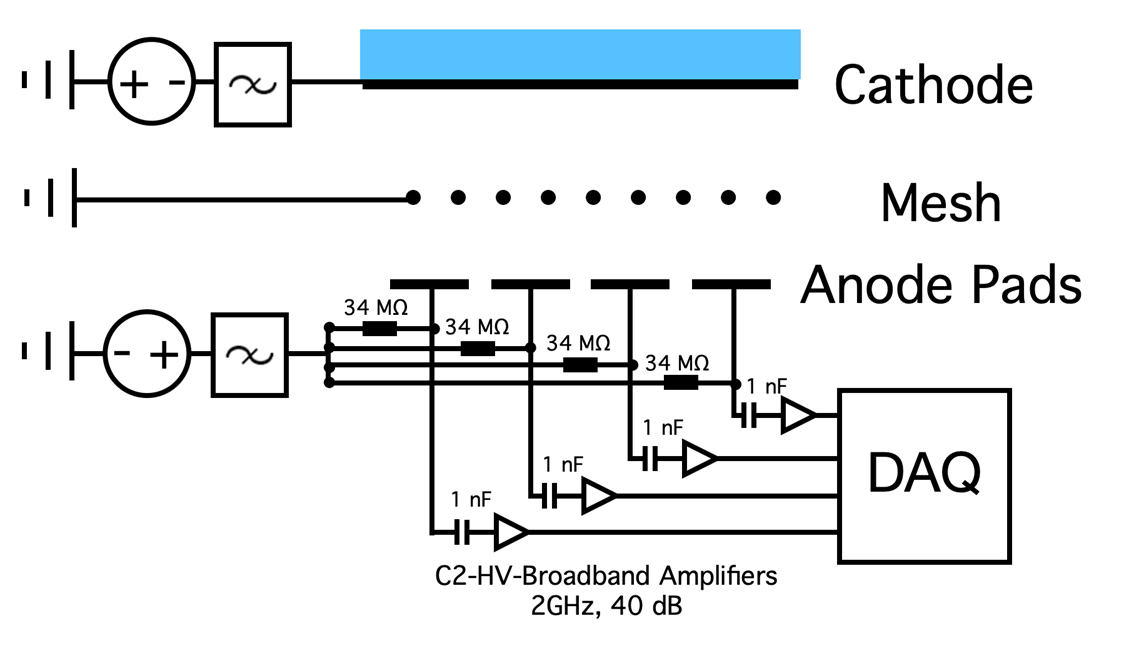}
\end{minipage}

\caption{(left) Sketch of the PICOSEC-Micromegas working principle. (right)
Electrical diagram of the detector power supply and signal read-out. The signal of each anode pad is individually amplified with a CIVIDEC C2-HV broadband amplifier.
}
\label{fig:conceptsketch}
\end{figure}

The main idea of the PICOSEC-Micromegas concept is to suppress the inevitable time jitter of the ionisation in a Micromegas \cite{micromegas}, due to different ionisation cluster positions. Figure~\ref{fig:conceptsketch} illustrates the PICOSEC-Micromegas detector concept. A Cherenkov radiator and a photocathode are placed in front of the gaseous volume. The passage of a charged particle through the Cherenkov radiator produces UV photons, which are then absorbed in the photocathode and primary electrons (photoelectrons) are created on the bottom surface of the photocathode. These electrons are subsequently preamplified in a first and then amplified in a second high-field stage, inducing a signal on the read-out plane.

%\begin{figure}[ht]
%\centering
%\includegraphics[width = \textwidth]{fig1.jpeg}
%\caption{ Sketch of the PICOSEC-Micromegas working principle}
%\label{fig:conceptsketch}
%\end{figure} 
    
The drift region thickness in the PICOSEC-Micromegas is reduced to the same order as the amplification gap. It is moreover operated with an electric field similar to that of the amplification gap. In this field configuration, a  preamplification of the electrons happens in the drift gap improving the time resolution as it reduces the drift time of the primary electrons \citep{pico24, model}. 

A PICOSEC-Micromegas detector with an active area of 3.5\,cm diameter has been developed with a $MgF_{2}$ Cherenkov radiator, 5\,cm in diameter. 
A layer of $CsI$, with a thickness of 18 nm has been used as UV semi-transparent photocathode. A thin $Cr$ layer, with a thickness of about 3 nm, has been used as conductive interface layer between radiator and $CsI$. The UV transmission of the $MgF_{2}$ radiator at 160 nm is about 80\%. An additional 50\%  of transmission  decrease 
has been measured at the same wavelength  when the $Cr$ layer is present.

A bulk Micromegas \cite{bulk} with $184\,\textrm{\selectlanguage{greek}m\selectlanguage{english}m}$ drift gap  and $128\,\textrm{\selectlanguage{greek}m\selectlanguage{english}m}$ amplification gap   is used for this detector with the mesh directly connected to ground on the PCB. 
A calendered mesh with 18 $\textrm{\selectlanguage{greek}m\selectlanguage{english}m}$  wire diameter, 45 $\textrm{\selectlanguage{greek}m\selectlanguage{english}m}$ opening
and about 30 $\textrm{\selectlanguage{greek}m\selectlanguage{english}m}$ thickness is used. 

\begin{figure}
\centering
\begin{minipage}{.48\textwidth}
\centering
\includegraphics[width=1.\textwidth]{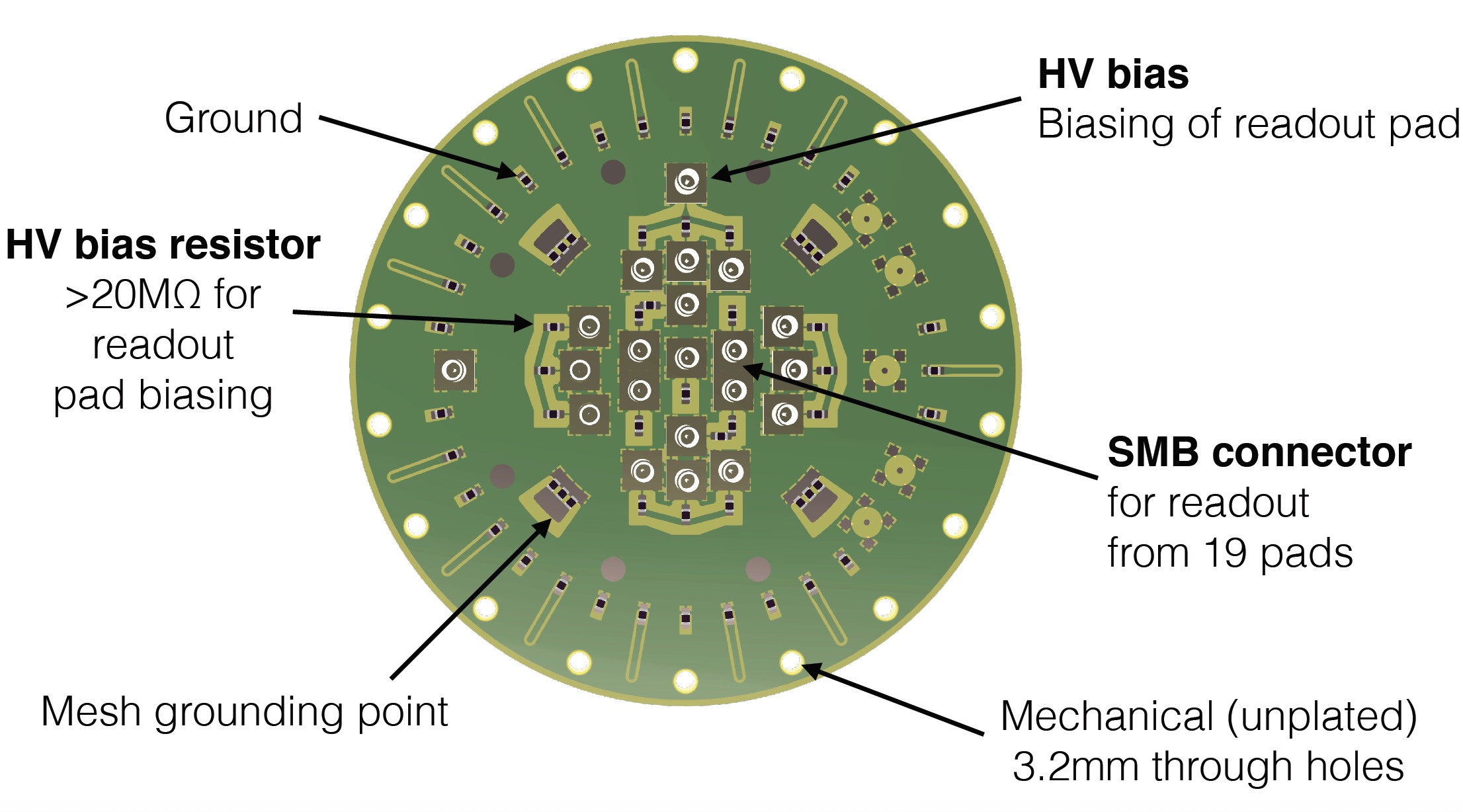}
\end{minipage}%
\begin{minipage}{.48\textwidth}
\centering
\includegraphics[width=1.\textwidth]{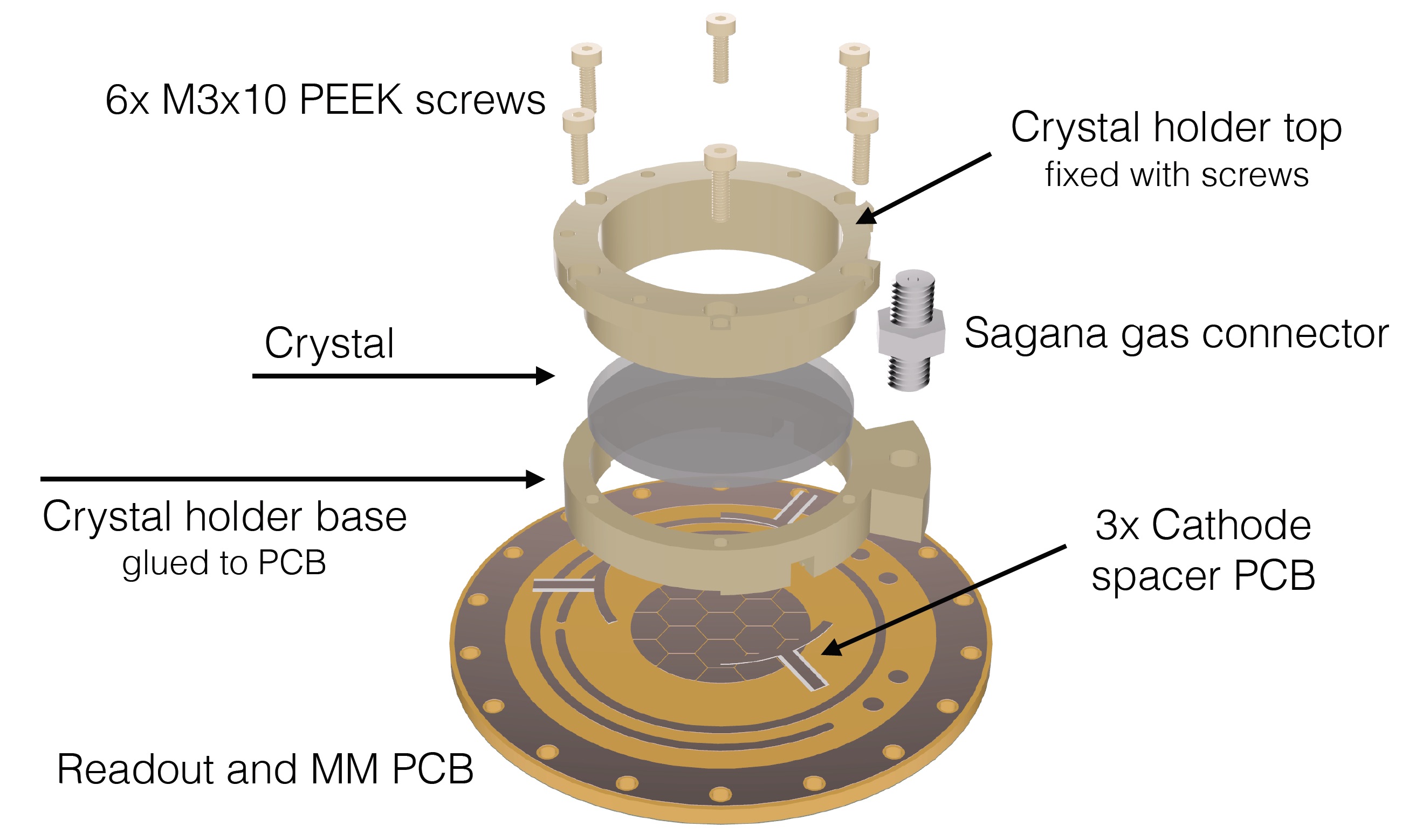}
\end{minipage}

\caption{(left) Technical drawing of the PCB. (right)
Technical sketch of the mechanics that holds the Cherenkov crystal with photocathode in place.}
\label{fig:pcb_cad}
\end{figure}

This is the first PICOSEC-Micromegas detector with a segmented read-out anode divided into 19 hexagonal pads with a diameter of 1\,cm each. The adjacent edges  of neighboring pads are separated by 200 $\textrm{\selectlanguage{greek}m\selectlanguage{english}m}$  gap. The left sketch in Fig. \ref{fig:pcb_cad} shows a technical drawing of the read-out PCB  and the right sketch in the same figure shows the mounting mechanism of the Cherenkov radiator on top of the PCB. Only the seven inner  hexagons are printed in full size as the detector boundary curtails the outer hexagons. The left photograph in Fig. \ref{fig:pad_geom} shows a top view picture of the detector during assembly in the clean-room. The pad structure with the full pads in the centre and the cut pads on the border are visible.

The PCB with the Micromegas is mounted on one flange that seals the chamber hermetically. Over- and under-pressure  operation as well as mechanical stress during assembly have therefore a direct impact on the PCB flatness. As the PCB is sealing the detector, the signal for each pad is individually routed to the back of the PCB by an individual SMB (SubMiniature version B)  connector.  The high voltage is instead supplied through one common connector for all pads. The high voltage is decoupled to each pad connector by a resistor ($>20\,\textrm{M}\Omega$) to allow a separate read-out of some pads while powering the whole detector.

Unlike the previous PICOSEC-Micromegas \citep{pico24} that had a window in the outer gas vessel, this detector is completely sealed and has no windows. In order to calibrate the detector, a UV-LED is  mounted inside the chamber with the bias voltage  fed through the PCB.

\

\section{Experimental setup and waveform processing} \label{exptest}
The time response of the multi-pad PICOSEC-Micromegas to 150 GeV muons was measured at the CERN SPS H4
secondary beamline. A similar experimental setup   as the one used in previous measurements \citep{pico24} allows the characterization
of the detector prototype, situated as shown in Fig. \ref{fig:exp_setup}.
Two trigger scintillators of 5 $\times$ 5 mm$^2$ operate in anti-coincidence with a veto scintillator whose aperture (hole) matches the same area (small area trigger). This trigger configuration efficiently selects muons that
do not undergo scattering and suppresses triggers from particle showers. Another set of trigger scintillators is available covering a 5 $\times$ 5 cm$^2$ area (large area trigger). This trigger is selected to study several pads in parallel. 
One PHOTEK MCP - PMT, model PMT240/Q/BI/NG77   \cite{mcpdata},  operated at -4800 V, provides the time reference; its entrance window (9 mm Fused Silica) is placed perpendicular to the beam
and serves as a Cherenkov radiator. In a previous work \cite{mcp}, we  determined that the  MCP  provides a reference time with an accuracy of  $\sim$7 ps in the inner 11\,mm diameter.
A telescope of three tracking GEM detectors with two-dimensional
strip readout is used to reconstruct the trajectory of each muon with
a combinatorial Kalman filter based algorithm, and to determine its
impact position at the PICOSEC-Micromegas photocathode with an accuracy of about 50 $\textrm{\selectlanguage{greek}m\selectlanguage{english}m}$.

 \begin{figure}
\centering
\begin{minipage}{.48\textwidth}
\centering
\includegraphics[width=1.\textwidth]{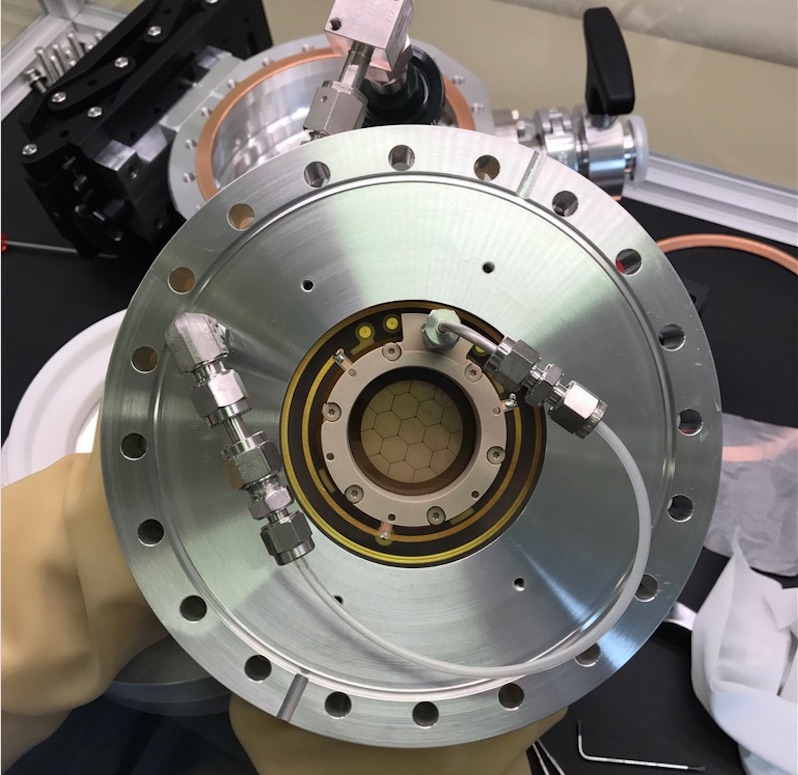}
\end{minipage}%
\begin{minipage}{.48\textwidth}
\centering
\includegraphics[width=1.\textwidth]{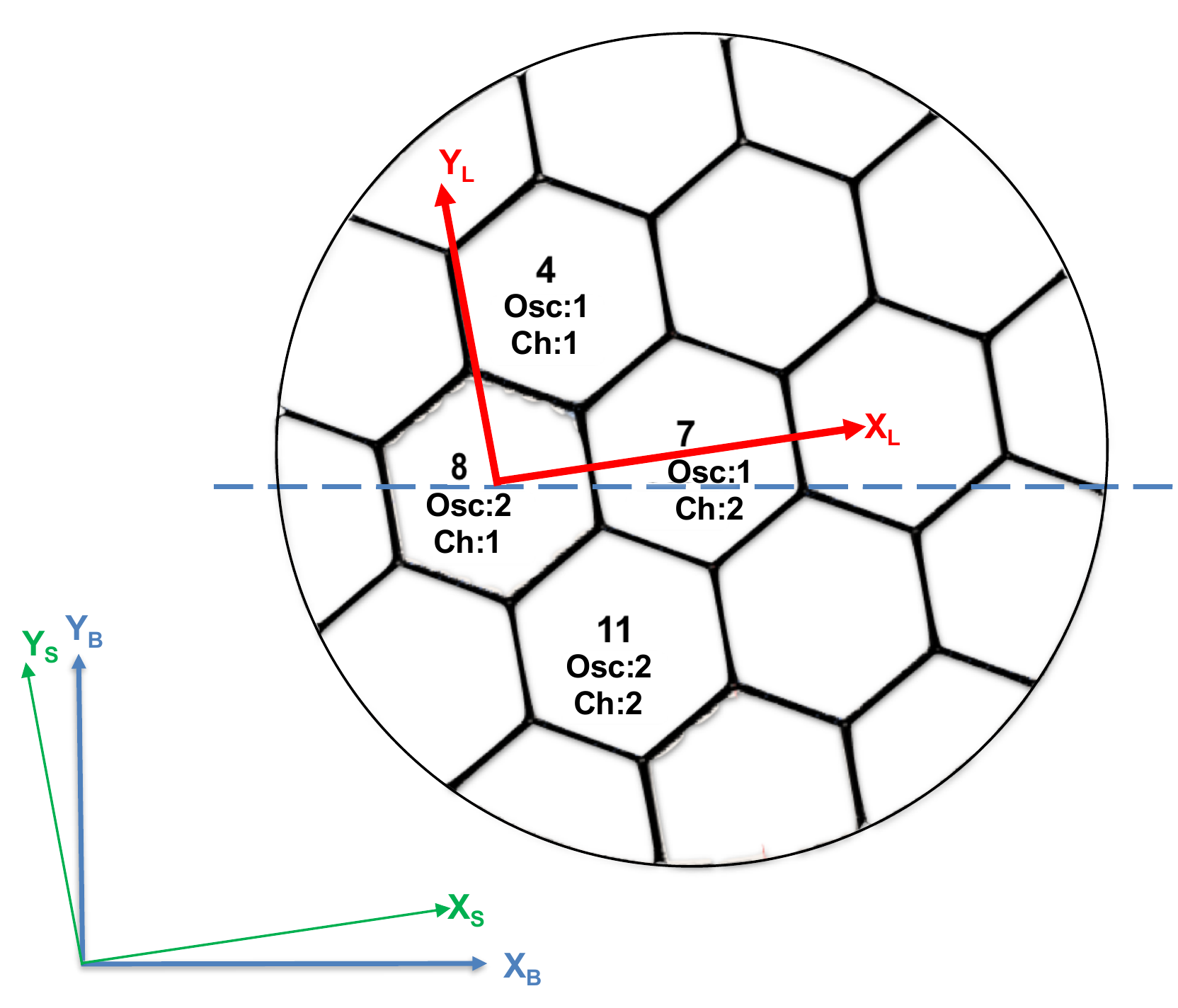}
\end{minipage}
\caption{(left) Photograph of the multi-pad chamber during assembly in the
clean room. The hexagonal pad structure of the readout is visible in the centre. (right)
A schematic diagram of the anode segmentation. Notice that there is a gap between adjacent pad edges, represented by the thick black lines.  The pads No. 4, 7, 8 and 11 are fully instrumented and their signals are digitized by the oscilloscope channels as indicated. The red axes, labelled as $X_{L}$ and $Y_{L}$, represent the local coordinate frame, while the blue axes, labelled as $X_{B}$ and $Y_{B}$, represent the global tracking coordinate frame (or else beam-frame). The green axes, labelled as $X_{S}$ and $Y_{S}$, represent the symmetry frame, which is used in the alignment procedure as described in the text.}
\label{fig:pad_geom}
\end{figure}

\

The signals of the four instrumented PICOSEC-Micromegas pads  go through  CIVIDEC \citep{cividec} preamplifiers before
being digitized and recorded (together with the MCP signal) by two LECROY WR8104 oscilloscopes \citep{lecroy} operated at 1.0 GHz analogue bandwidth and at a sampling rate of 10 GSamples/s. Specifically, the signals from pads No. 4 and 7 are digitized by channels Ch:1 and Ch:2  of the first oscilloscope (Osc:1), while the signals  from pads No. 8 and 11 are digitized by  Ch:1 and Ch:2  of the second oscilloscope (Osc:2), respectively (Fig. \ref{fig:pad_geom}). The MCP signal, after being split by a 50 Ohm splitter, is digitized by the third channel of both oscilloscopes, while
the tracking (GEM
detector) data are recorded simultaneously  in an APV25 based SRS
DAQ \citep{srs}. The SRS DAQ system of the tracking detectors is triggered by the scintillator triggers, definning the acceptance area of the beam particles. To ensure event alignment in the two DAQ systems, the
internal SRS event number is sent as a bit stream to the fourth channels of both oscilloscopes and is used to trigger the acquisition of the PICOSEC-Micromegas waveforms.

The PICOSEC-Micromegas detector was operated with the COMPASS\footnote{The term ``COMPASS gas'' refers to the mixture $80\%\, \textrm{Ne}, 10\%\, \textrm{C}_2 \textrm{H}_6, 10\%\, \textrm{CF}_4$, as used by the COMPASS Collaboration.} gas at atmospheric pressure and at different anode and drift
voltages, with  the  detector gain  exceeding $10^5$. The data used in this work are collected during a long run, when more than 500,000 triggers were collected with the detector operated under stable condition, with 300 V anode and 450 V drift voltages.
\\

\begin{figure}[t]
    \centering
    \includegraphics[width=0.7\textwidth]{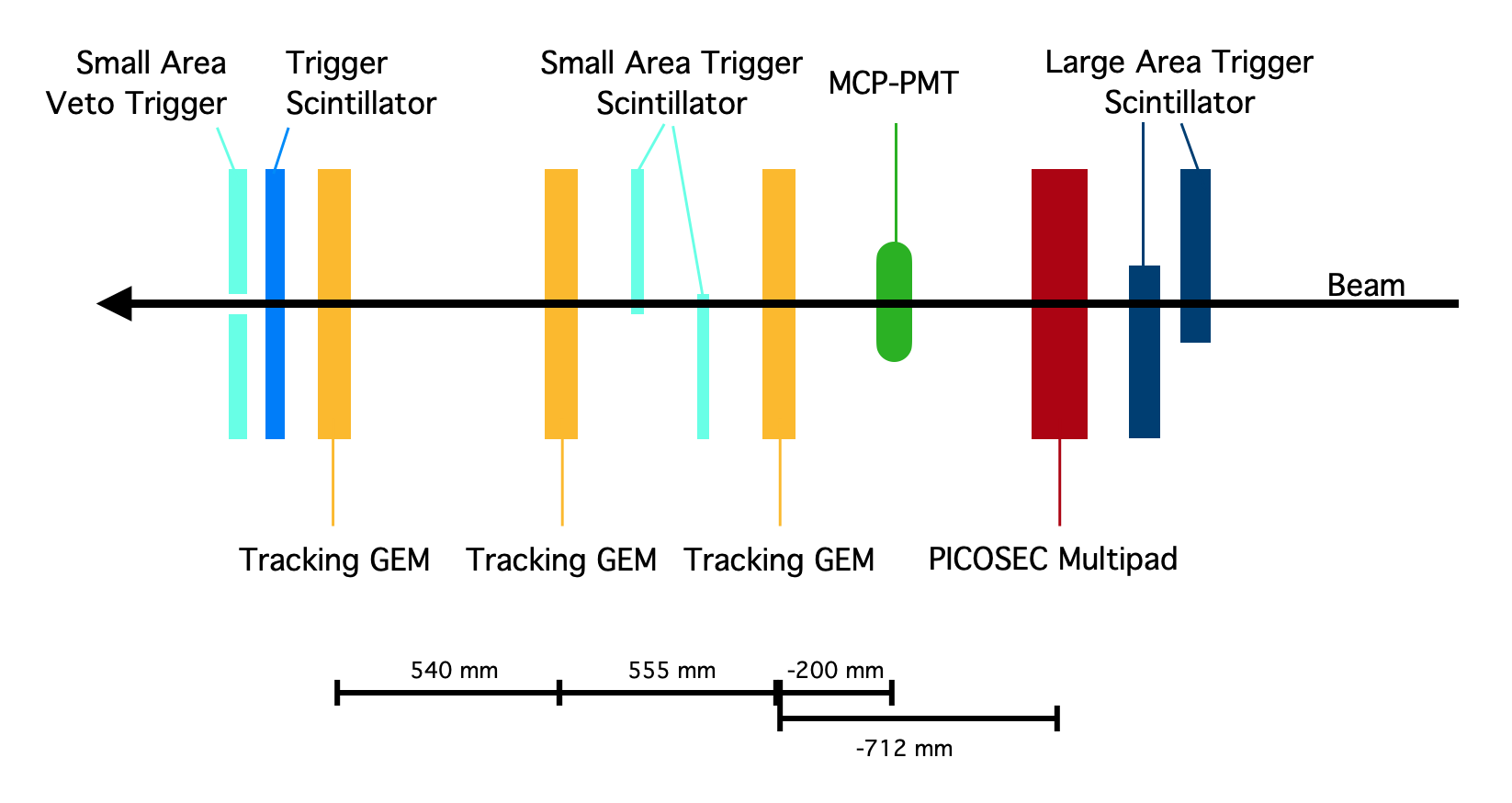}
    \caption{ Layout of the experimental setup (not to scale) during the beam tests. The incoming beam enters from the right side of the figure; events are triggered by either the coincidence of two 5 $\times$ 5 mm$^{2}$ scintillators in anti-coincidence with a ‘‘veto’’ scintillator (small area trigger) or by the coincidence of two 5 $\times$ 5 cm$^{2}$ scintillators (large area trigger). Three GEM detectors provide tracking information of the incoming charged
particles, and the MCP-PMTs provide the timing reference for the PICOSEC-Micromegas measurements. Details are given in the text.} 
    \label{fig:exp_setup}
\end{figure}

The digitized PICOSEC-Micromegas waveforms were processed according to the following steps as in \citep{pico24}: first by determining the baseline offset
and noise level using the 75 ns precursor of the pulse, then measuring the ‘‘electron-peak’’ amplitude ($V_{max}$)
as the difference between the maximum of the waveform and the baseline. Finally a logistic function is fitted to the leading
edge of the electron-peak to determine the CF20 (20\% constant fraction) time as the time when  the fitted function equals  20\% of $V_{max}$.
  
For the MCP digitized waveforms  a simpler approach is followed, as these signals are very fast and
almost noise-free: after the calculation of the
pulse baseline and amplitude, a linear interpolation between two points
around CF20 is used to extract the temporal position of the signal.

The ‘‘Signal Arrival Time’’ (SAT)  is  defined
as the difference between the  CF20 time of the PICOSEC-Micromegas pad and that of the MCP waveform, both digitized by the same oscilloscope. In this way, the SAT measurements are immune to any phase misalignment between the two oscilloscope clocks.

The ‘‘electron-peak charge’’ ($Q_{e}$) is defined by integrating a function that fits the electron-peak waveform.
Digitizations between the start and the end points of the electron-peak waveform (i.e. the first points situated
before and after the maximum whose amplitude is less than one standard
deviation away from the baseline offset\footnote{For those pulses with no clear
separation between the electron-peak and the ion-tail, the end point
has been alternatively defined as the time when the pulse derivative
changes sign.}) are fitted by a function, f(t),  defined as the difference of two generalized logistic functions:
\begin{equation}
f(t)=\dfrac{P_{0}}{\left[ 1+e^{-P_{1}(t-P_{2}) }\right]^{P_{3}} }- 
\dfrac{P_{0}}{\left[ 1+e^{-P_{4}(t-P_{5}) }\right]^{P_{6}} }
\label{eq:1}
\end{equation}
The terms, $ P_{0}, P_{1}, P_{2}, P_{3}, P_{4}, P_{5}$ and $P_{6}$ are free parameters in the fit of the electron-peak digitizations ($\sim$40 degrees of freedom). Eq. (\ref{eq:1}) is then numerically integrated  and the
 resulting value is transformed to Coulombs, using
the input impedance of 50 Ohm. The  electron-peak charge-to-amplitude
ratio, for all the instrumented PICOSEC-Micromegas pads,  is measured to be around 0.029 pC/mV.

\section{Detector aligment} \label{alignment}

We  reported \citep{mpgd2019} that the uncertainty in measuring the arrival time of MIPs  follows an ``inverse-square-root'' dependence on the number of photoelectrons, which initiate avalanches in the effective volume of the detector. In the case of a PICOSEC-Micromegas detector with segmented anode (Fig. \ref{fig:pad_geom}), the timing information provided by each pad should depend on the way that the  induced avalanches  are shared between the effective area of neighboring pads.  This sharing is determined by the  MIP impact point on the detector plane relative to the centres of the pads.  Thus, a precise knowledge of the detector position, relative to the global tracking coordinate system (hereafter called beam-frame), is necessary for evaluating the timing performance of the PICOSEC-Micromegas prototype.

\begin{figure}
\centering
\begin{minipage}{.48\textwidth}
\includegraphics[width=1.\textwidth]{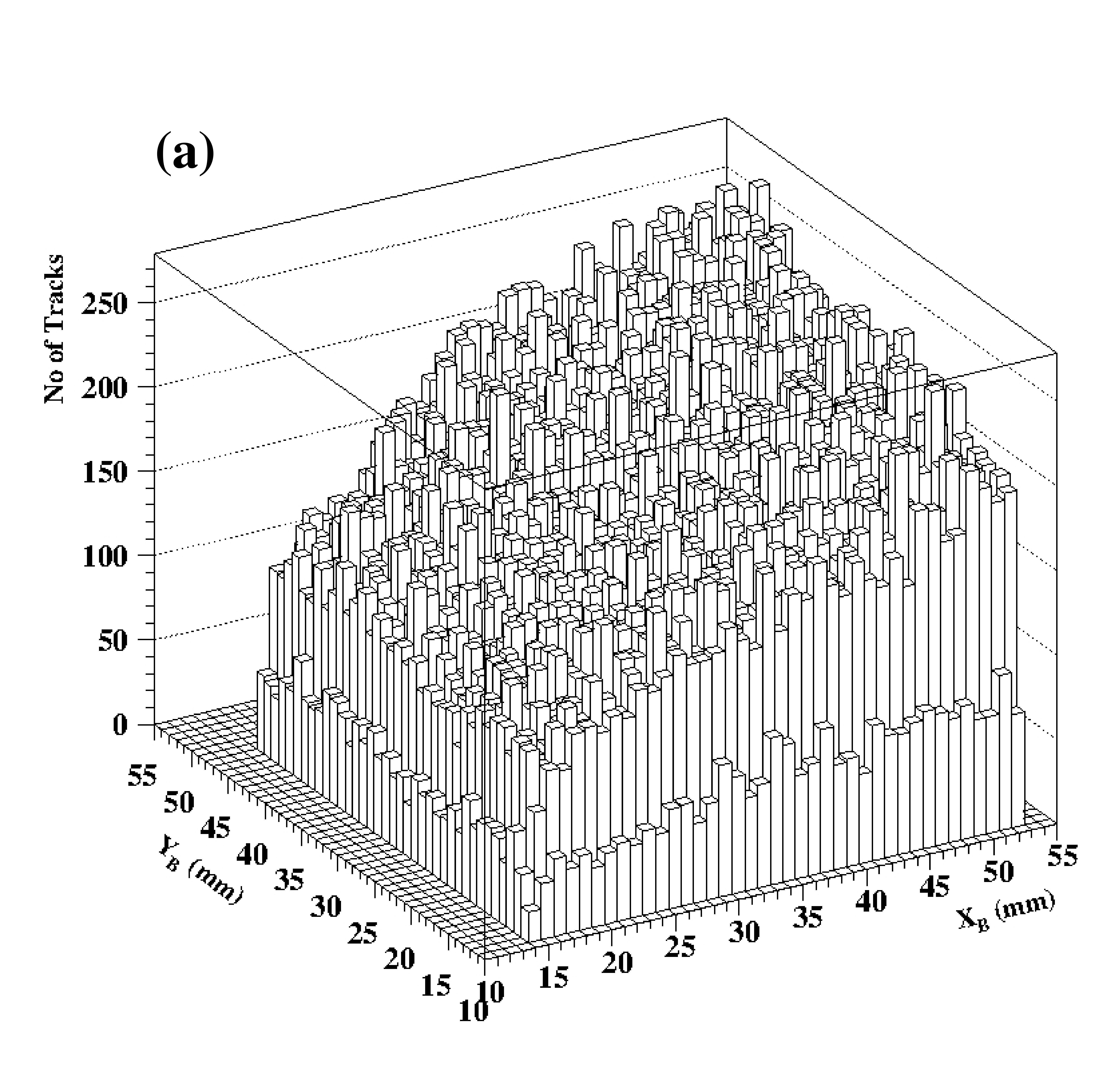}
\end{minipage}\\
\centering
\begin{minipage}{.48\textwidth}
\includegraphics[width=1.\textwidth]{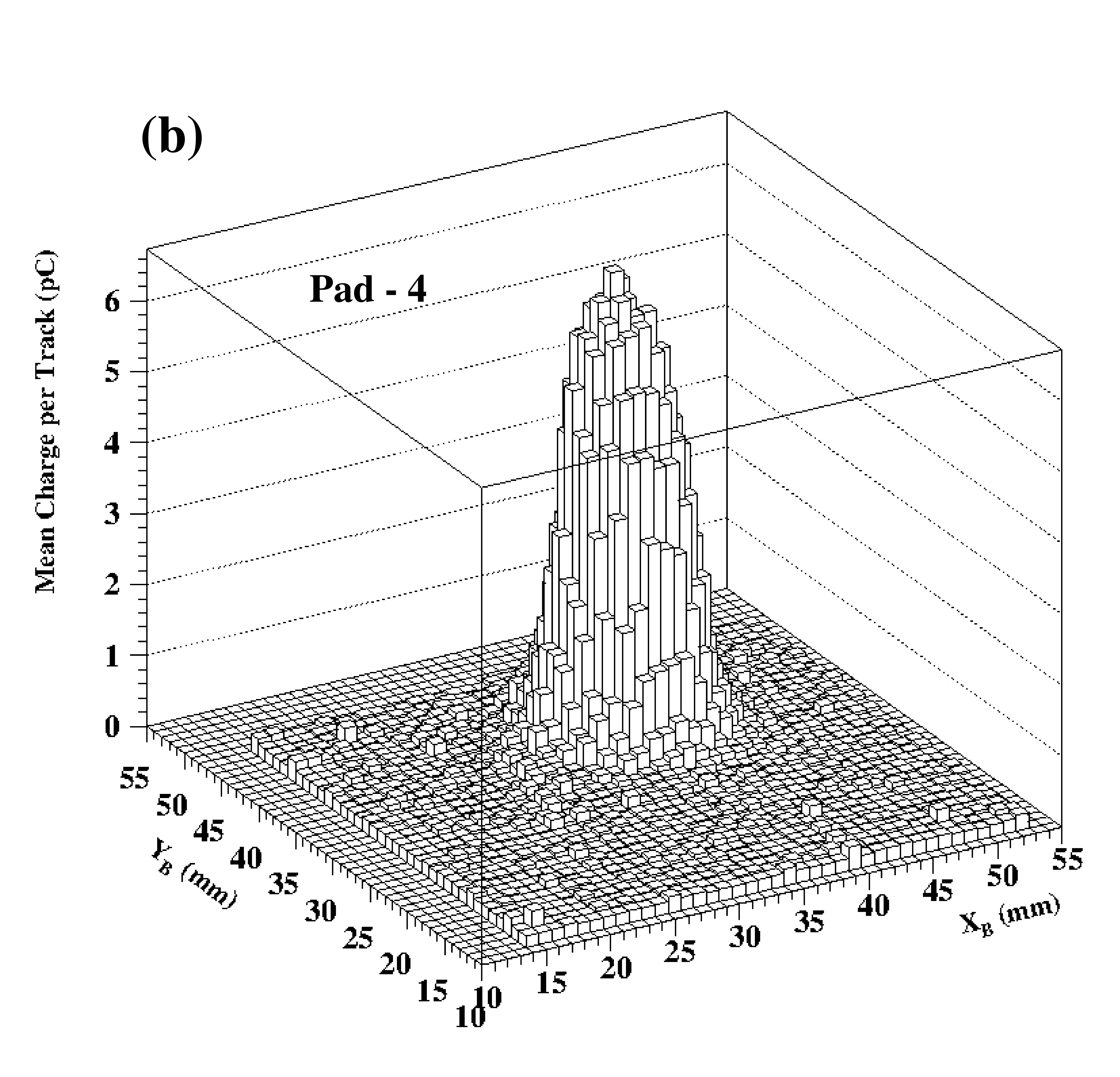}
\end{minipage}
\centering
\begin{minipage}{.48\textwidth}
\includegraphics[width=1.\textwidth]{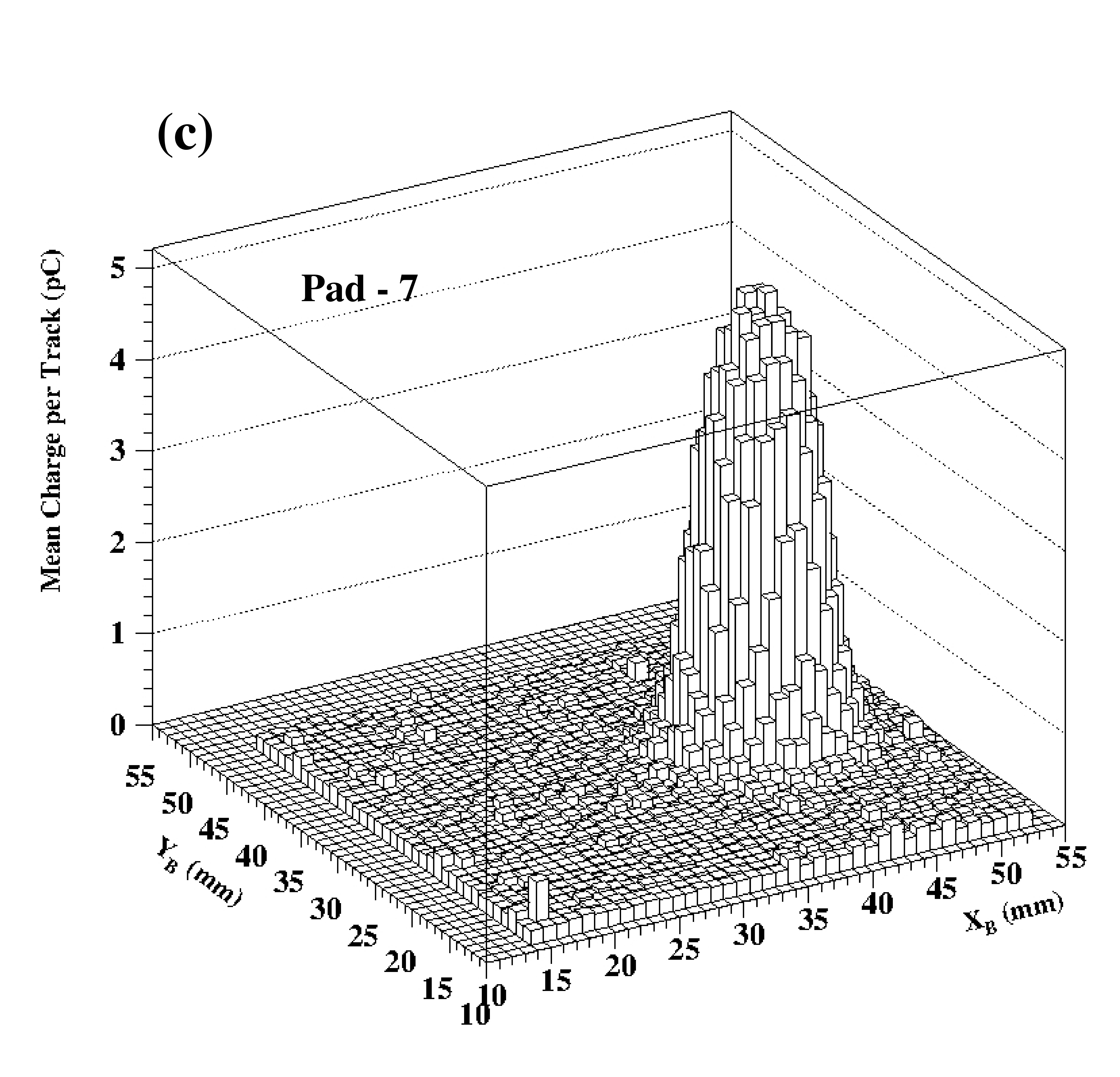}
\end{minipage}
\centering
\begin{minipage}{.48\textwidth}
\includegraphics[width=1.\textwidth]{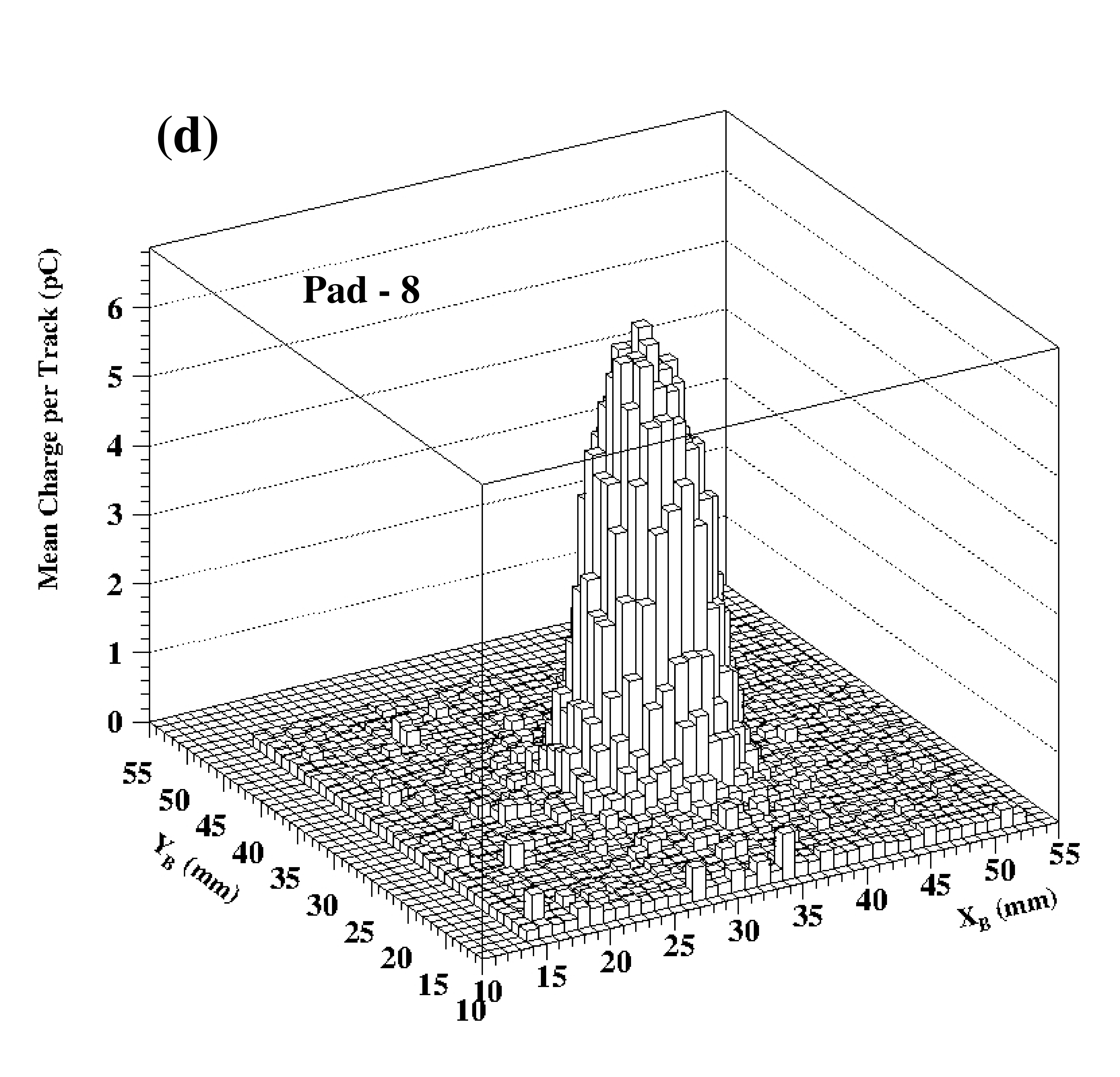}
\end{minipage}
\centering
\begin{minipage}{.48\textwidth}
\includegraphics[width=1.\textwidth]{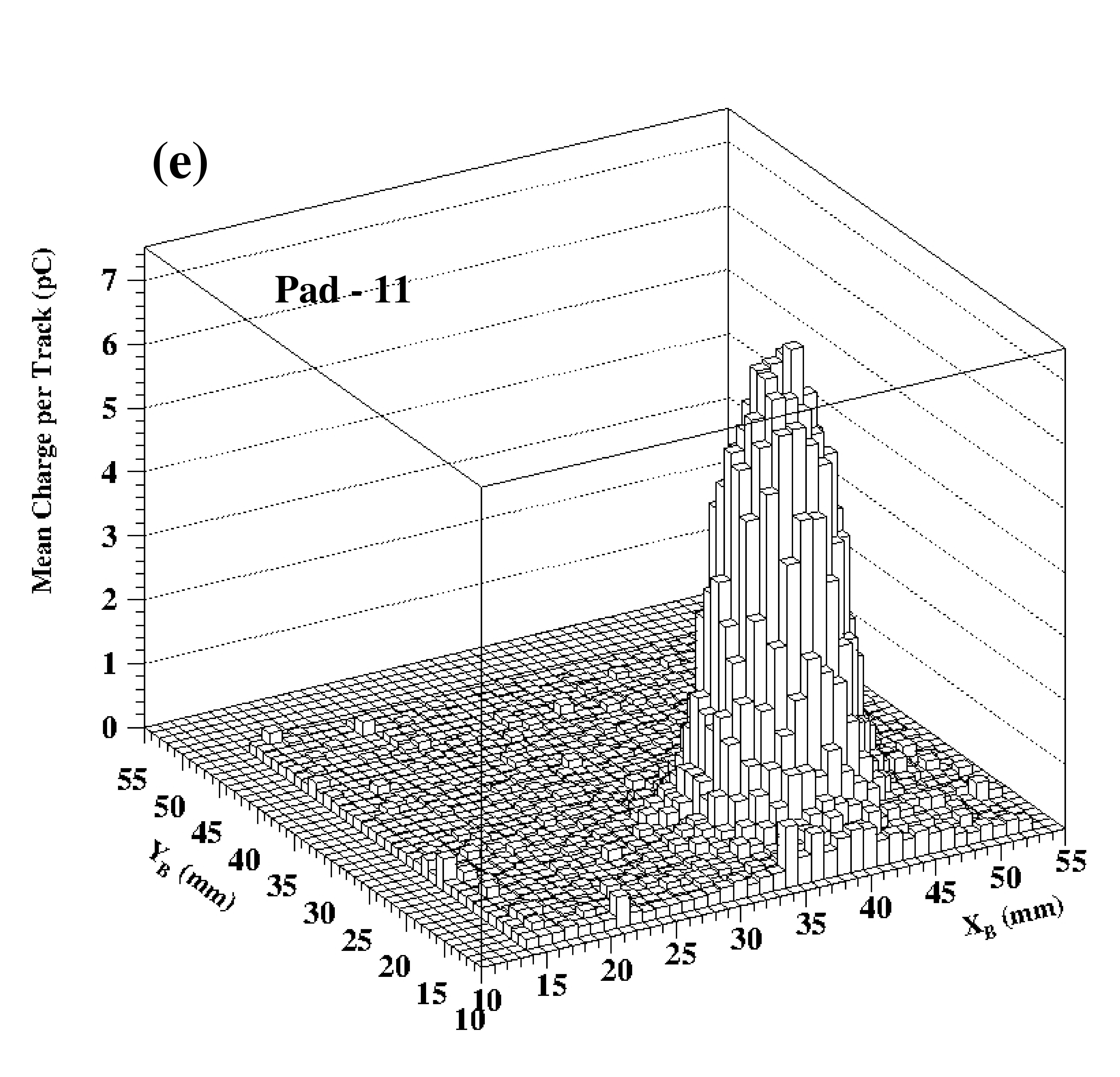}
\end{minipage}
\caption{The 2D distribution in plot-a represents  the beam profile that illuminates the area around  the instrumented PICOSEC-Micromegas pads. The distributions below show the mean value of the electron-peak charge versus the $\{ x_{B},y_{B} \}$ coordinates  where the MIPs traverse the detector.  The  distributions in plots b, c, d and e correspond to the response of pad No. 4, pad No. 7, pad No. 8 and pad No. 11, respectively. }
\label{fig:3dbeam}
\end{figure}

Special care was taken in placing the  PICOSEC-Micromegas detector in the experimental set-up with its radiator window perpendicular to the beam. The positions of the PICOSEC-Micromegas pad centres in the beam-frame  are estimated off-line, by analysing  the spatial distributions of the electron-peak charge ($Q_{e}$).
\\
 A first estimation of  a pad position is provided by the  dependence of the average $Q_{e}$, induced by MIPs that impact the PICOSEC-Micromegas  at   $\{ x_{B},y_{B} \}$  coordinates  in the beam-frame.  In principle, the  centroid of each 2D-histogram shown in Fig. \ref{fig:3dbeam}  provides an  estimation of the respective pad center position, which is expressed in terms of the corresponding abscissa  and bin contents as:
\begin{equation}
x_{c}=\dfrac{\sum\limits_{i=1,n}\langle Q_{e} \rangle_{i}x_{B}^{i}}{\sum\limits_{i=1,n}\langle Q_{e} \rangle_{i}}\quad , \quad y_{c}=\dfrac{\sum\limits_{i=1,n}\langle Q_{e} \rangle_{i}y_{B}^{i}}{\sum\limits_{i=1,n}\langle Q_{e} \rangle_{i}} \quad
\label{eq:2}
\end{equation}
where $n$ is the number of bins, each bin corresponding to a $ 1 \times 1$ mm$^{2}$  area around a point with $\{ x_{B}^{i}, y_{B}^{i}\}$ coordinates at the beam-frame and  $\langle Q_{e} \rangle_{i}$ is the mean electron-peak charge for tracks with impact points in the $i^{th}$ bin.

\begin{figure}[t]
    \centering
    \includegraphics[width=1.\textwidth]{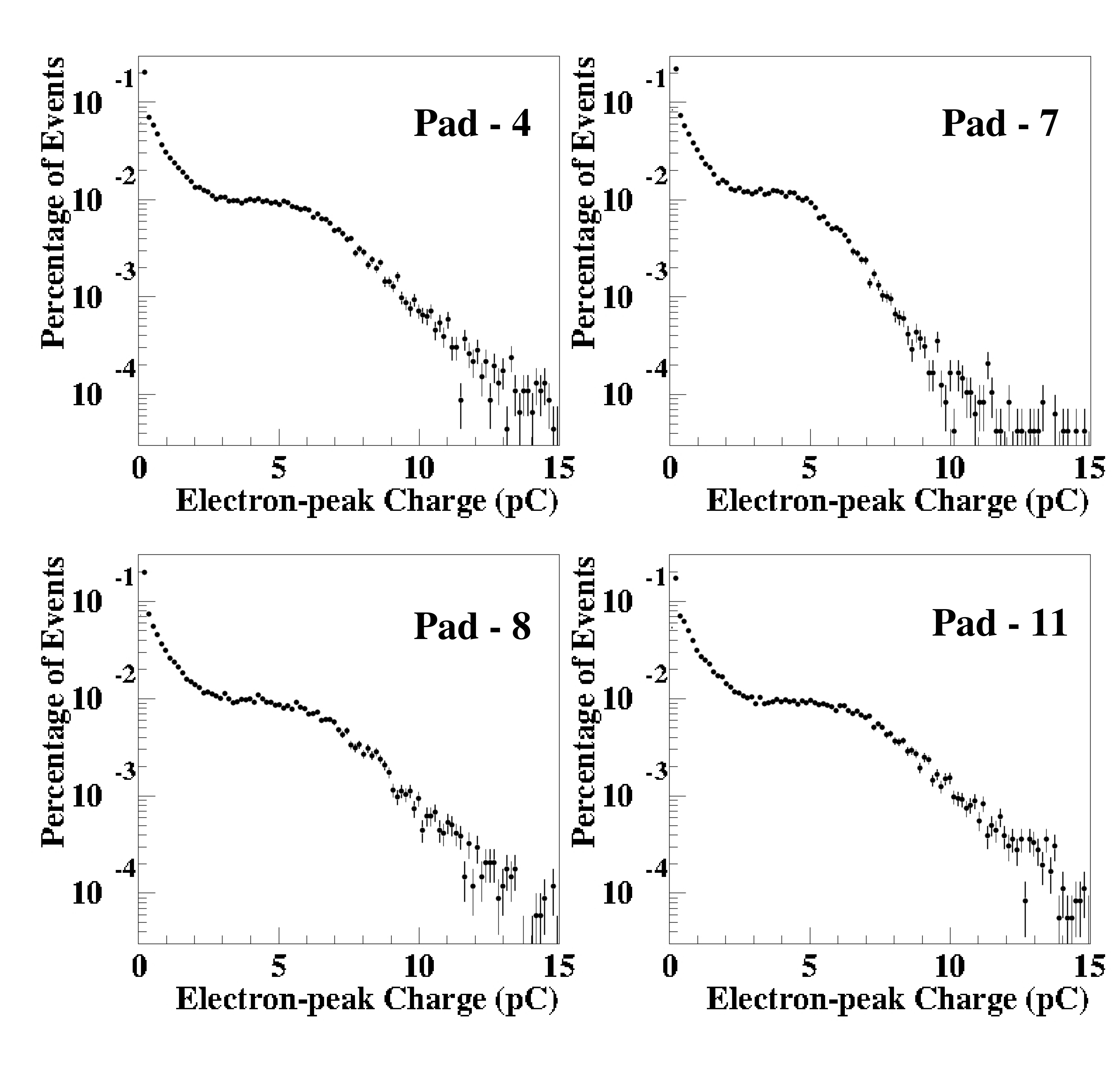}
    \caption{Electron-peak charge distributions of the instrumented pads, corresponding to tracks passing within 6 mm from the respective mean charge centroids (see text).} 
    \label{fig:charge_raw}
\end{figure}

As it is shown in Fig. \ref{fig:charge_raw}, the $Q_{e}$ distributions related to  the selected tracks are dominated by small pulses and noise; nevertheless, the 2D charge-weighted beam profiles  (Fig. \ref{fig:3dbeam}) clearly indicate the position of the pad centres. Furthermore, since the beam illuminates the whole active area of all PICOSEC-Micromegas  pads\footnote{  Also because, a) the hexagonal pad shape is 2D symmetric and b) the   Cherenkov photons on the photocathode are symmetrically distributed  around the MIP track.} the estimation by Eq. (\ref{eq:2}) is unbiased.  However, because the two-dimensional binning of the events reduces the statistical accuracy of the estimation, Eq. (\ref{eq:2}) is only utilized to approximate the starting values, $\{ x_{c}^{k}, y_{c}^{k}  \}$ (hereafter, k=1,2,3,4  stands for pad No. 4, 7, 8 and 11, respectively)  for a more accurate estimation procedure.\\

% which also takes  into account the constraints provided by the  the anode segmentation geometry.

We define the  local-frame  as the coordinate system with its origin at the centre of pad No. 8 and its axes, $X_{L}$ and $Y_{L}$, parallel to the principal, symmetry axes of the hexagonal pads (Fig. \ref{fig:pad_geom}).  In the local-frame, the pad centre coordinates  ($\{ x_{L}^{k}(d), y_{L}^{k}(d) \} , k=1,2,3,4$)  are completely defined in terms of the hexagonal pad dimensions and the width, d, of the  gap between neighboring pads. Then, the pad centres coordinates ($\{ x_{B}^{k}, y_{B}^{k} \}, k=1,2,3,4)$ in the beam-frame are expressed in terms of the local-frame variables as:
\begin{equation}
\begin{split}
x_{B}^{k}(x_{o}, y_{o}, \vartheta, d)=(x_{L}^{k}(d)-x_{o})cos(\vartheta)+(y_{L}^{k}(d)-y_{o})sin(\vartheta) \\
y_{B}^{k}(x_{o}, y_{o}, \vartheta, d)=(y_{L}^{k}(d)-y_{o})cos(\vartheta)-(x_{L}^{k}(d)-x_{o})sin(\vartheta)
\end{split}
\label{eq:3}
\end{equation}
where $\{ x_{o}, y_{o} \}$ and $\vartheta$ are the coordinates of the beam-frame origin  and the azimuth angle of the $X_{B}$ axis in the local-frame, respectively.

We use  the  coordinates of the centroids as ``data points'' (i.e. $ x_{data}^{k}=x_{c}^{k}$ and $y_{data}^{k}=y_{c}^{k} $ for $k=1,2,3,4$) for an initial evaluation of the  $ x_{o}$, $y_{o}$ , $\vartheta$ and d parameter values,  by minimizing the following $\chi ^{2}$ estimator:

\begin{equation}
\chi ^{2}=\sum\limits_{k=1,4}\dfrac{(x_{data}^{k}-x_{B}^{k}(x_{o}, y_{o}, \vartheta, d))^{2} }{(\sigma_{x}^{k})^{2}} 
+\sum\limits_{k=1,4}\dfrac{(y_{data}^{k}-y_{B}^{k}(x_{o}, y_{o}, \vartheta, d))^{2} }{(\sigma_{y}^{k})^{2}}
\label{eq:4}
\end{equation}
where $x_{B}^{k}(x_{o}, y_{o}, \vartheta, d)$ and $y_{B}^{k}(x_{o}, y_{o}, \vartheta, d)$ are given by Eq. (\ref{eq:3}), whilst $\sigma_{x}^{k}$ and $\sigma_{y}^{k}$ are approximated by propagating the $\langle Q_{e} \rangle_{i}$ statistical errors in Eqs. (\ref{eq:2})\footnote{ That is $\sigma^{2}_{x}=\dfrac{\sum\limits_{i=1,n}\delta^{2}_{i}\left( x_{c}-x_{B}^{i}\right)^{2} }{\sum\limits_{i=1,n}\langle Q_{e} \rangle_{i}}$ and $\sigma^{2}_{y}=\dfrac{\sum\limits_{i=1,n}\delta^{2}_{i}\left( y_{c}-y_{B}^{i}\right)^{2} }{\sum\limits_{i=1,n}\langle Q_{e} \rangle_{i}}$, where $\delta_{i}$ denotes the statistical error in evaluating $\langle Q_{e} \rangle_{i}$}. 
The estimated parameter values ($ \hat{x}_{o}, \hat{y}_{o} , \hat{\vartheta}, \hat{d}$) also define the pad centre positions in the beam-frame via Eq. (\ref{eq:3}).
\\

The above estimations of the  pad centre positions and of the  azimuth angle, $\hat{\vartheta}$, are used as input values to a two-step, iterative algorithm. In the first step, each pad is treated  independently; its centre coordinates are estimated by fitting the charge-weighted, beam profiles projected to the pad principal, symmetry axes. The constraints arising from the anode segmentation geometry  (Eq. (\ref{eq:3})) are imposed  in the second step, by minimizing the $\chi^{2}$ expression of Eq. (\ref{eq:4}) using the results of the first step as ``data points''.

\begin{figure}
\centering
\begin{minipage}{.48\textwidth}
\includegraphics[width=1.\textwidth]{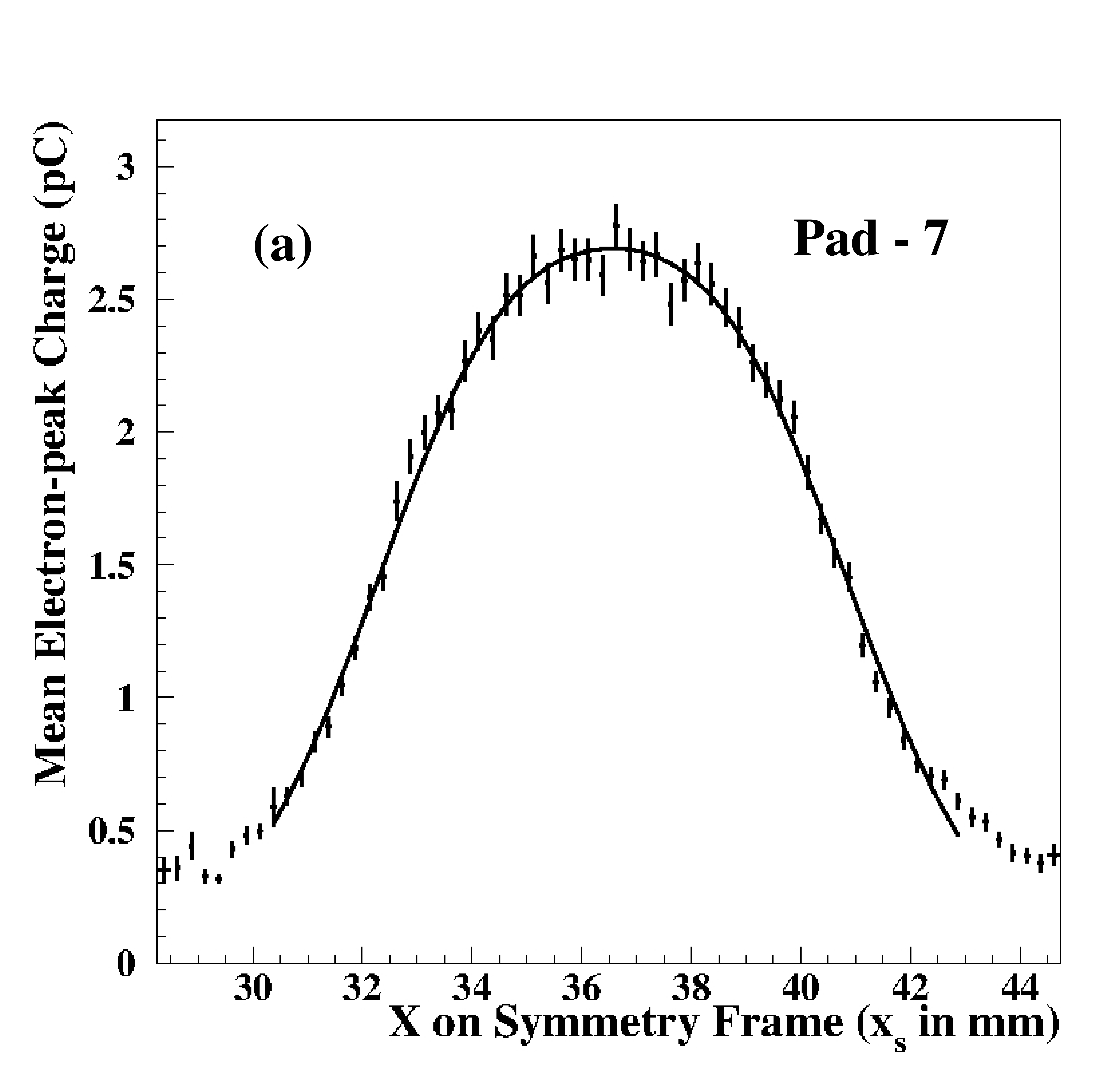}
\end{minipage}
\centering
\begin{minipage}{.48\textwidth}
\includegraphics[width=1.\textwidth]{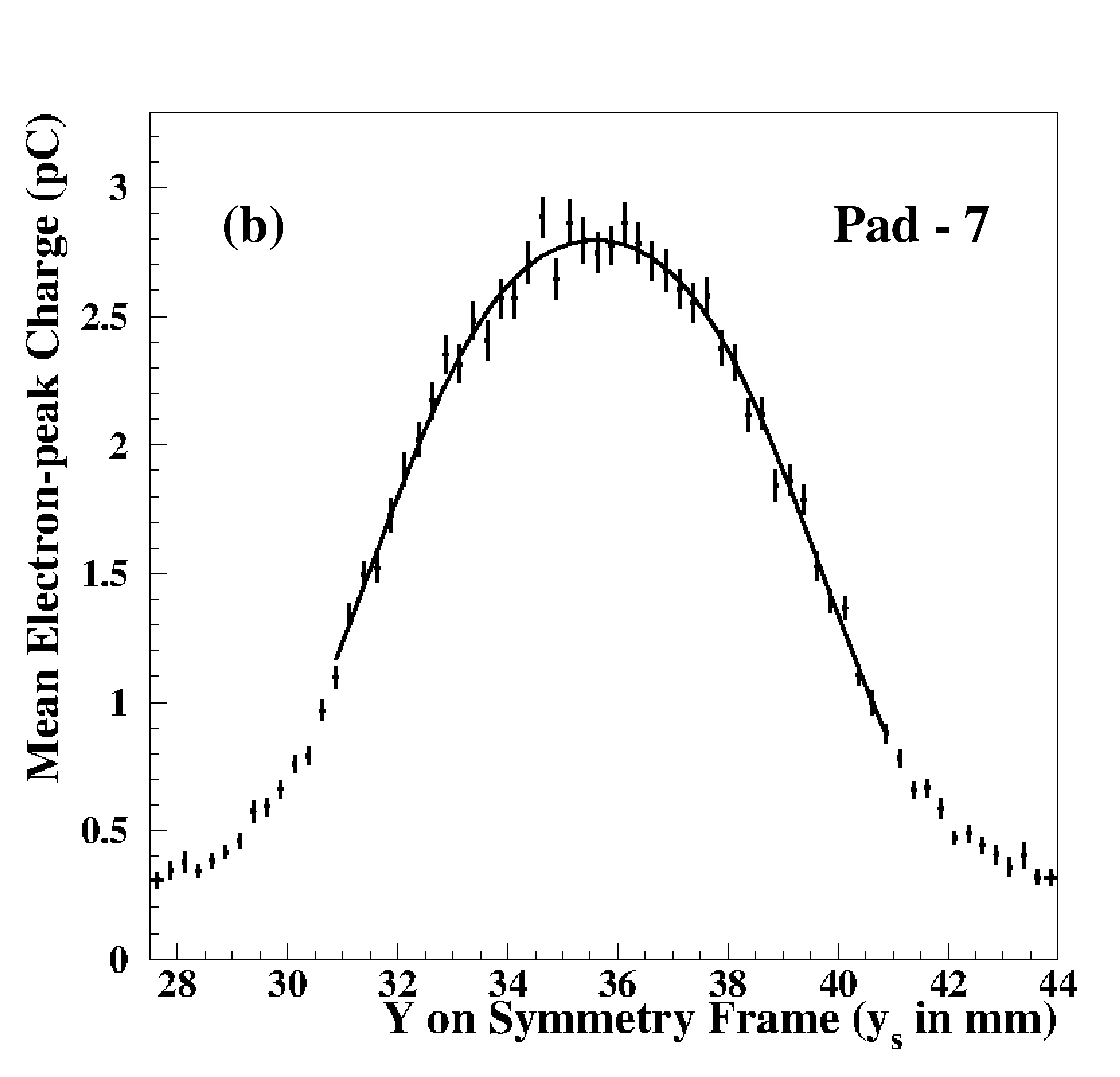}
\end{minipage}
\centering
\begin{minipage}{.48\textwidth}
\includegraphics[width=1.\textwidth]{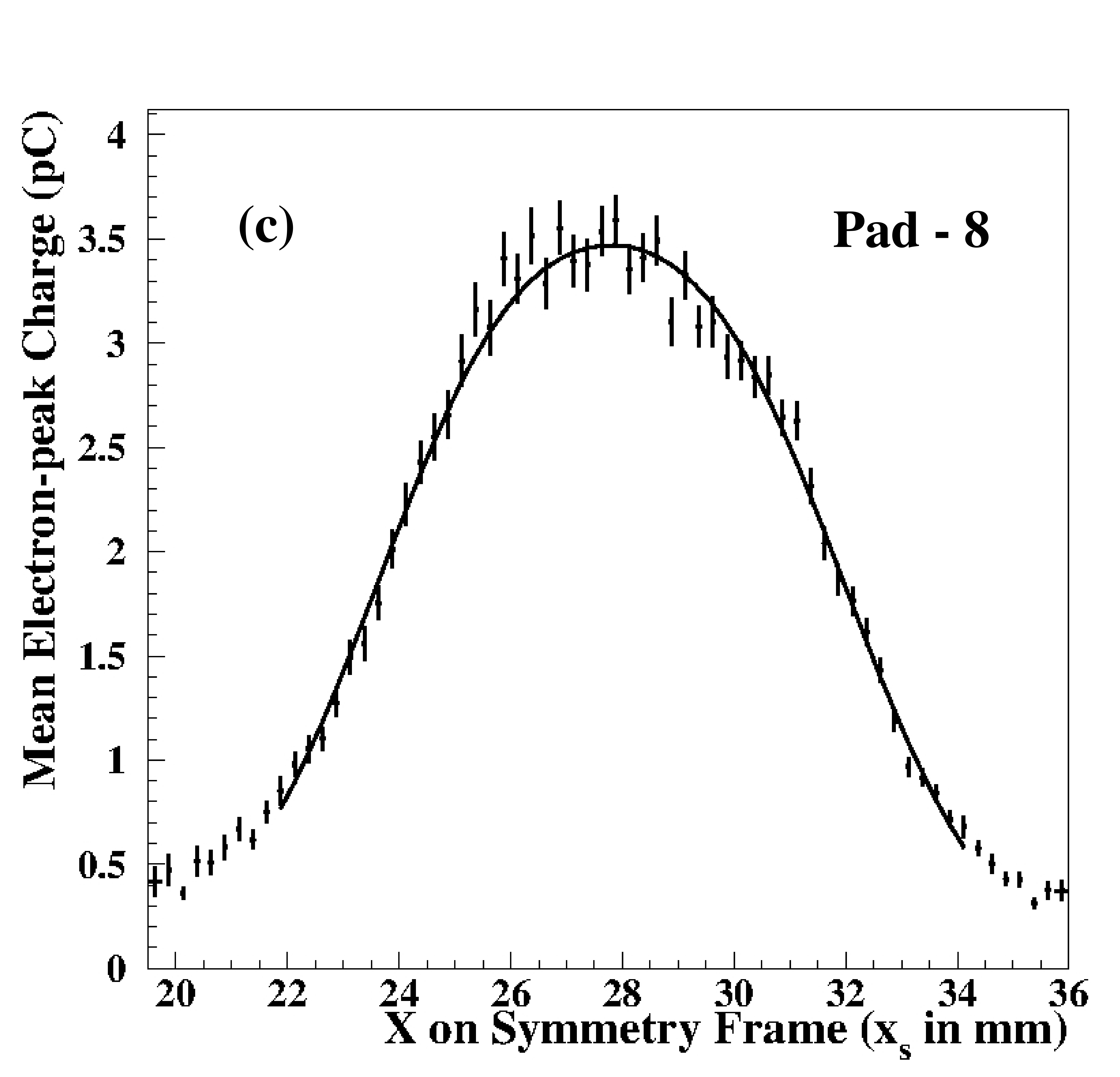}
\end{minipage}
\centering
\begin{minipage}{.48\textwidth}
\includegraphics[width=1.\textwidth]{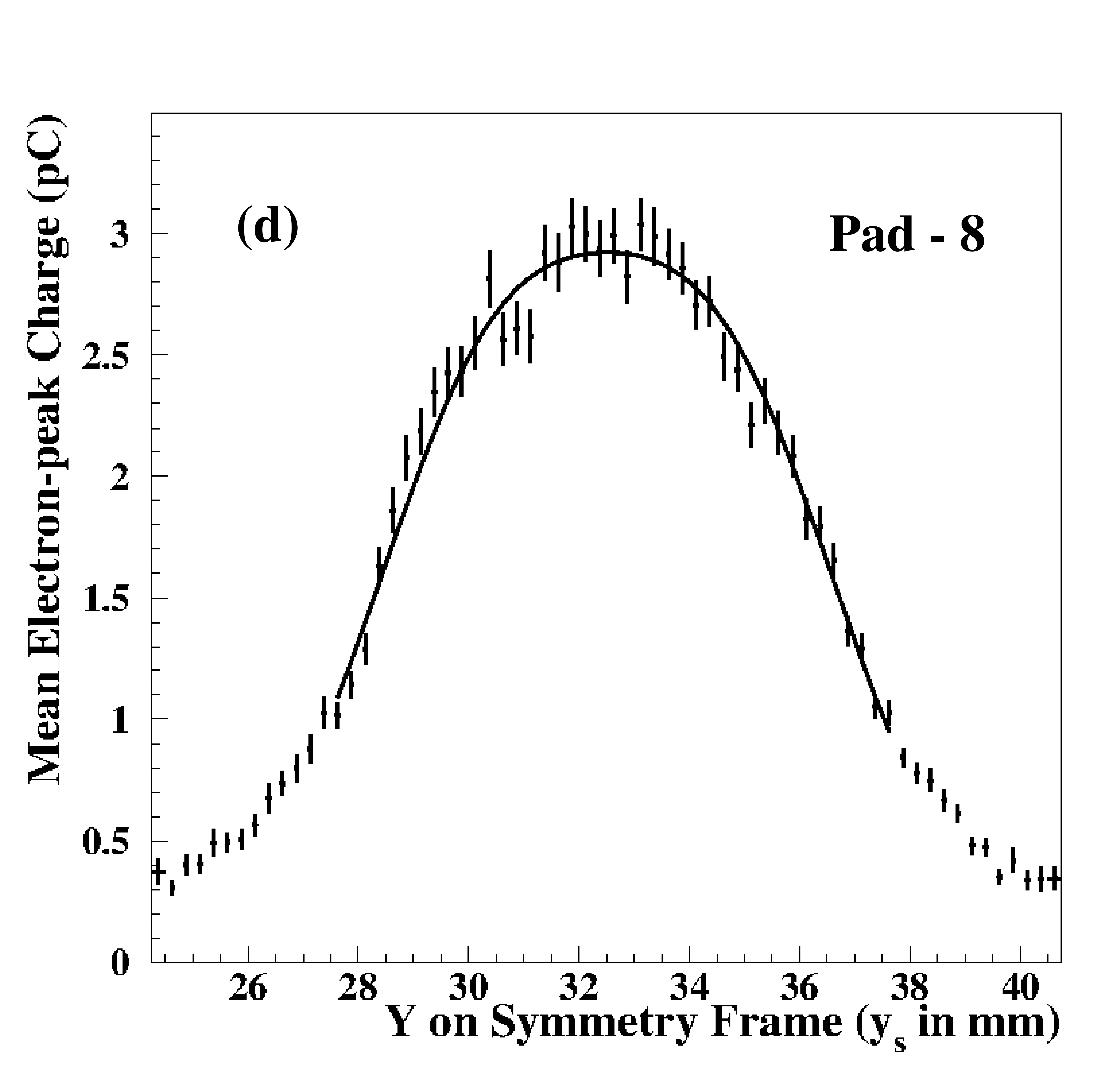}
\end{minipage}
\caption{The mean value of the electron-peak charge, when the pad responds to MIPs entering  the PICOSEC-Micromegas radiator at points with $x_{S}$ (plots a and c) or $y_{S}$ (plots b and d) coordinate, relative to the  symmetry axes (see text). The upper and lower rows correspond to pads No. 7 and 8 respectively.  }
\label{fig:1dprj}
\end{figure}

\begin{description}
\item[First Step:] Use the previously estimated values of the pad centre coordinates to select MIPs that pass through a disk of 6 mm radius around the pad centre. Use the previously estimated value of the angle $\vartheta$ ($\vartheta=\hat{\vartheta}_{I}$)  to transform the beam coordinates to the symmetry axes, $X_{S}$ and $Y_{S}$  (Fig. \ref{fig:pad_geom}):
\begin{equation}
\begin{split}
x_{S}^{k}=x_{B}^{k}cos(-\vartheta)+x_{B}^{k}sin(-\vartheta) \\
y_{S}^{k}=y_{B}^{k}cos(-\vartheta)-x_{B}^{k}sin(-\vartheta)
\end{split}
\label{eq:5}
\end{equation}
Determine  the   $Q_{e}$ weighted beam profiles, projected along the $X_{S}$ and  the $Y_{S}$ symmetry axes, i.e. the average $Q_{e}$ of the pad response to a MIP versus the respective $x_{S}$ or $y_{S}$ coordinate of the track impact point.  Due to the symmetry of the anode segments with respect to the $X_{S}$ and $Y_{S}$ axes  and the azimuthal symmetry of the Cherenkov disk, the above 1D profiles are symmetric around the respective centre coordinate (Fig. \ref{fig:1dprj}). We explore this symmetry by fitting  Eq. (\ref{eq:6})  to the central part of the respective $X_{S}$ and $Y_{S}$ , $Q_{e}$-weighted, beam profiles. 
\begin{equation}
\begin{split}
\left\langle Q_{e}\right\rangle_{X_{S}}=N_{X}\left[ e^{-\dfrac{(x_{S}-\mu_{x1})^{2}}{2\varrho_{x}^{2}}} + e^{-\dfrac{(x_{S}-\mu_{x2})^{2}}{2\varrho_{x}^{2}} }                      \right] \\
\left\langle Q_{e}\right\rangle_{Y_{S}}=N_{Y}\left[ e^{-\dfrac{(y_{S}-\mu_{y1})^{2}}{2\varrho_{y}^{2}}} + e^{-\dfrac{(y_{S}-\mu_{y2})^{2}}{2\varrho_{y}^{2}} }                      \right]
\end{split}
\label{eq:6}
\end{equation}
Accordingly, the pad centre coordinates  at the $X_{S}$ and $Y_{S}$  axes are defined as, $0.5(\hat{\mu}_{x1}+\hat{\mu}_{x2})$ and $0.5(\hat{\mu}_{y1}+ \hat{\mu}_{y2})$, respectively.
%\footnote{Several other symmetric functional forms have been tried to fit the 1D, $Q_{e}$-weighted, beam %profiles, without any significant change to the results. }. 
The corresponding coordinates at the beam-frame are evaluated by the inverse transformation of Eq. (\ref{eq:5}). The procedure is applied for all instrumented pads of the PICOSEC-Micromegas prototype, resulting in  independent estimations of the  $x_{B}^{k}$ and $y_{B}^{k}$ ($k=1,2,3,4$) centre coordinates of all pads.

\item[Second Step:] Using  the   first step results (and the related estimation uncertainties)  as ``data points'',  the values of the  $ x_{o}$, $y_{o}$, $\vartheta$ and d parameters are estimated by minimizing the $\chi ^{2}$ expression defined in Eq. (\ref{eq:4}). The pad positions in the beam-frame are then evaluated by inserting the above estimated values in Eq. (\ref{eq:3}). 

\end{description}
The iterative process is considered complete when  the pad positions found in the second step are the same, within the estimated statistical errors, with the results of the preceding first step. Otherwise, the values of the pad centres coordinates and of the angle  $\vartheta$, found in this trial, are used as input values to the first step, for a new iteration.
\\

\begin{figure}
\centering
\begin{minipage}{.48\textwidth}
\includegraphics[width=1.\textwidth]{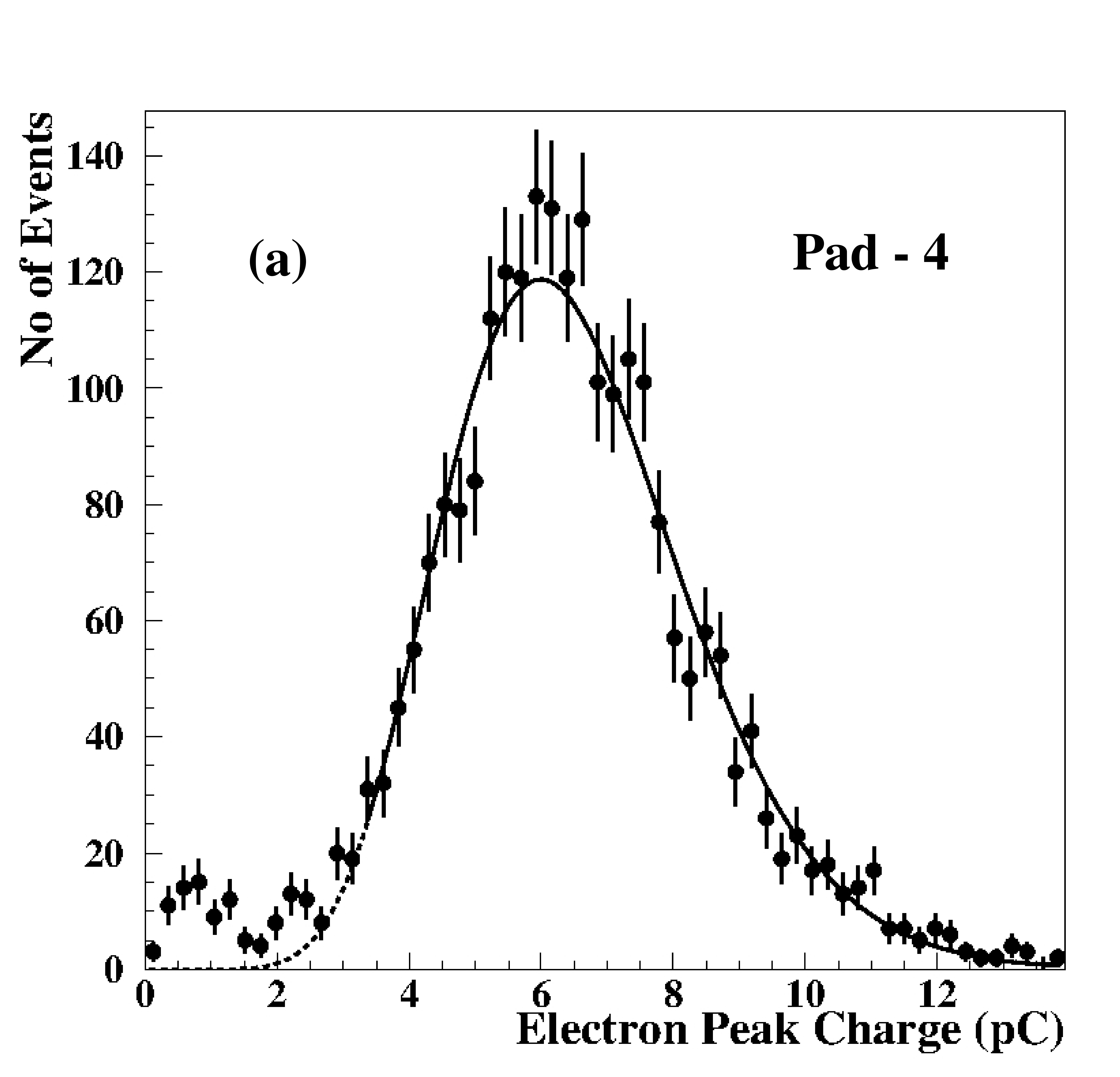}
\end{minipage}
\centering
\begin{minipage}{.48\textwidth}
\includegraphics[width=1.\textwidth]{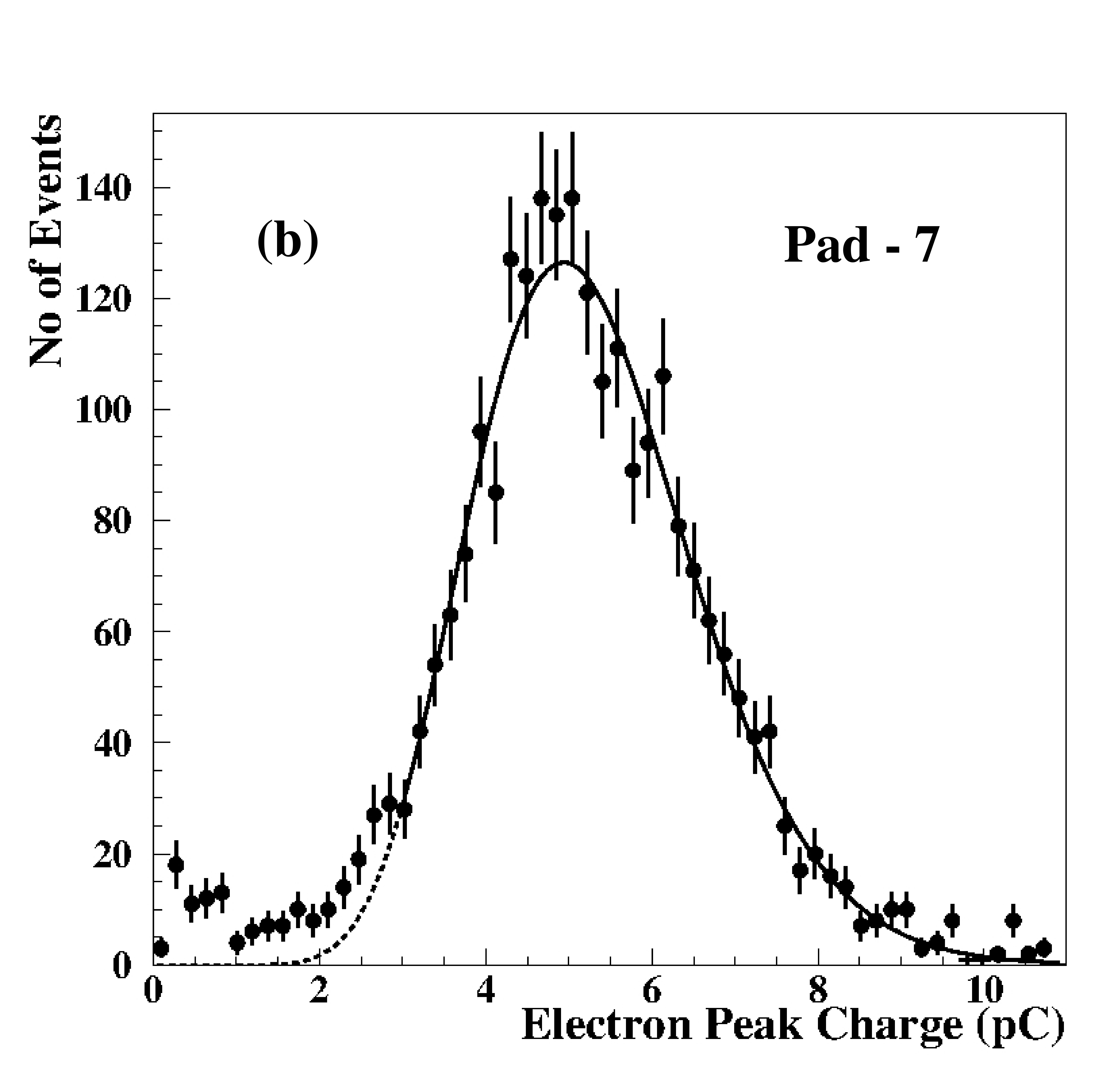}
\end{minipage}
\centering
\begin{minipage}{.48\textwidth}
\includegraphics[width=1.\textwidth]{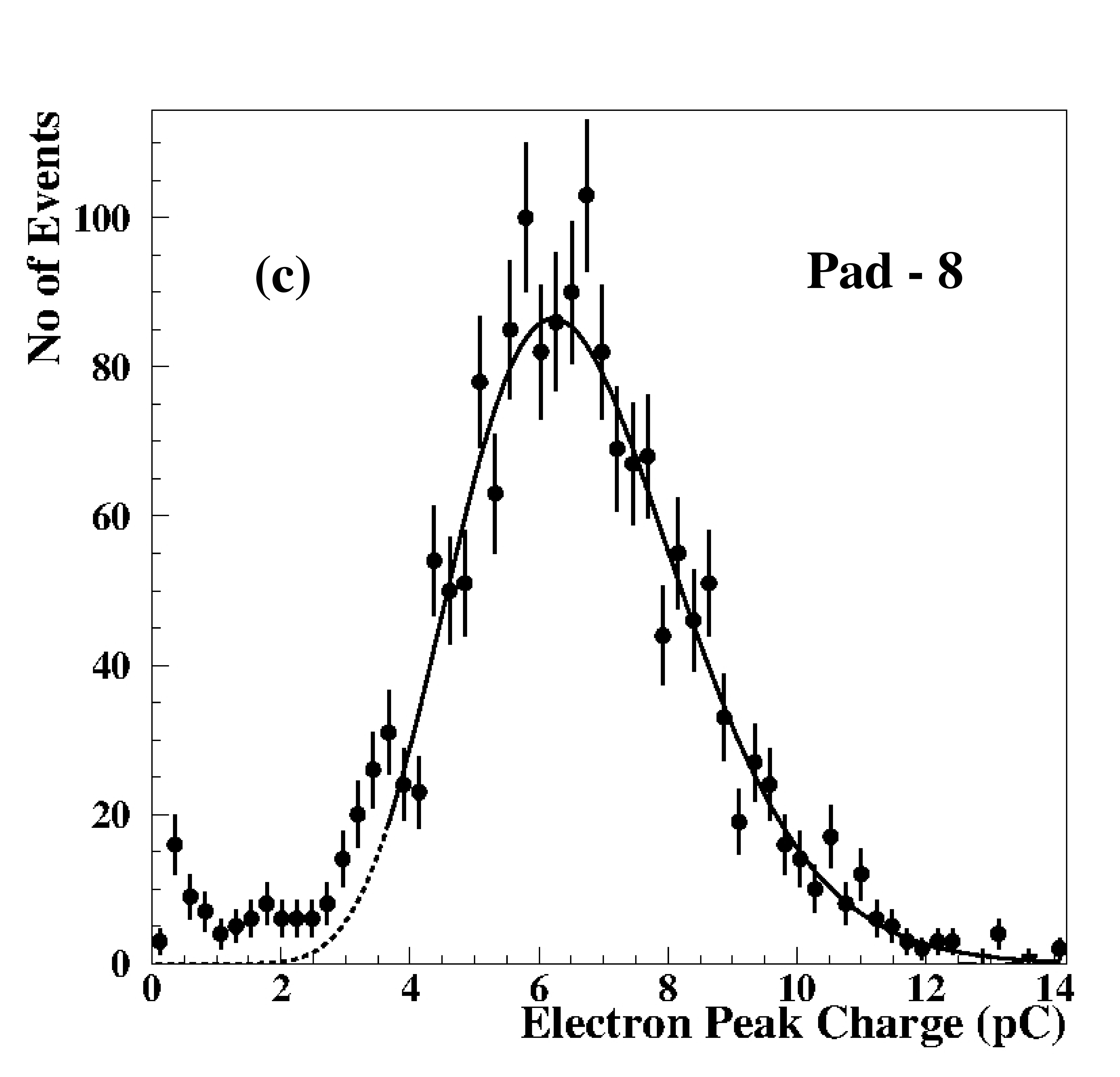}
\end{minipage}
\centering
\begin{minipage}{.48\textwidth}
\includegraphics[width=1.\textwidth]{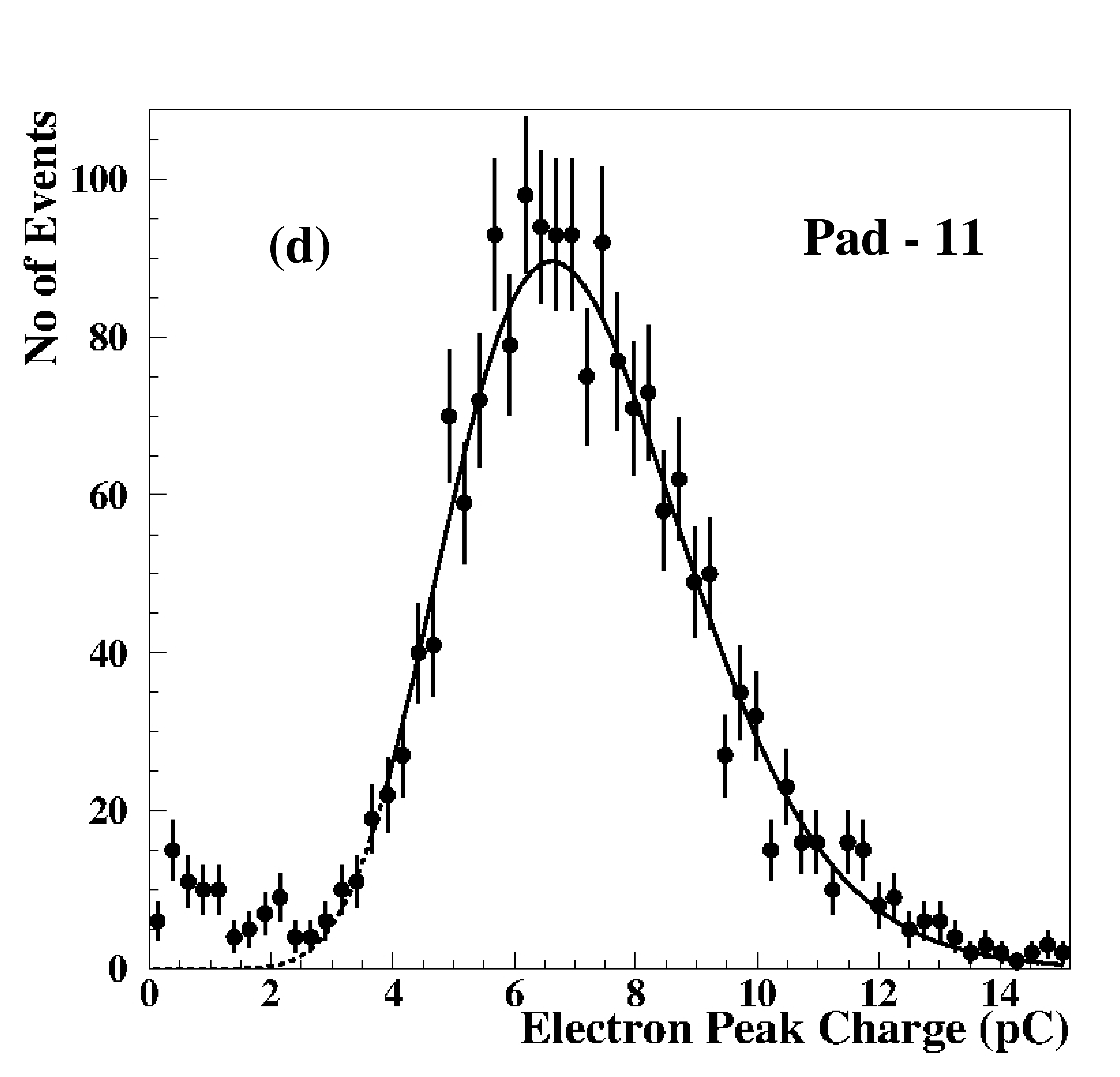}
\end{minipage}
\caption{Electron-peak charge distributions of pads No. 4 (plot a), No. 7 (plot b), No. 8 (plot c) and No. 11 (plot d), when respond to MIPs passing closer than 2 mm from the respective pad centre. The solid lines represent fits with a Gamma Distribution function, with estimated mean values equal to 6.54 ($\pm$ 0.04) pC, 5.29 ($\pm$ 0.03) pC, 6.70 ($\pm$ 0.05) pC and 7.20 ($\pm$ 0.05) pC for pads No. 4, 7, 8 and 11, respectively. }
\label{fig:in_q}
\end{figure}

We  checked for possible systematic biases of the above  method by: a) employing other symmetric functional forms instead of Eq. (\ref{eq:6}), b)  selecting  to fit different central regions of the 1D $Q_{e}$-weighted beam profiles,  and  c)  changing (by three sigmas) the starting values provided by the initial stage.  The alignment algorithm converges each time after the second iteration, resulting consistently to almost the same estimation of the pad centre coordinates. The minimum $\chi^{2}$ per degree of freedom (p.d.f.) remains always around one, whilst the estimation errors, including the above  systematic variations, are found to be less than $100\,\textrm{\selectlanguage{greek}m\selectlanguage{english}m}$. However, the estimated value of the parameter d  is found to be   $430 \pm 70\,\textrm{\selectlanguage{greek}m\selectlanguage{english}m}$, which is large compared to the design value ($\sim 200\,\textrm{\selectlanguage{greek}m\selectlanguage{english}m}$). By setting  the parameter d equal to  $200\, \textrm{\selectlanguage{greek}m\selectlanguage{english}m}$  the algorithm converges to coordinate values  that differ by less than $100\,\textrm{\selectlanguage{greek}m\selectlanguage{english}m}$ from the respective values when d was treated as a free parameter, whilst the   minimum $\chi^{2}$ p.d.f.  is getting  more than three times larger. 
\\

\begin{figure}[t]
    \centering
    \includegraphics[width=1.\textwidth]{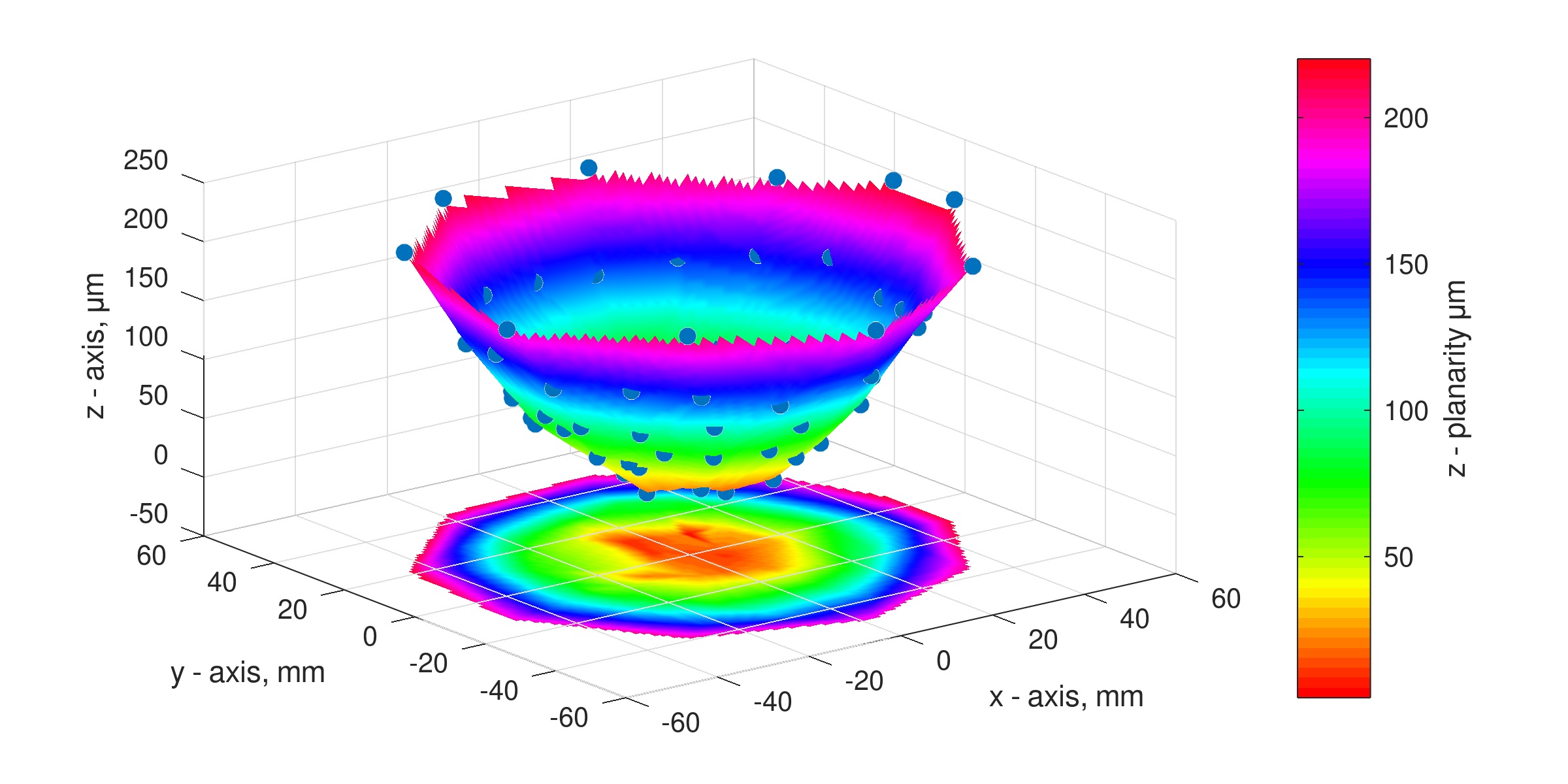}
    \caption{The radial deformation of the multi-pad PICOSEC-Micromegas anode has been measured with a ZEISS O-INSPECT 863 (length measurement error in $\textrm{\selectlanguage{greek}m\selectlanguage{english}m}$: 2.2 + L/150mm) at the CERN Metrology service. The shown measurements have been performed on the PCB installed on the aluminium flange.} 
    \label{fig:deform}
\end{figure}

Fig. \ref{fig:in_q} presents the $Q_{e}$ distributions of the instrumented PICOSEC-Micromegas pads, responding to MIPs with impact points within 2 mm from the respective pad centres, i.e. the vast majority  of the Cherenkov induced signal is contained in a single pad. As evident in Fig. \ref{fig:charge_raw}, the central pad No. 7 has a  lower gain than the other pads (hereafter pads No. 4, 8 and 11 are referred as peripheral pads). We  verified with laboratory tests that this  difference is due to the spatial variation of the detector gain and it is independent of the external electronics. Microscopic inspection of the  anode, revealed a non-flat, sagging surface  with the lower tip close to the centre of pad No. 7, as it is shown in Fig. \ref{fig:deform}. As the micromesh is attached to the anode by the pillars, the depth of the amplification gap remains constant over the entire anode surface. However, as the cathode is planar, the  preamplification (drift) gap is larger in the region above pad No. 7 than above the other pads. Therefore, despite the constant drift voltage, the drift field is lower above pad No. 7 as compared to the region above the peripheral pads, resulting in a gain variation as indicated in Fig. \ref{fig:in_q}. Furthermore, as indicated in Fig. \ref{fig:deform},  the drift field   would also vary across the surface of any peripheral pad, i.e. a preamplification avalanche that develops  further away from pad No. 7 feels a higher drift field than  those developing closer to the central pad. Such a gain asymmetry affects the  alignment of the peripheral pads by introducing biases outwards from pad No. 7. These biases result in an overestimation of the gap width, d, between neighboring pads, as reported earlier. Although, the drift field variation across the anode surface causes minor effects on the detector alignment, it has significant consequences on the timing performance,  discussed in the next Section.

\section{Timing characteristics and flatness corrections of individual pads} \label{spad}

PICOSEC-Micromegas is a standalone detector which performs precise timing measurements  without requiring  external information on the track of the incoming particle. 
%It is also desirable  for the timing accuracy of the detector to be independent of how the Cherenkov photon disk is distributed among neighboring,  anode segments.  
In the case that the Cherenkov ring is shared among neighboring pads, a proper combination of the individual  timing measurements (see Section \ref{multicomb}), corrected for SAT shifts and weighted by their expected accuracy, provides precise timing of the incoming particles. However, the a priori knowledge of the individual measurements expected  accuracy and time shifts  is highly important for the combined timing. Since the timing accuracy and the SAT shift of a single pad depend explicitly on $Q_{e}$  \citep{mpgd2019,pico24, model} such dependences have to be quantified in a calibration phase prior to the detector operation.  
%It is highly desirable, especially for large area detectors, that all  anode segments of a PICOSEC-Micromegas module share the same timing characteristics. Such a uniformity simplifies the  calibration procedure. 
%\footnote{The timing resolution of a single pad depends on the number of photoelectrons that induce the signal on the pad \citep{mpgd2019}. Furthermore, the contribution of any single photoelectron to the timing resolution depends on the number of secondary electrons produced in the preamplification  avalanche which initiates \citep{pico24, model}. }
\\

\begin{figure}
\centering
\begin{minipage}{.48\textwidth}
\includegraphics[width=1.\textwidth]{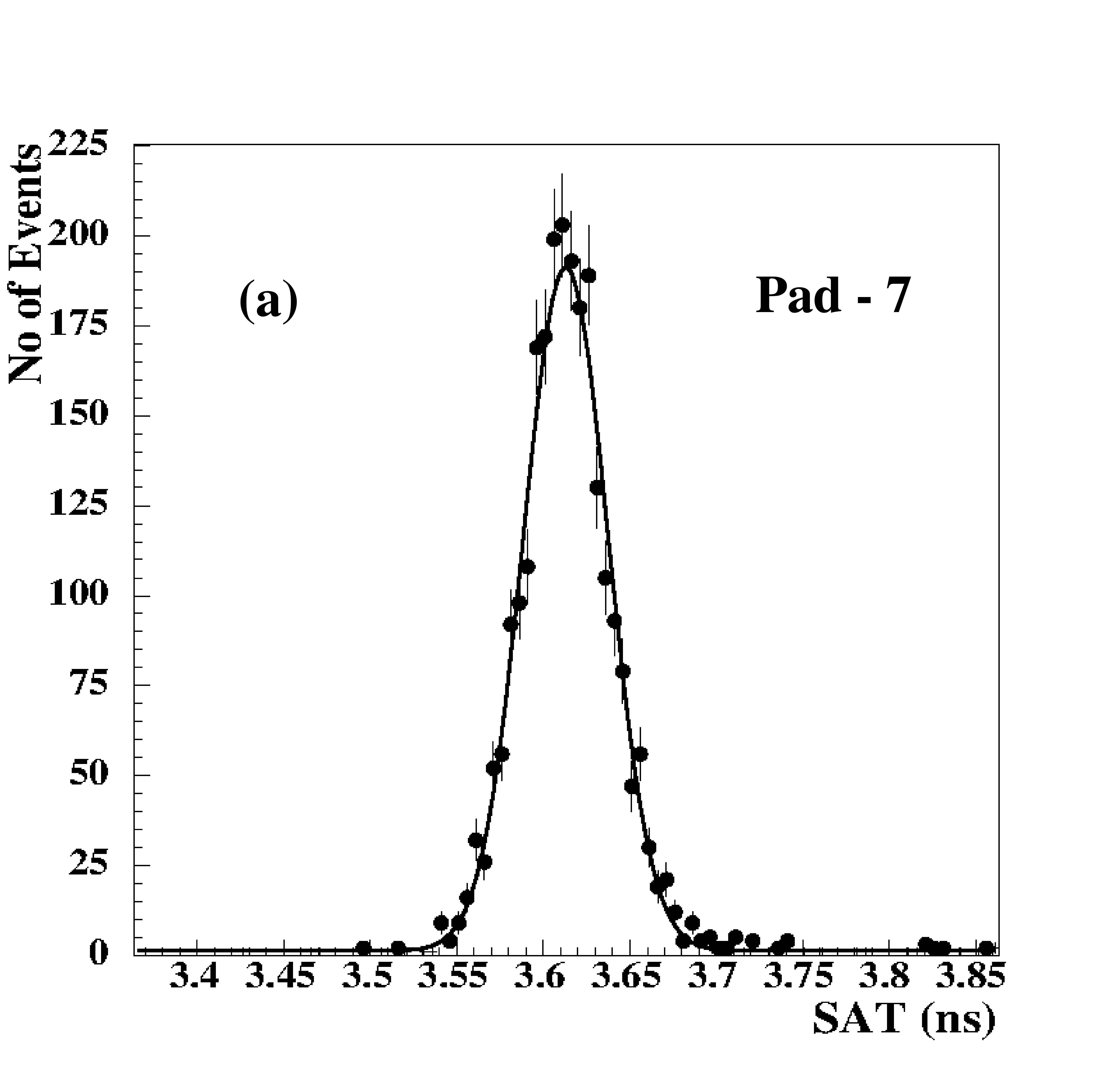}
\end{minipage}
\centering
\begin{minipage}{.48\textwidth}
\includegraphics[width=1.\textwidth]{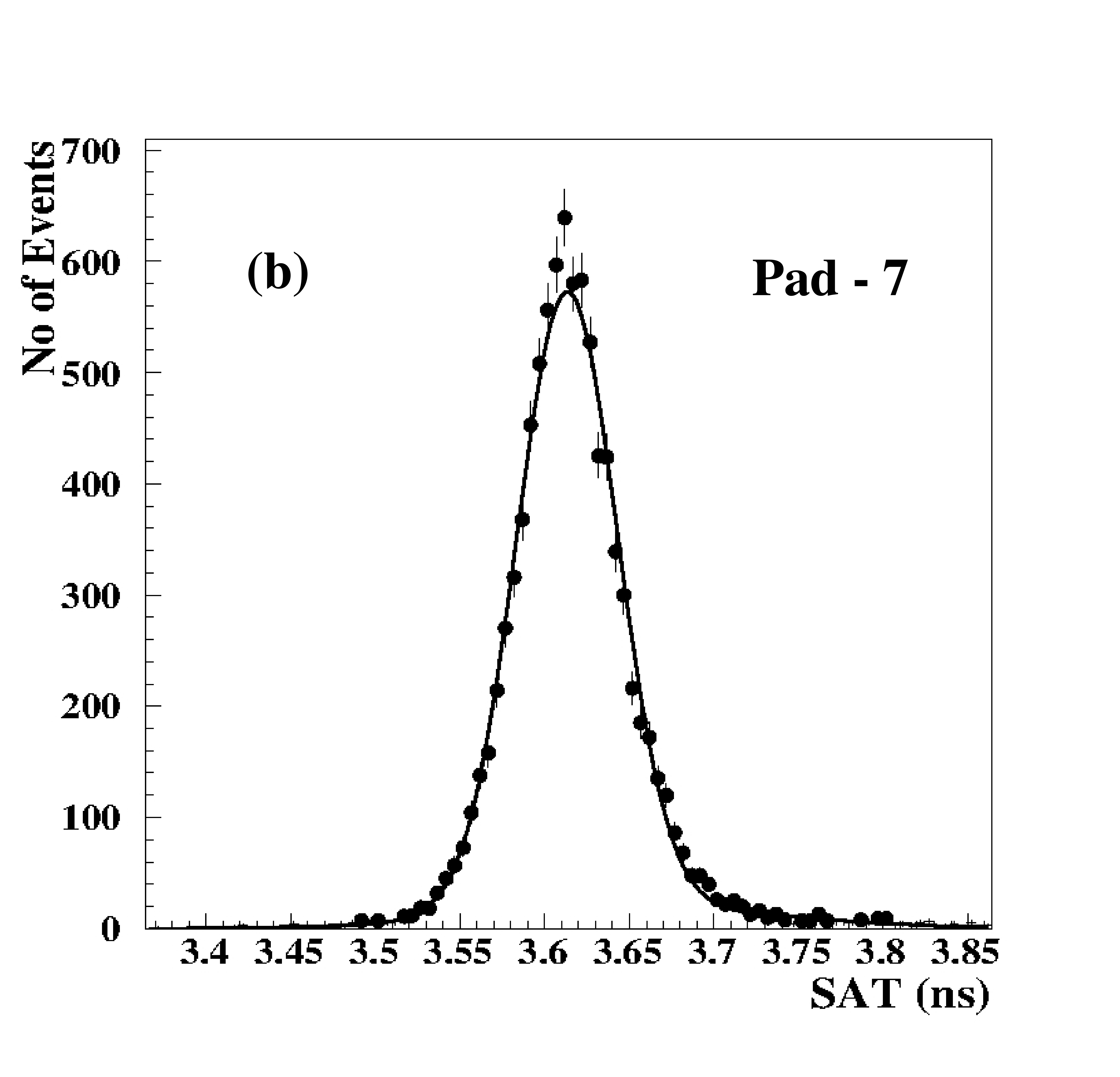}
\end{minipage}
\centering
\begin{minipage}{.48\textwidth}
\includegraphics[width=1.\textwidth]{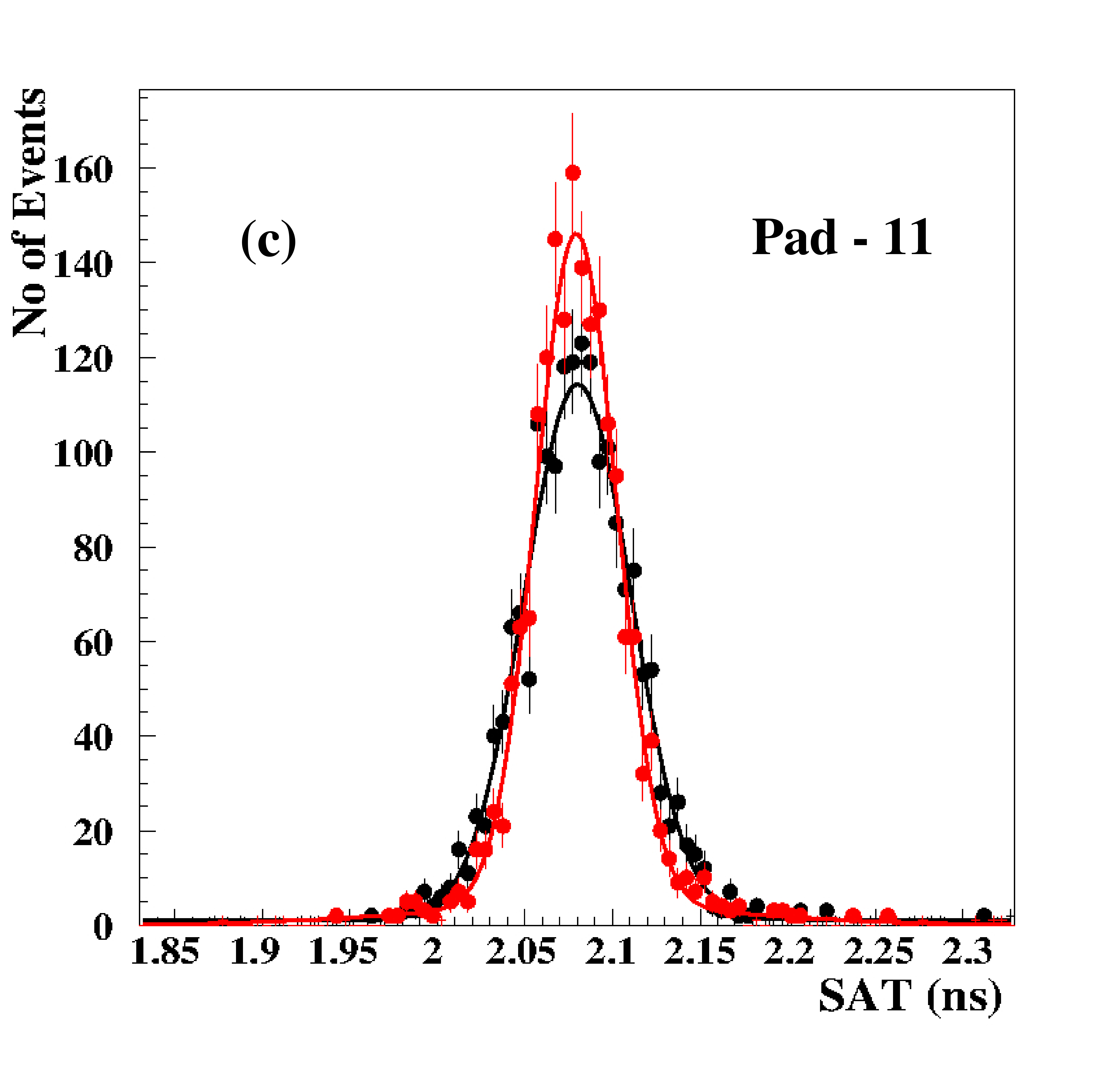}
\end{minipage}
\centering
\begin{minipage}{.48\textwidth}
\includegraphics[width=1.\textwidth]{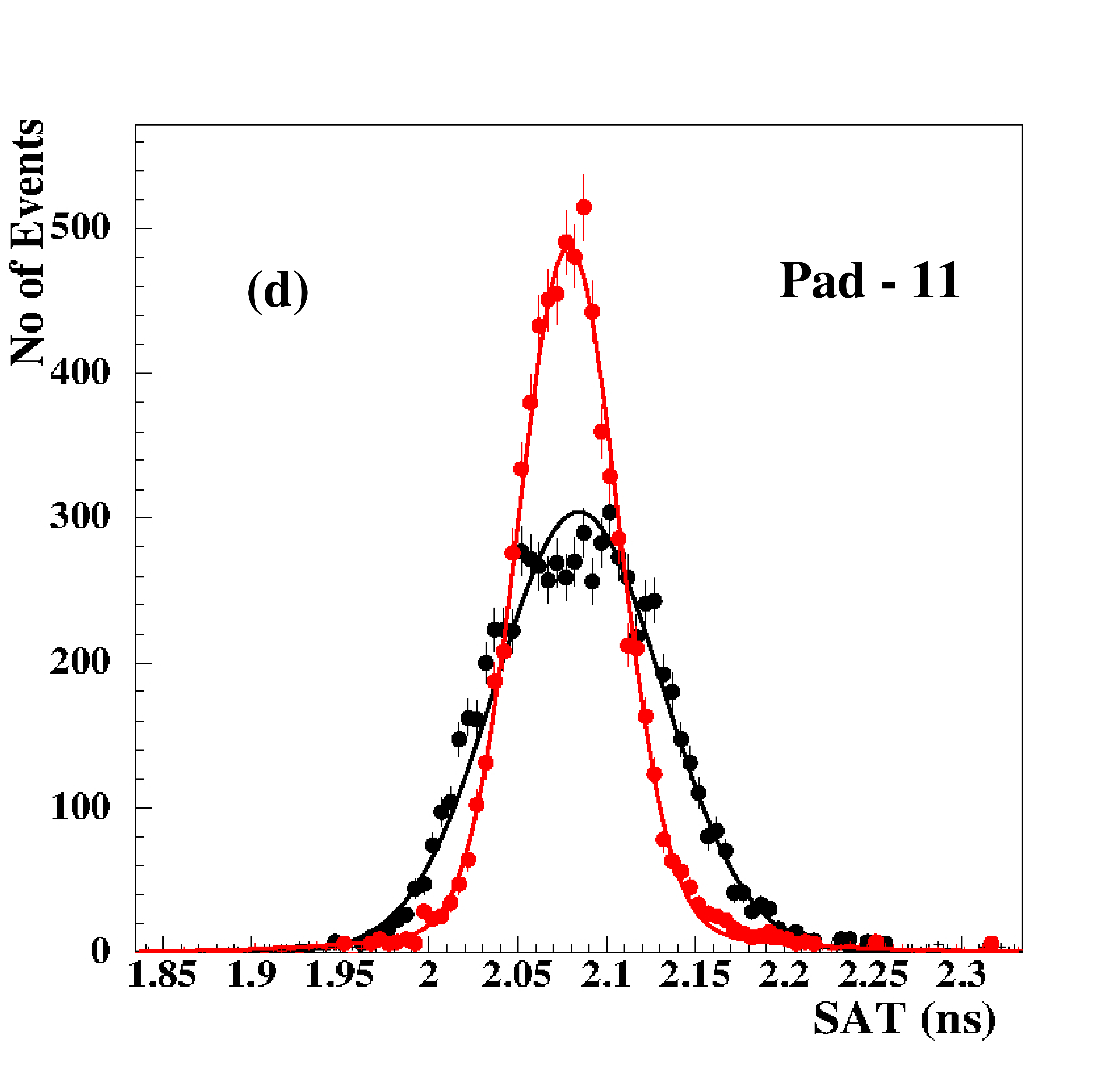}
\end{minipage}
\caption{ SAT  distributions of pad No. 7 (plots a and b) and pad No. 11 (plots c and d) signals, corresponding to incoming MIPs passing within 2 mm  (plots a and c) or between 2 mm and 4.33 mm (plots b and d) from the respective pad centres. The black  points represent raw SAT measurements while the red points show the same measurement distributions after applying the flatness corrections described later in Section \ref{spad}. The solid lines represent fits to the data points with the sum of two Gaussian functions sharing the same mean value. }
\label{fig:inner_periph}
\end{figure}

 Because  the drift field of the  prototype detector under study varies across the anode surface, the individual pads are expected to exhibit  different timing behaviour. Specifically, due to the shape of the geometrical deformation discussed in  Section \ref{alignment}, the timing performance of the central pad  is almost unperturbed but the variation of the drift-gap across the peripheral pads  causes inhomogeneities to their timing precision and SAT. In this Section, we describe corrections that recover  (almost) uniform timing characteristics across the entire anode surface.

 \begin{figure}
\centering
\begin{minipage}{.48\textwidth}
\includegraphics[width=1.\textwidth]{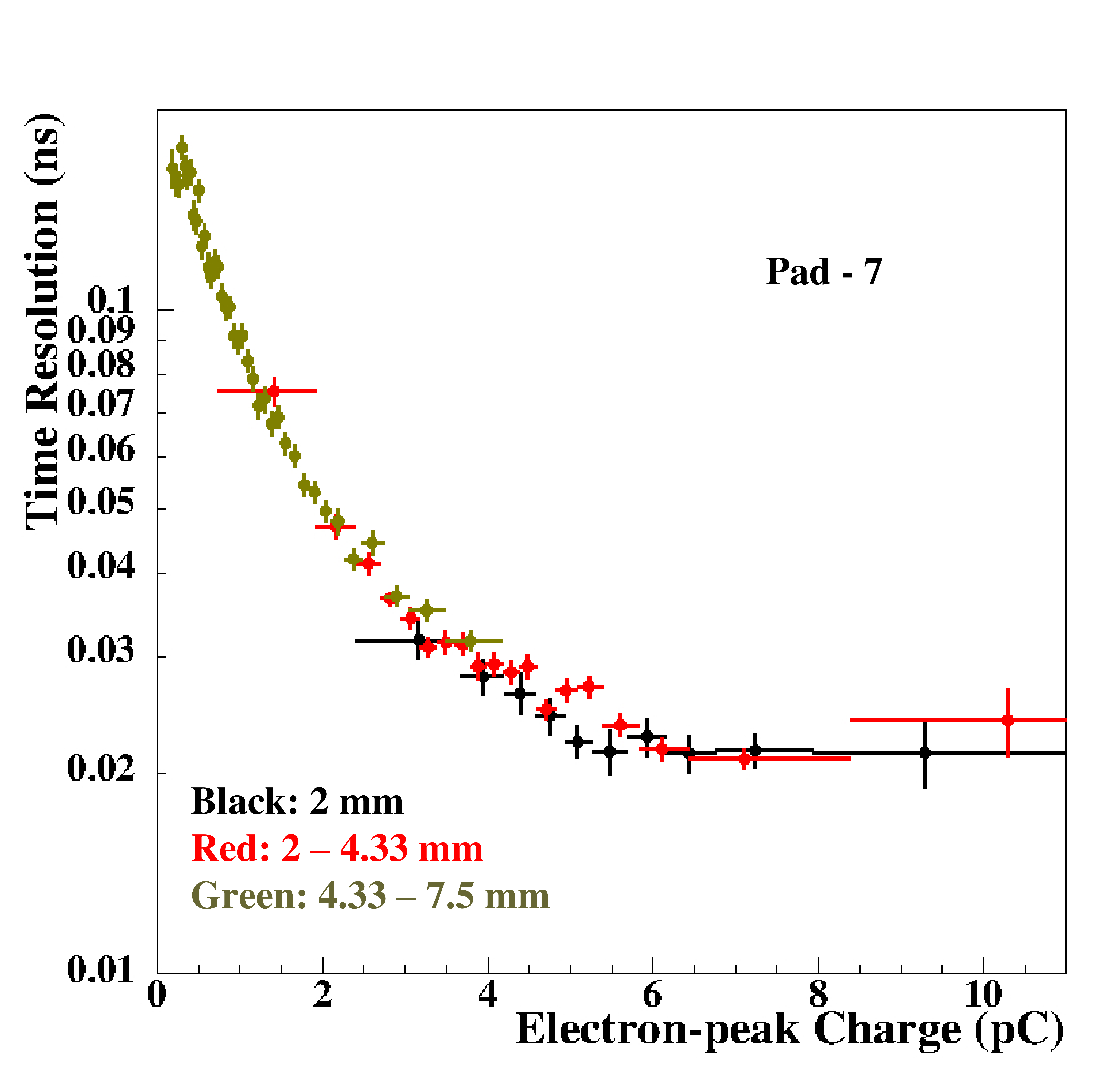}
\end{minipage}
\centering
\begin{minipage}{.48\textwidth}
\includegraphics[width=1.\textwidth]{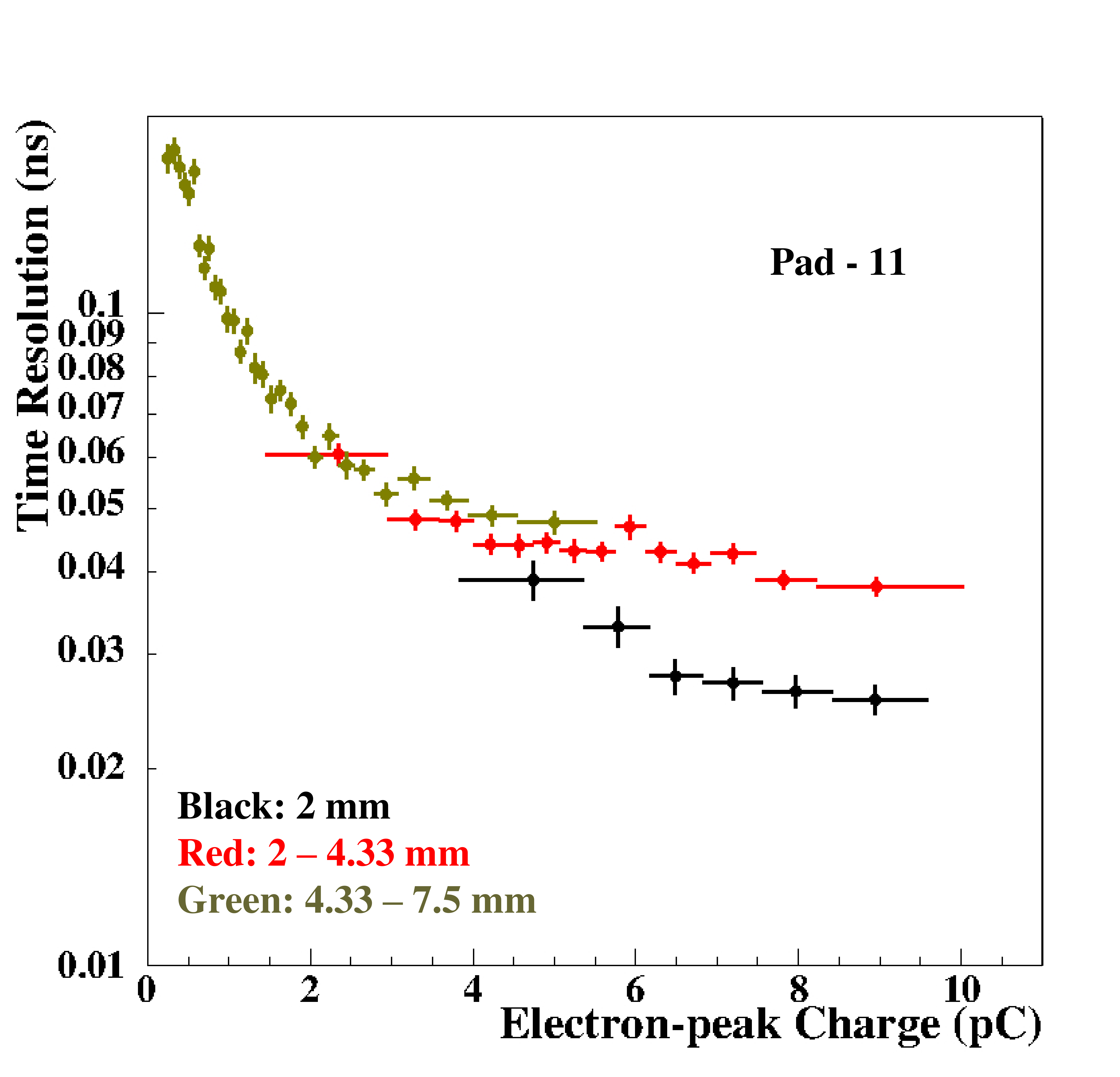}
\end{minipage}
\caption{The time resolution of a single pad versus the respective $Q_{e}$ for MIPs passing within 2 mm (black), between 2 mm and 4.33 mm (red) and between 4.33 mm and 7.5 mm (green) from the pad centre, respectively. The left plot corresponds to the central pad No. 7 and the right plot to the peripheral pad No. 11. }
\label{fig:dist_resol}
\end{figure}

Fig. \ref{fig:inner_periph} shows  SAT distributions for the central pad (pad No. 7) and  one of the peripheral pads (pad No. 11) when the tracks of incoming MIPs  pass within 2 mm  and between 2 mm and 4.33 mm\footnote{The radius of the inscribed circle into the hexagonal pad is 4.33 mm.}  of the respective pad centres. In the first case almost all the  photoelectrons produced by a MIP contribute to the signal of the pad, while in the second case  the pad signal is induced by a fraction   of the  photoelectrons. The time resolution of the central pad, quantified as the spread (RMS) of the corresponding  SAT distribution, is 26.5 $\pm$ 0.5 ps  and 34.0 $\pm$ 0.6 ps in the first and in the second case, respectively.  The peripheral pads  perform with worse time resolution, i.e. $\sim$33.5 ps and $\sim$49 ps on average, for the above two categories of incoming MIP tracks\footnote{For MIPs passing 2 mm around the centres of  pads  No. 4, No. 8 and No. 11 the respective SAT spreads  are 35.0 $\pm$ 0.6 ps,  33.5 $\pm$ 0.6 ps and 32.5 $\pm$ 0.6 ps. For MIPs that pass between 2 mm and 4.33 mm from the pads centres the SAT spreads are 50.0 $\pm$ 1.0 ps,  48. $\pm$ 0.8 ps and 48.5 $\pm$ 1.0 ps respectively.}, respectively.

Moreover, the time resolution of the central pad is  indeed an exclusive  function of   $Q_{e}$,  irrespectively of the  MIP impact point distance from the pad centre, as demonstrated in the left plot of Fig. \ref{fig:dist_resol}. On the contrary,   the time resolution of all the peripheral pads  depends on both, $Q_{e}$ and the MIP track proximity  to the pad centre. The right plot of Fig. \ref{fig:dist_resol} demonstrates this dependence  for pad No. 11, which is typical for any of the peripheral pads.  MIPs passing closer than 2 mm from the centre of a peripheral pad are timed with better accuracy than MIPs passing  further away, even though in both cases the induced signals have the same $Q_{e}$. In the following, we show that the  above,  undesirable, timing performance of the peripheral pads is mainly caused by the spatial inhomogeneity of the drift velocity, following the drift field variation across the anode surface.
\\

\begin{figure}[t]
    \centering
    \includegraphics[width=0.5\textwidth]{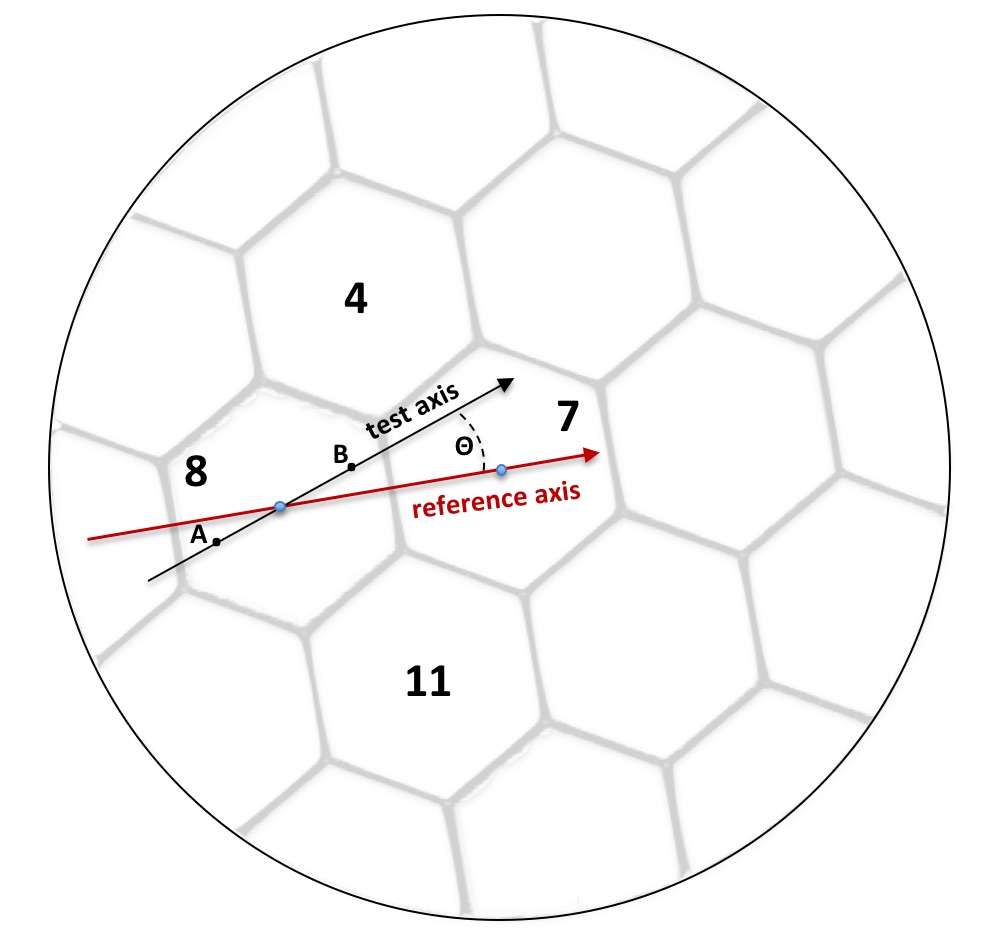}
    \caption{Illustration of the test and reference axes used for investigating the symmetry properties (see text) of the  mean SAT and  the time resolution of a single pad.} 
    \label{fig:testaxis}
\end{figure}

\begin{figure}

\centering
\begin{minipage}{1.\textwidth}
\includegraphics[width=1.\textwidth]{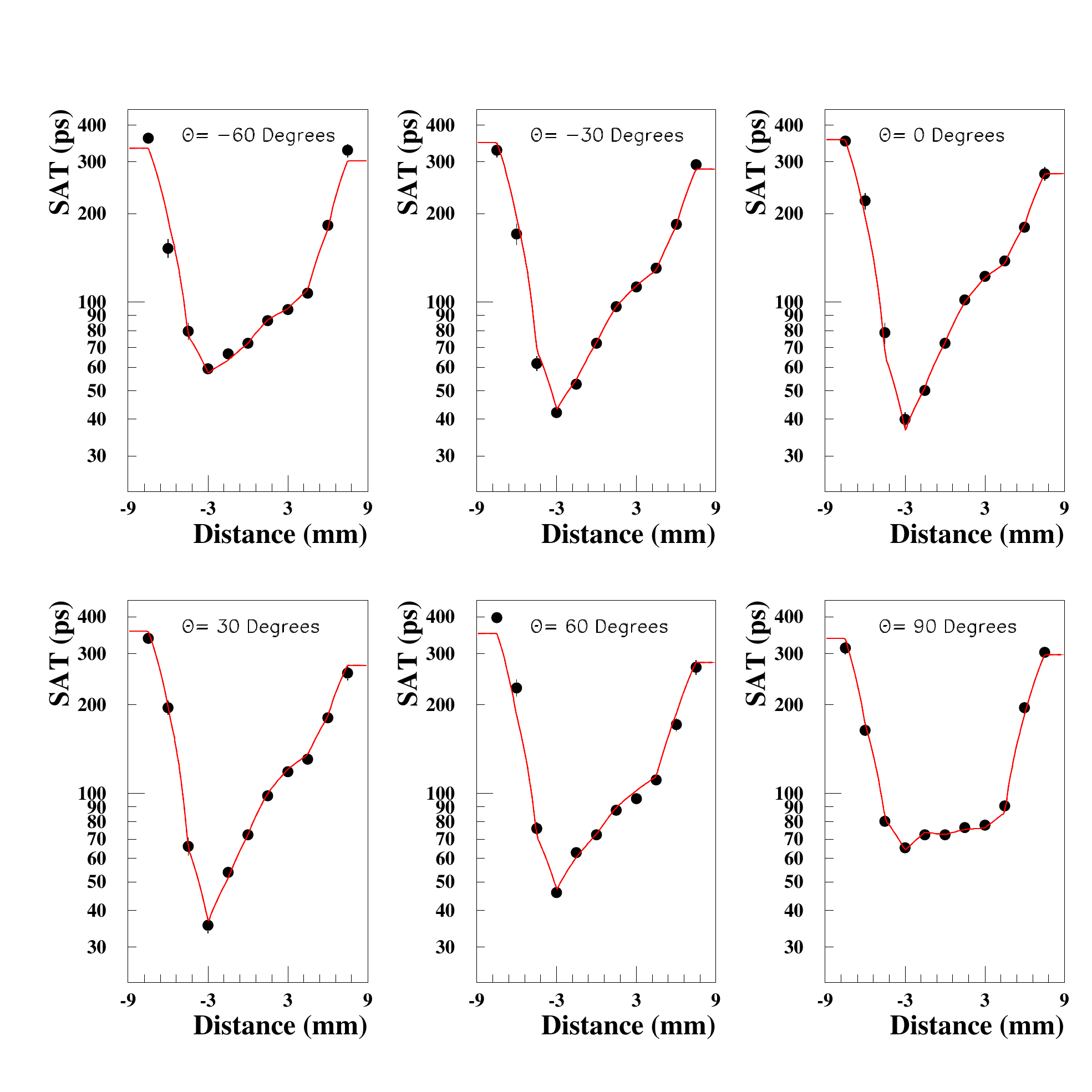}
\end{minipage}
\caption{ The mean signal arrival time (SAT) induced by MIPs to pad No. 8 versus the  coordinates of the respective seeds on the test axis.  Each point in the graphs corresponds to tracks passing within 0.5 mm around a seed, as it is described in the text. Every plot is marked by a value of the angle $\Theta$, which is the angle between the respective test-axis and the reference axis for pad No. 8. The solid lines represent empirical parametrizations. (For a better illustration, an offset of 2.019 ns  has been subtracted from all the raw SAT measurements used in the plots)}
\label{fig:slew_as}
\end{figure}

Consider any axis passing through the centre of a pad (hereafter called test axis) and two points  on this axis, A and B, on either side and at equal distances from the pad centre, as shown in Fig. \ref{fig:testaxis}.   In an ideal case, the timing  information carried by the  pad signals  induced by MIPs passing from either of the above points should be identical. We found that the  time resolution and the mean SAT values of the central pad  respect  the above, mirror symmetry. However,  the mean SAT of the peripheral pads manifestly violates the  symmetry with respect to the pad centres, as shown in Fig. \ref{fig:slew_as} for the peripheral pad No. 8. 

We define a reference axis, collinear to the line segment connecting  the peripheral (in this case the pad No. 8) with the central pad centres and directing towards to the centre of the central pad, as illustrated in Fig. \ref{fig:testaxis}. In the rightmost, upper plot of  Fig. \ref{fig:slew_as}, the chosen test-axis  coincides with the reference axis for pad No. 8.  The mean SAT values, corresponding to MIPs  passing within  0.5 mm  around  certain points of the test-axis  (hereafter called seeds),  are plotted versus the coordinate of the respective  seed along the test-axis, i.e. the signed distance  from the pad centre. The peripheral pad signals induced by MIPs that pass around seeds with negative coordinates arrive faster than signals  induced by tracks passing by their mirror symmetric (positive) seeds. 

We  examined the above mean SAT asymmetry along several test-axes, keeping always the axis origin at the centre of the peripheral pad and turning the axis orientation  by an angle, $\Theta$, with respect to the reference axis (see Fig. \ref{fig:testaxis}). The  plots in Fig. \ref{fig:slew_as} present mean SAT asymmetries along six test-axes,  where positive (negative) $\Theta$ values indicate anti-clockwise (clockwise) turns. For all peripheral pads, the asymmetry is larger at small absolute values of the  angle $\Theta$ while at $\Theta = 90^{o}$ is significantly reduced. This mean SAT asymmetry reflects the spatial variation of the drift velocity, which in turn maps the variation of the drift field due to the geometrical deformation of the anode and the non-uniformity of the drift gap thickness. An almost spherical anode surface (as discussed in Section \ref{alignment}), with its lower tip close to the centre of the central pad, is consistent with the observed SAT asymmetries, i.e. with the observation that the SAT asymmetry is maximal along the direction defined by the peripheral and central pad centres whilst is minimal along the perpendicular direction. 
 \begin{figure}

\centering
\begin{minipage}{1.\textwidth}
\includegraphics[width=1.\textwidth]{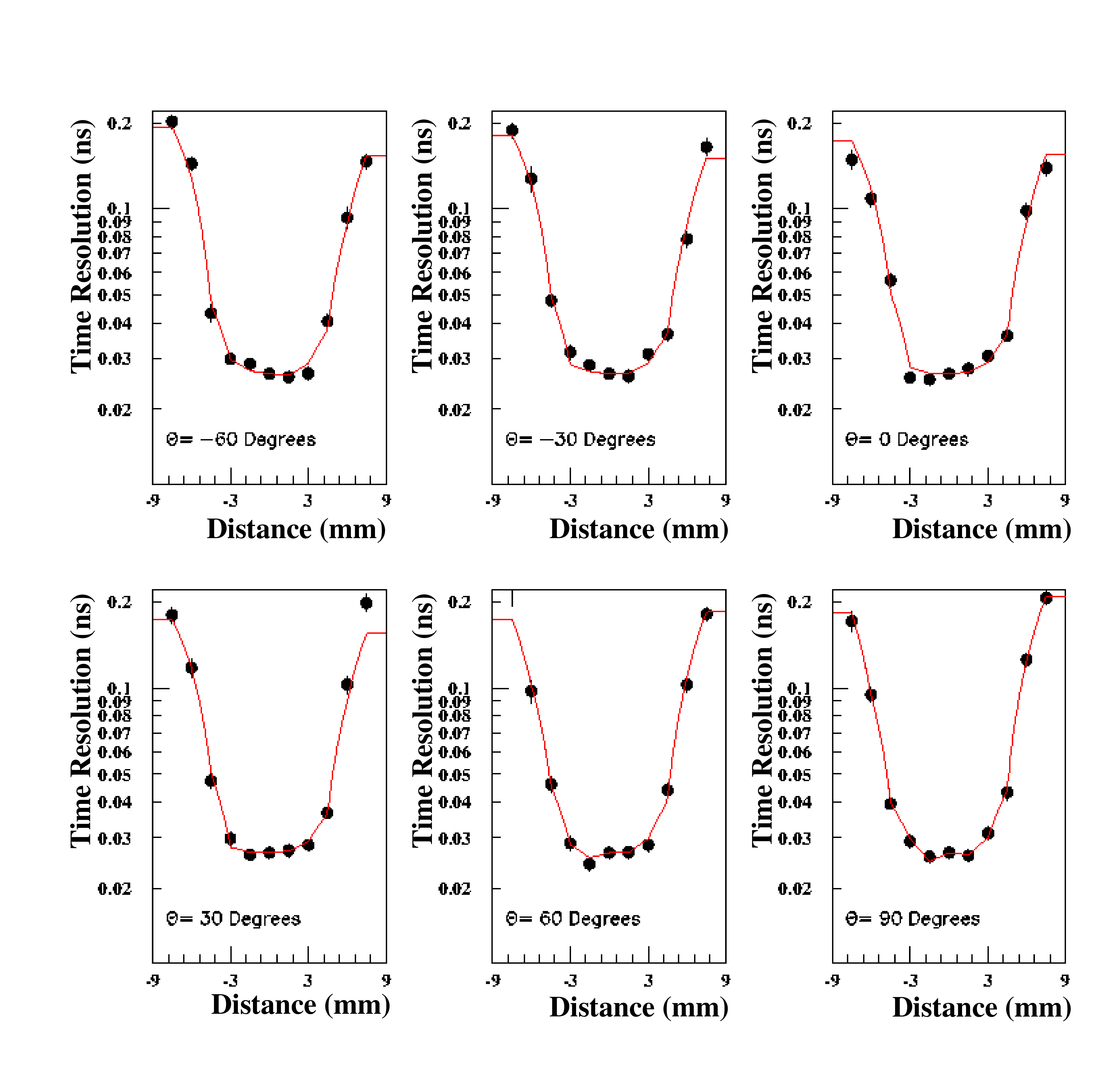}
\end{minipage}
\caption{The spread (RMS) of SAT distributions induced by MIPs to pad No. 4 versus the  coordinates of the respective seeds on the test axis.  Each point in the graphs corresponds to tracks passing within 0.5 mm around a seed, as it is described in the text. Every plot is marked by a value of the angle $\Theta$, which is the angle between the respective test-axis and the reference axis for pad No. 4. The solid lines represent empirical parametrizations.}
\label{fig:res_as}
\end{figure}

However,  by examining the RMS values of the SAT distributions (i.e. the time resolution) along the same test-axes we found  that, unlike the mean SAT asymmetry, the time resolution of the peripheral pads  are, in a good approximation, symmetric around the pad centres  for all angles $\Theta$, as shown in Fig. \ref{fig:res_as}. 
\\

The observed timing characteristics, shown in Figs. \ref{fig:slew_as} and \ref{fig:res_as},  indicate that the spatial variation of the drift gap thickness\footnote{Notice that a change in the drift gap thickness besides of changing the drift field also affects the maximum  available length for the  preamplification avalanches evolution.},  although it affects the uniformity of the detector drift field which causes variations on the drift velocity, it does not have a major influence on the symmetry of the time resolution around the pad centre. However,  the imposed spatial variation on the drift velocity causes local systematic shifts on the SAT.  
\\

\begin{figure}
\centering
\begin{minipage}{.48\textwidth}
\includegraphics[width=1.\textwidth]{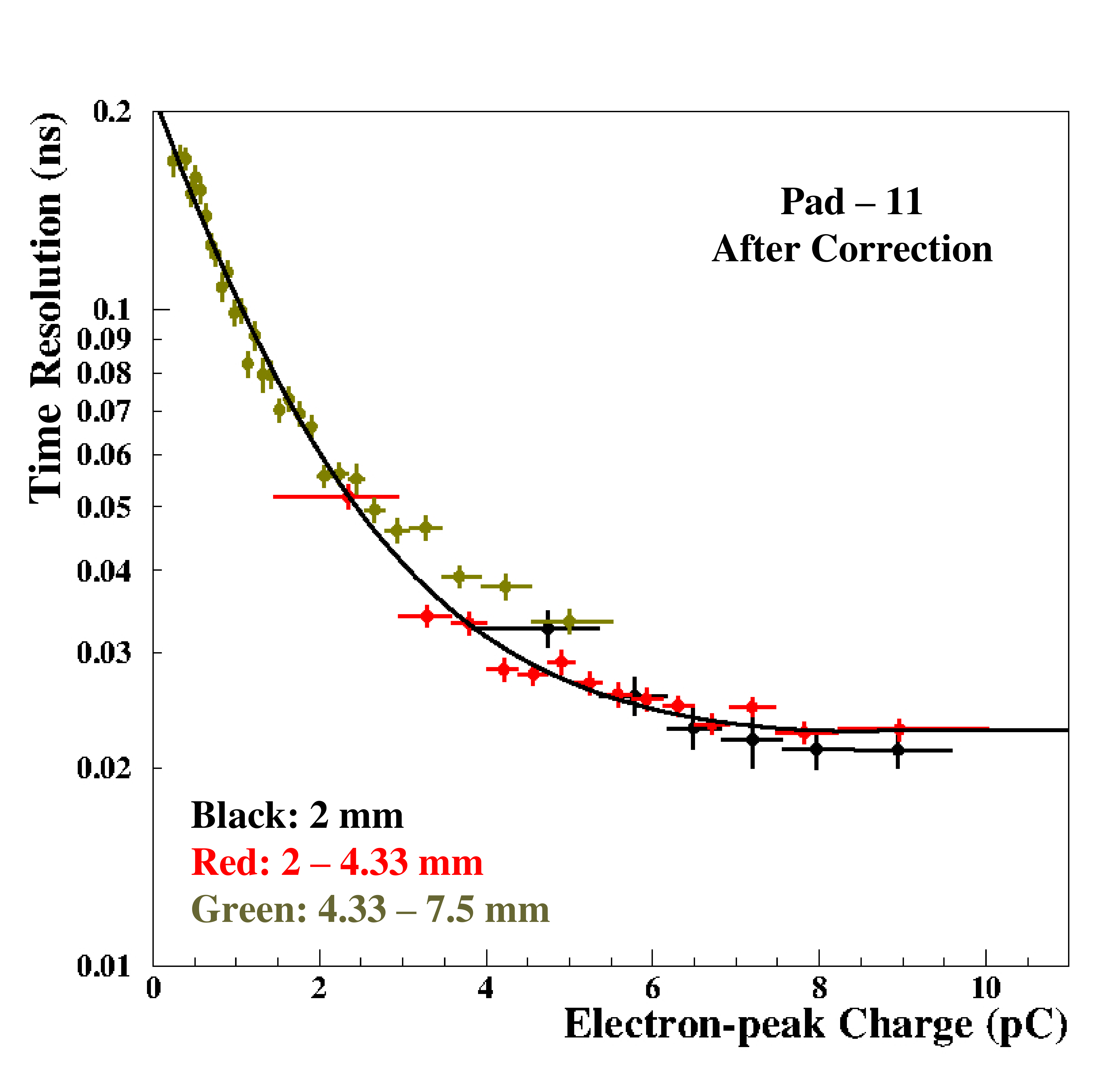}
\end{minipage}
\centering
\begin{minipage}{.48\textwidth}
\includegraphics[width=1.\textwidth]{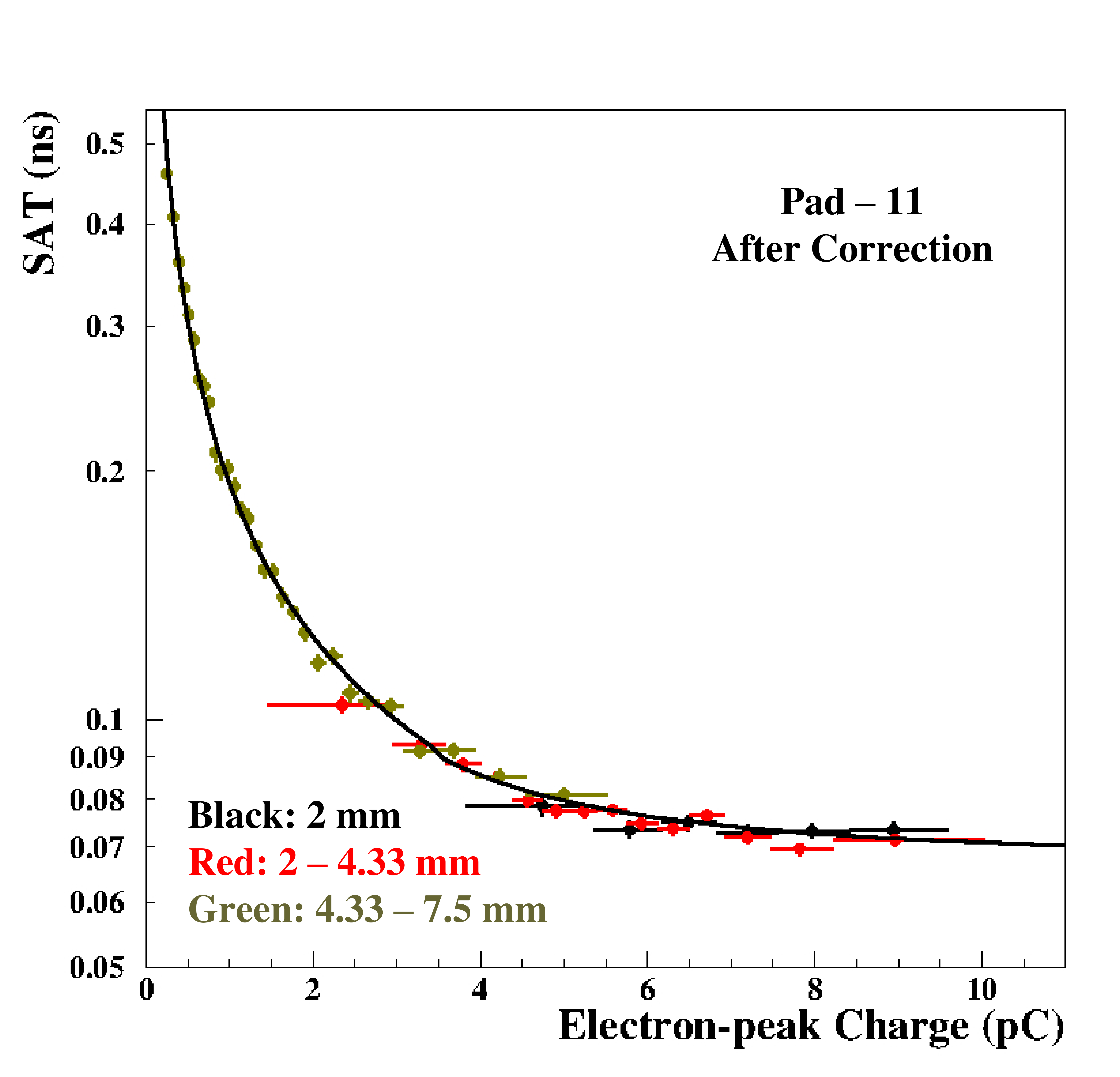}
\end{minipage}
\caption{ (left) The time resolution versus  $Q_{e}$, after applying  the flatness correction to signals of pad No. 11.  (right)  The mean SAT versus  $Q_{e}$, after applying  the flatness correction to signals of pad No. 11. For better illustration, an offset of 2 ns has been subtracted from the SAT values. Both plots follow the same colour code as in Fig. \ref{fig:dist_resol} to denote the regions of proximity of the  respective MIP impact points to the pad centre. The solid curves represent fits to the data points from all proximity regions.}
\label{fig:after_corr}
\end{figure}

We developed  corrections (hereafter called flatness corrections) in order to remove the timing biases caused by the geometrical deformation. For each peripheral pad (k), we  parametrized the  mean SAT values, e.g. those presented in Fig. \ref{fig:slew_as}, as a function of the  cylindrical coordinates of the respective seed in the pad-frame\footnote{The pad-frame  is a coordinate system with its origin at the centre of the pad and  the reference axis defines the X axis of the frame. Notice that the azimuth angle in this frame coincides with the angle $\Theta$.}, $S^{k}\left( r,\Theta\right) $. A correction factor, $\Delta^{k}\left( r,\Theta\right) = S^{k}\left( r,\Theta\right) - S^{k}\left( r,\Theta = 90^{\circ}\right)$, is defined based on the observation that the SAT is almost mirror symmetric along the axis with $\Theta= 90^{\circ}$. Then, the flatness correction is applied to the arrival times of all signals of the $k^{th}$ pad,   as:
\begin{equation}
T_{f-corr.}^{k}= T_{SAT}^{k}\left( r,\Theta\right) - \Delta^{k}\left( r,\Theta\right)
\label{eq:7}
\end{equation}
where $T_{SAT}^{k}\left( r,\Theta\right)$ is the raw arrival time of the $k^{th}$ pad signal,  which is induced by a MIP that impacts the PICOSEC-Micromegas plane at $\left( r,\Theta\right)$, and $T_{f-corr.}^{k}$ is the SAT value corrected from  biases due to geometrical distortions.
It should be noticed that Eq. (\ref{eq:7}) is an empirical correction, which is based on several approximations, e.g. when we evaluate the parametrization  $S^{k}\left( r,\Theta\right) $ the coordinates  $\left( r,\Theta\right) $ refer to the seed points but when we use the flatness correction factors, $\Delta^{k}\left( r,\Theta\right)$, the coordinates refer to the specific impact point of the incoming MIP. Nevertheless, when the flatness correction  is applied to the data, the timing properties of the peripheral pads approach those  of the central pad. 
\\

\begin{figure}
\centering
\begin{minipage}{.48\textwidth}
\includegraphics[width=1.\textwidth]{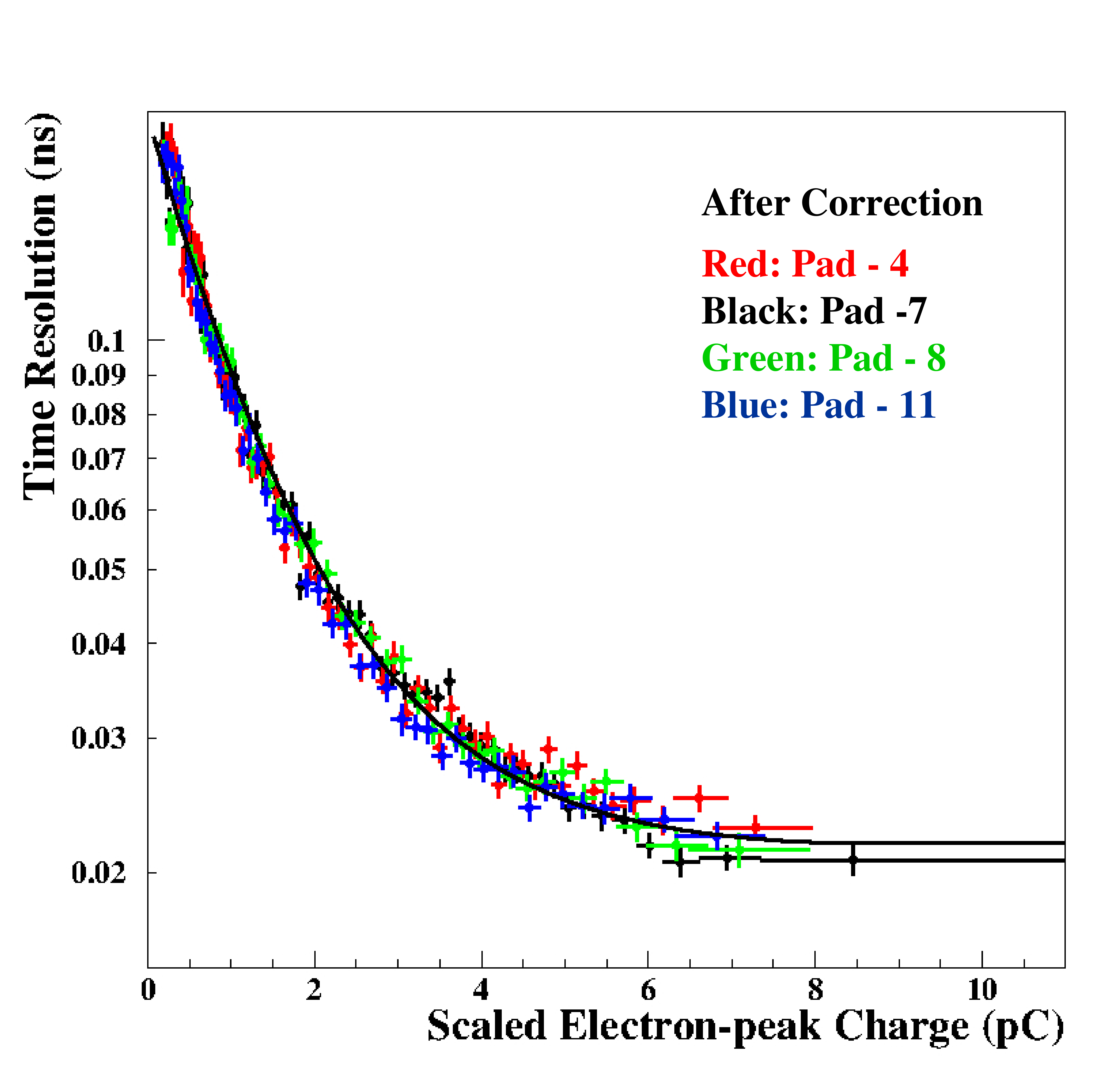}
\end{minipage}
\centering
\begin{minipage}{.48\textwidth}
\includegraphics[width=1.\textwidth]{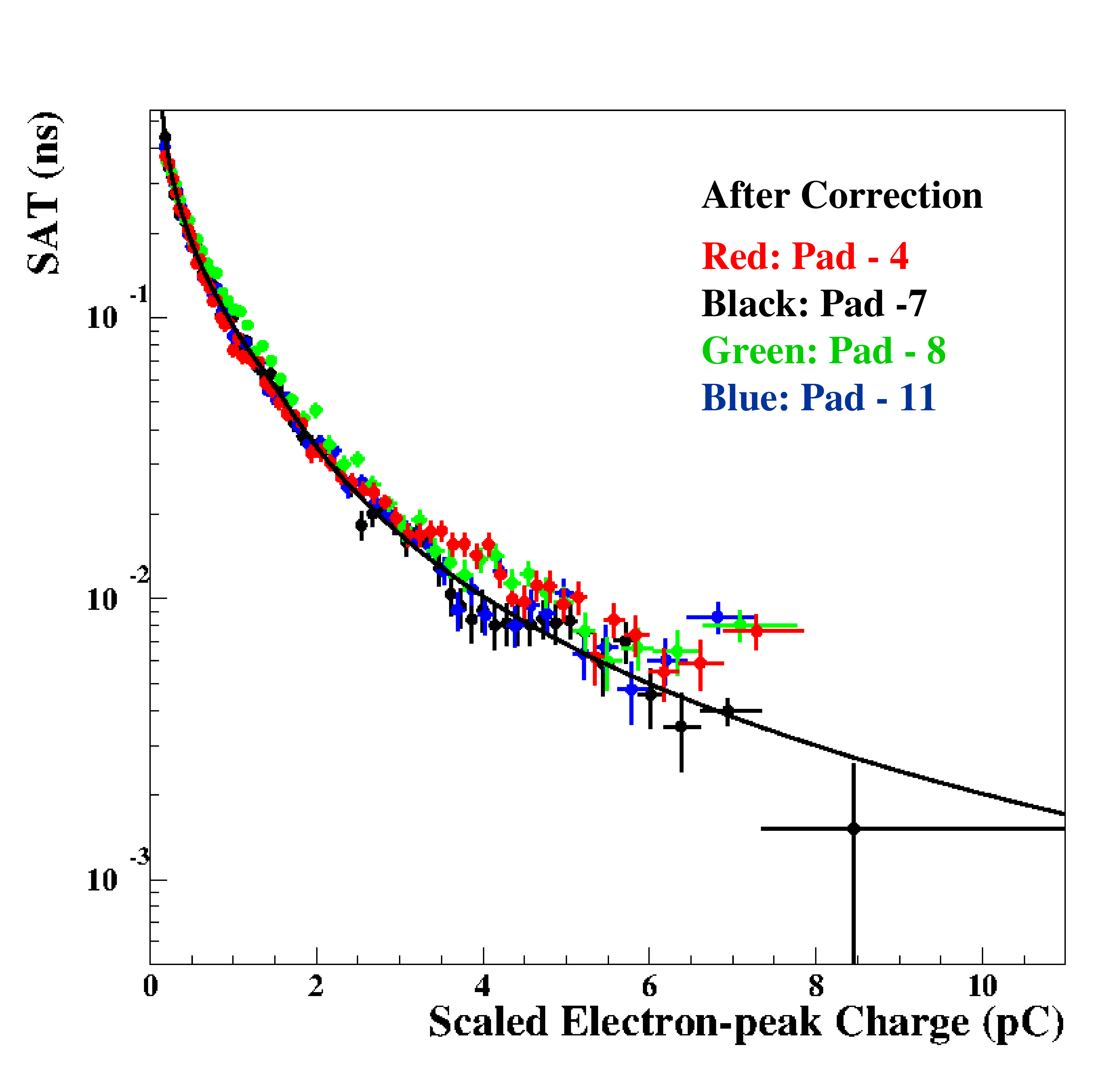}
\end{minipage}
\caption{The time resolution (left) and the mean SAT (right), after applying the flatness correction to the peripheral pad signals, as a function of the scaled electron-peak charge (see text). External time delays have been subtracted from the SAT measurements of each pad. The black points correspond to the central pad while the other, coloured points correspond to the peripheral pads. The solid curves represent fits of  the central pad data.}
\label{fig:scaled_all}
\end{figure}

The effect of the flatness correction   is demonstrated in Fig. \ref{fig:inner_periph}. The red points   represent,  restored by Eq. (\ref{eq:7}), SAT distributions of the peripheral pad No. 11. After the flatness correction the time resolution for MIPs passing through the inner and outer region around the pad centre is  27.0 $\pm$ 0.7 ps and 35.2 $\pm$ 0.5 ps, respectively,  in a very good agreement with the central pad performance. Furthermore, as Fig. \ref{fig:after_corr} demonstrates, after applying the flatness correction the time resolution  of the peripheral pads versus $Q_{e}$  depends exclusively on the electron-peak charge, irrespectively  of the proximity of the MIP track to the pad centre. Similarly, after applying the corrections, the  mean SAT  also depends exclusively on $Q_{e}$, independently on the MIP impact position. 

 In addition, by taking  into account the gain difference  between pads, i.e.  by scaling down\footnote{ The mean values of the electron-peak charge distributions, shown in Fig. \ref{fig:in_q}, are used to determine the respective scaling factors.}  the electron-peak charge of the peripheral pads, the time resolution and the mean SAT dependences on the (scaled) $Q_{e}$ becomes almost the same for all pads, as it is illustrated in Fig. \ref{fig:scaled_all}. 
\\

\begin{figure}
\centering
\begin{minipage}{.48\textwidth}
\includegraphics[width=1.\textwidth]{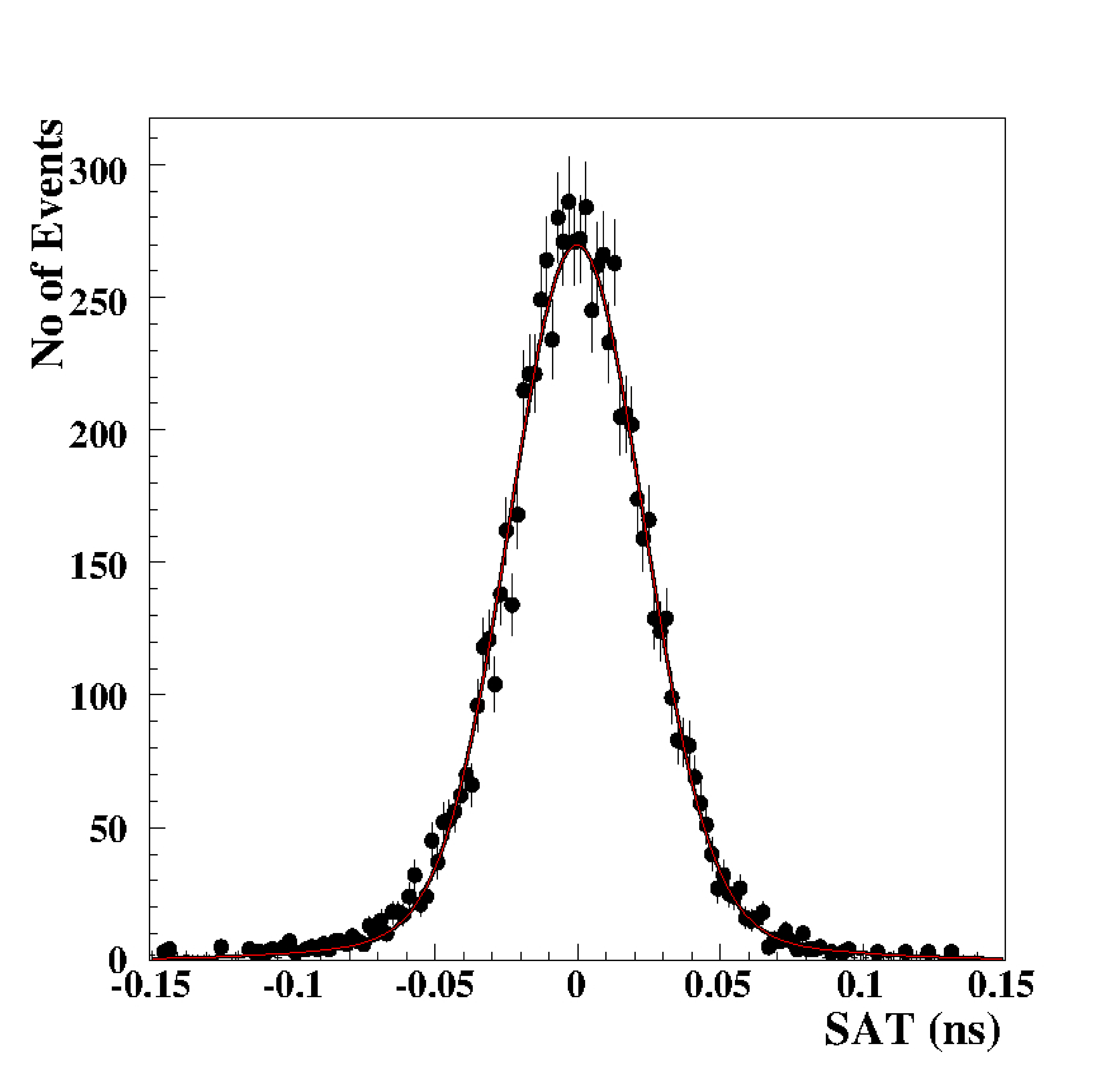}
\end{minipage}
\centering
\begin{minipage}{.48\textwidth}
\includegraphics[width=1.\textwidth]{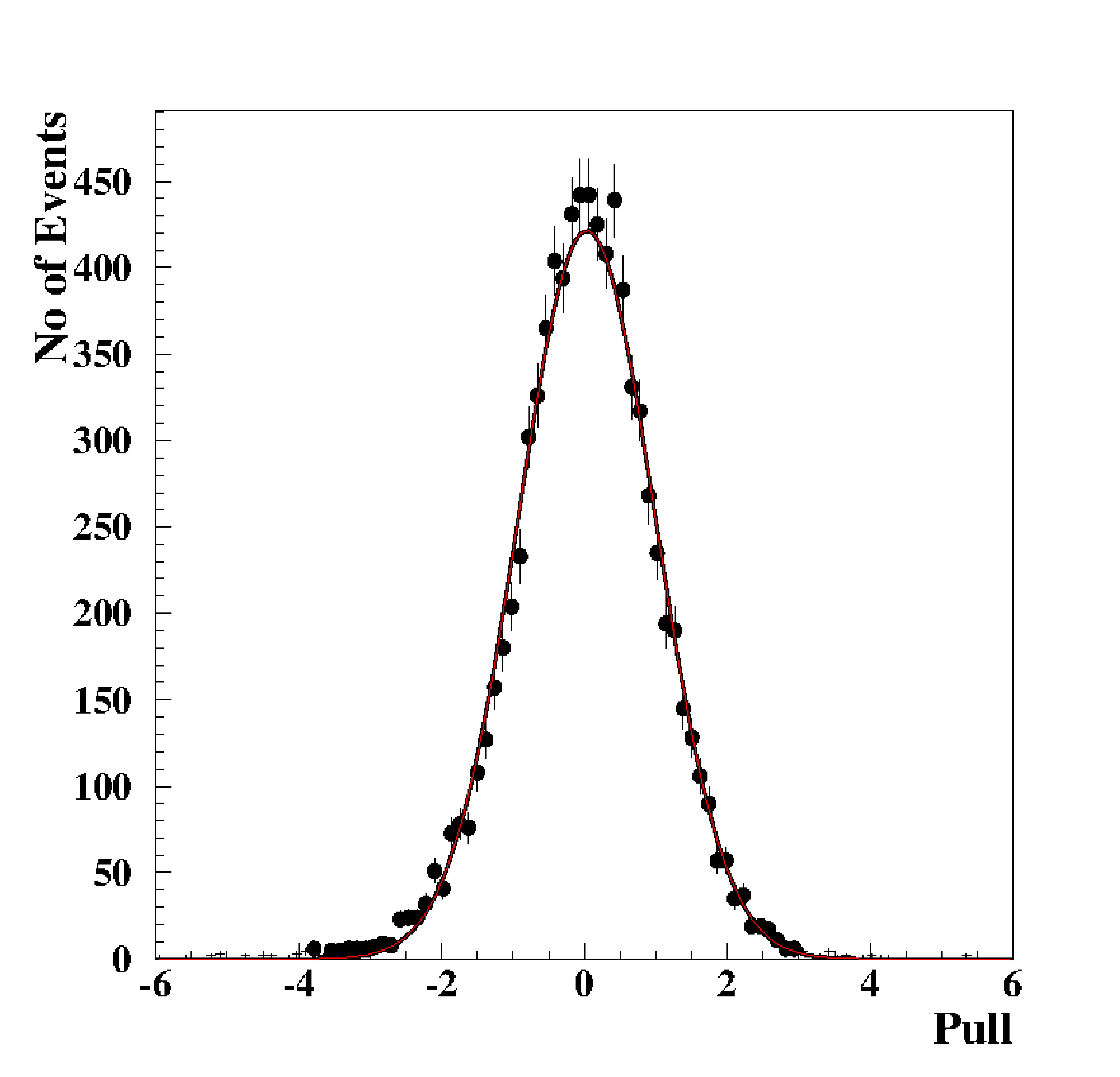}
\end{minipage}
\caption{ (left) Distribution of  fully corrected arrival time measurements of all pad signals induced   by MIPs passing within 2 mm from the respective pad centre. The solid line represent a double Gaussian fit to the data points with an RMS of $25.8 \pm 0.6$ ps. (right) Distribution of the fully corrected SAT measurements normalized to their expected error (pull). The solid line represents a Gaussian fit with  estimated mean and  $\sigma$ values consistent to 0 and 1 respectively.}
\label{fig:allpad_in}
\end{figure}

 Naturally, a  multi-pad PICOSEC-Micromegas detector with a perfectly flat anode will exhibit the above timing characteristics without the need of any flatness correction.  Nevertheless, due to the physical processes that produce the signal \citep{model}, the PICOSEC-Micromegas SAT depends on the electron-peak charge. Such a dependence should be evaluated a priori (calibration) and must be taken into account for precise timing measurements. In Fig.   \ref{fig:scaled_all}, the solid lines represent fits to the experimental points, providing also the functional description of the SAT (and of the time resolution) dependence on $Q_{e}$, i.e. $\tau\left( Q_{e}\right) $ (and $\sigma\left( Q_{e}\right) $). The left plot in Fig. \ref{fig:allpad_in} presents the distribution of fully corrected SAT values, i.e  $T_{f-corr.}^{k} -\tau\left( Q_{e}\right)$, for all the available single pad measurements, when the MIPs are passing within 2 mm from the respective  pad centre. The global time resolution of the prototype multi-pad-PICOSEC-Micromegas, i.e. includind all pads, provided by the spread (RMS) of this distribution, is  $25.8 \pm 0.6$ ps. The right plot illustrates the pull (i.e. $  \frac{T_{f-corr.}^{k} -\tau\left( Q_{e}\right)}{\sigma\left( Q_{e}\right)}$) distribution which is consistent with a standard normal distribution, signifying also the consistency in the  evaluation of the expected resolution of a pad as a function of $Q_{e}$.

\

\section{Combining the timing information of several pads} \label{multicomb}

We consider cases when the Cherenkov ring of a MIP produces signals on more than one ($M > 1$) pad. After applying the flatness correction we assume that the pad signals share the same timing properties, as it is shown in Fig \ref{fig:scaled_all}. The dependence of the mean SAT  and of the time resolution on the respective $Q_{e}$ have been calibrated and quantified by the functions $\tau\left( Q_{e}\right)$ and $\sigma\left( Q_{e}\right)$, respectively. Henceforth, we assume that each of the pad signals carries an independent information on the MIP arrival time. Then, we combine the single-pad measurements to form the following $\chi^{2}$ estimator:

\begin{figure}
\centering
\begin{minipage}{.33\textwidth}
\includegraphics[width=1.\textwidth]{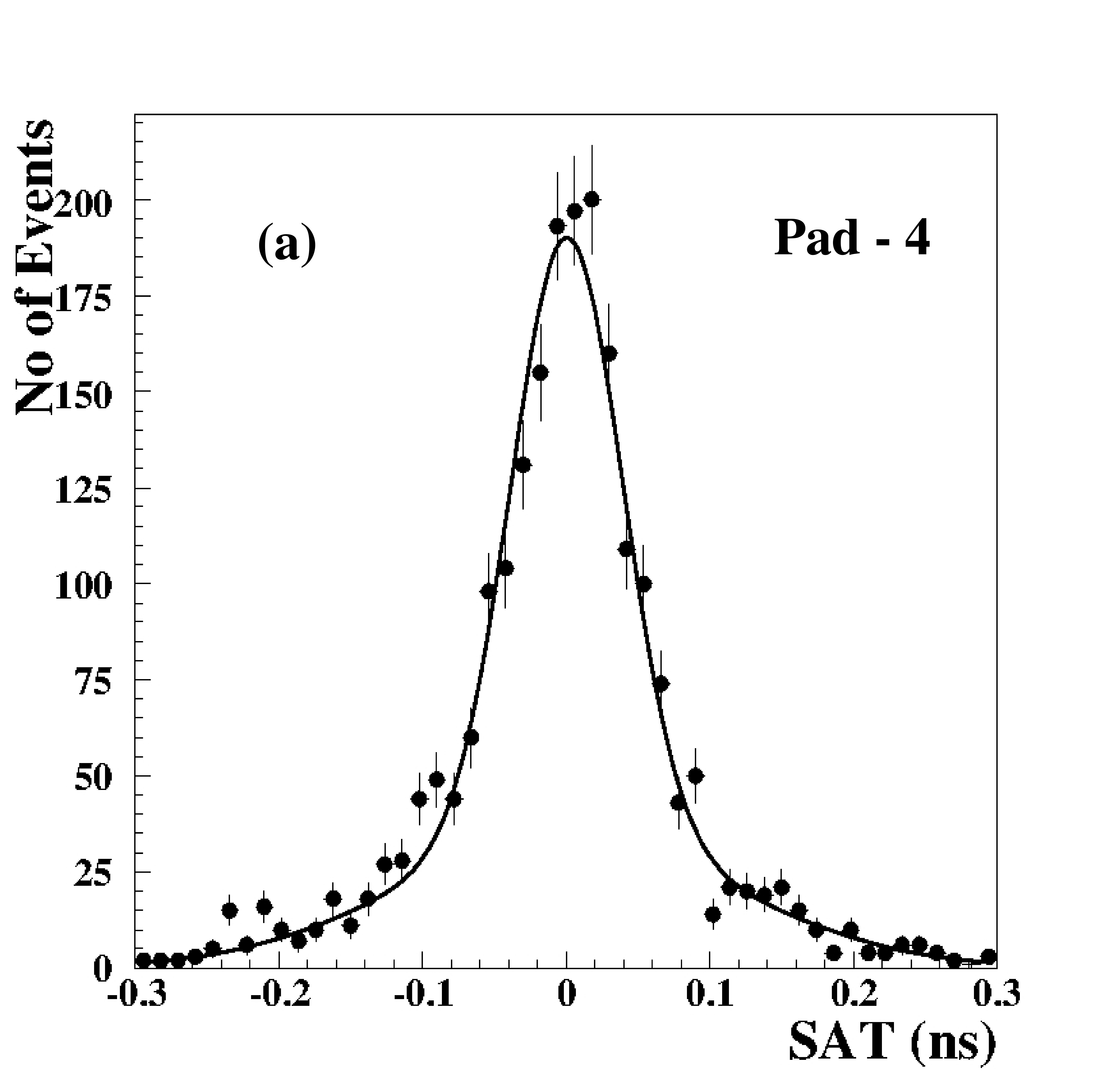}
\end{minipage}\hfill
\centering
\begin{minipage}{.33\textwidth}
\includegraphics[width=1.\textwidth]{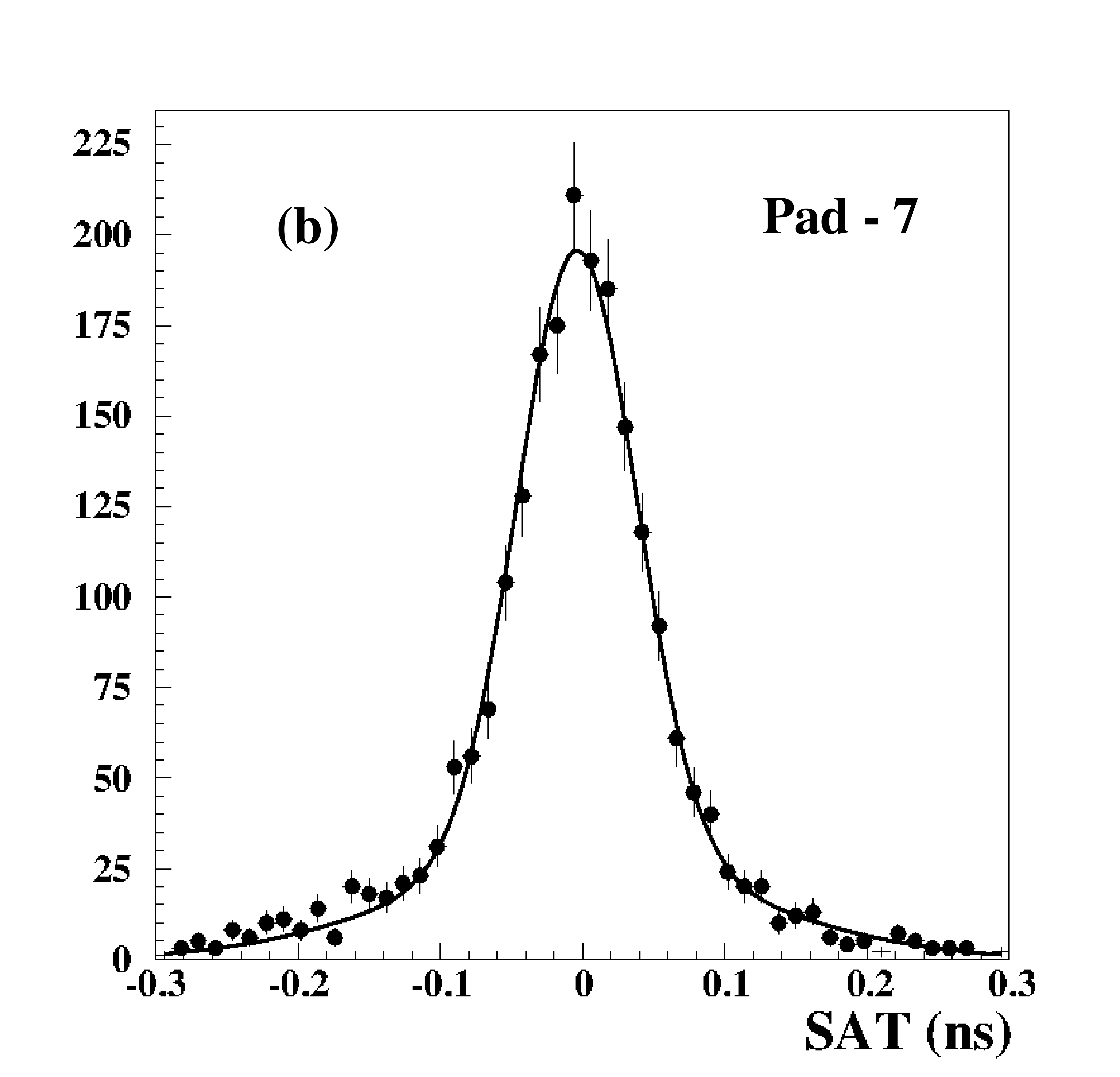}
\end{minipage}\hfill
\centering
\begin{minipage}{.33\textwidth}
\includegraphics[width=1.\textwidth]{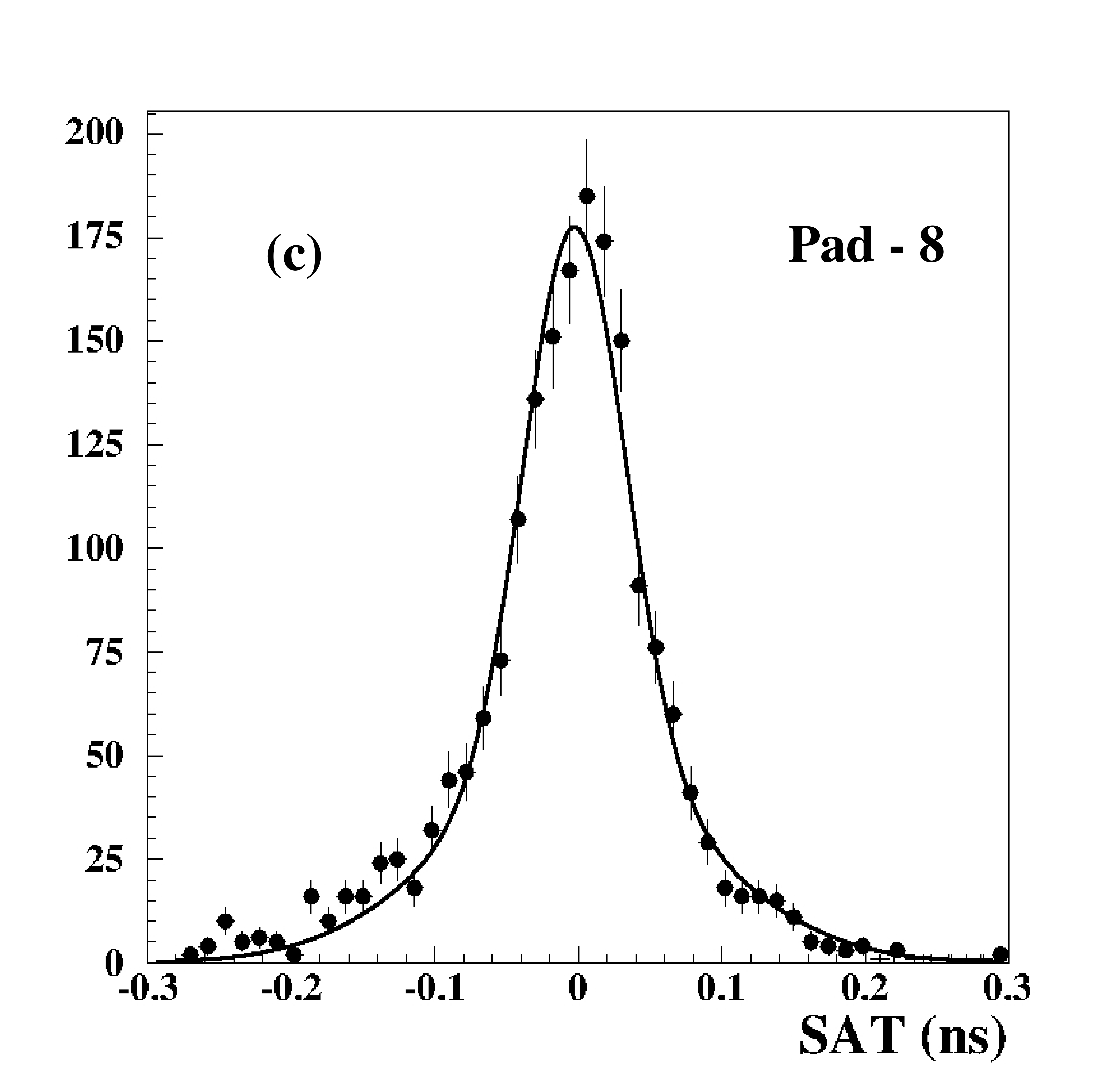}
\end{minipage}\hfill

\centering
\begin{minipage}{.33\textwidth}
\includegraphics[width=1.\textwidth]{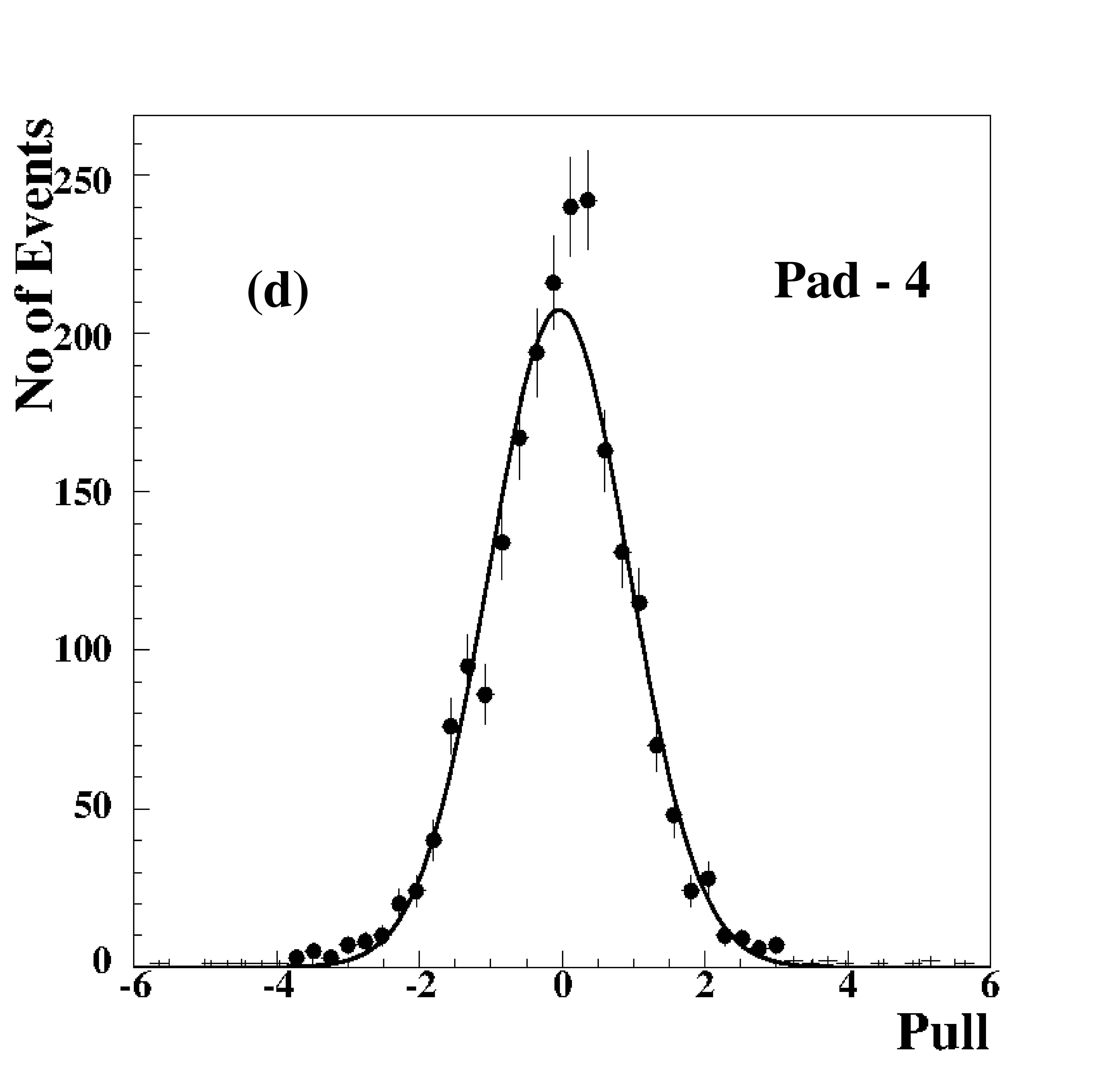}
\end{minipage}\hfill
\centering
\begin{minipage}{.33\textwidth}
\includegraphics[width=1.\textwidth]{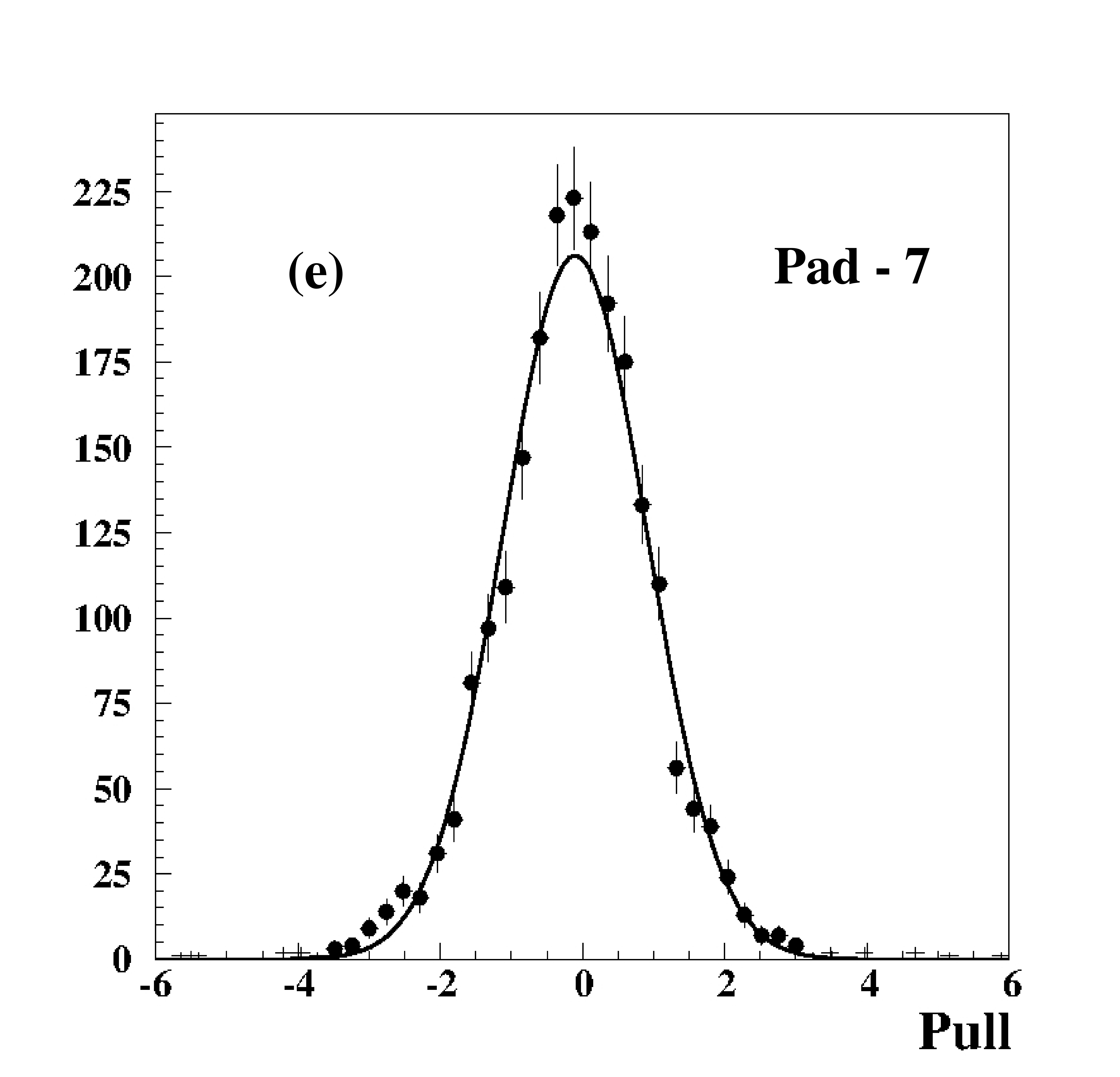}
\end{minipage}\hfill
\centering
\begin{minipage}{.33\textwidth}
\includegraphics[width=1.\textwidth]{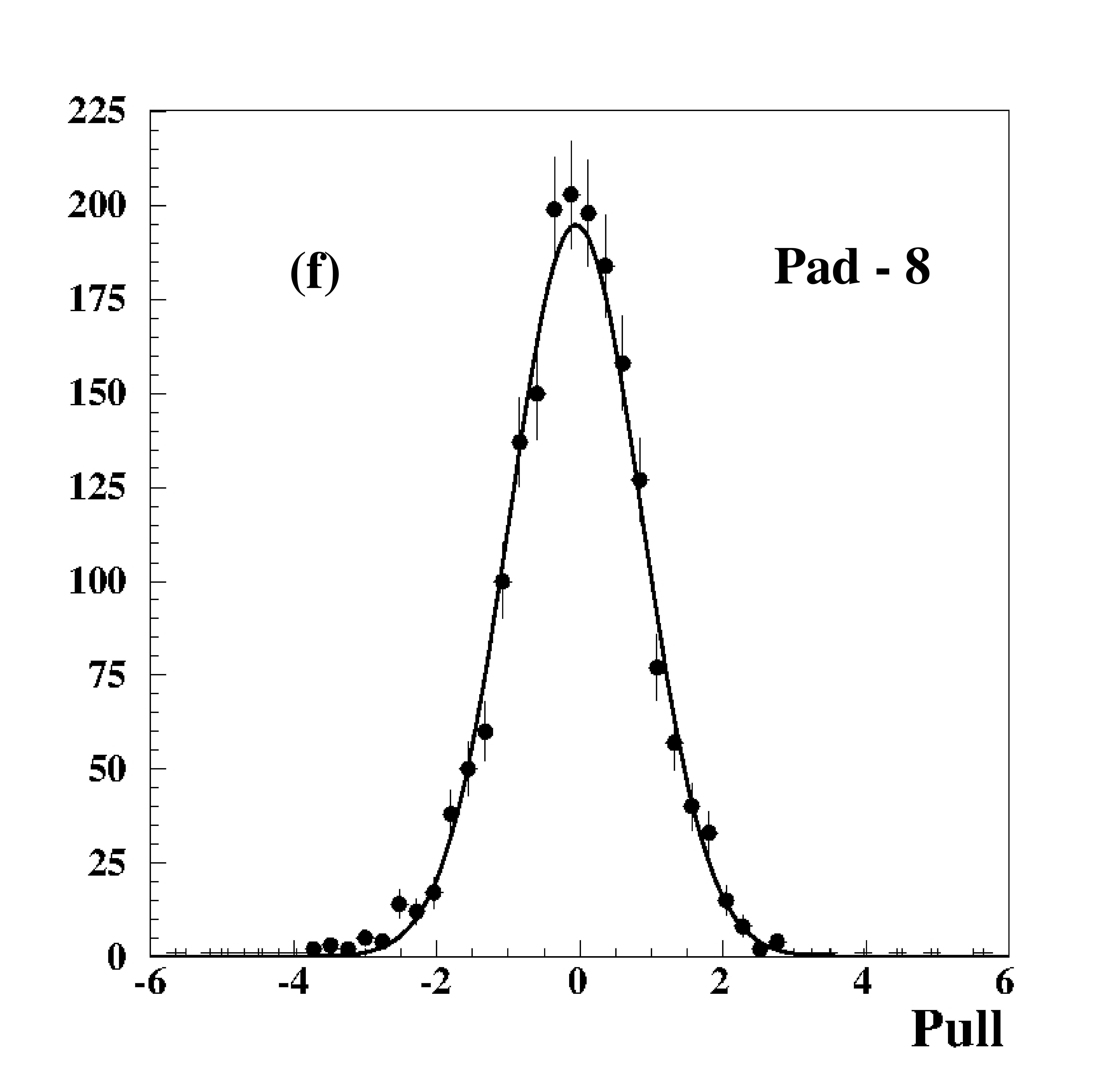}
\end{minipage}\hfill
\caption{ Distributions of fully corrected arrival time measurements of signals (SAT) induced on the pads No. 4 (plot a), No. 7 (plot b) and No. 8 (plot c) by MIPs passing within 2 mm of their common corner (see text). The solid lines represent fits of the data points with the sum of two  Gaussian functions. The mean value of the distributions is consistent with zero, as it is expected after correcting for shifts by subtracting the calibration function  $ \tau\left( Q_{e}^{m}\right)$. The  corresponding RMS values are  71.3 $\pm$ 2.5, 66.5.0 $\pm$ 2.5 and 68.0 $\pm$ 2.5 ps for pads No. 4, 7 and 8, respectively. The plots d, e and f show the corresponding  pull distributions of the fully corrected single-pad SAT measurements normalized to the expected measurement errors. The solid lines represent Gaussian fits, which are consistent with standard normal distribution functions.   }
\label{fig:pad_contr}
\end{figure}

\begin{equation}
\chi^{2}=\sum\limits_{m=1,M}\dfrac{\left(  T_{comb.}-     \left[ T_{f-corr.}^{m} -\tau\left( Q_{e}^{m}\right) \right]  \right)^{2} }{     \sigma^{2}\left( Q_{e}^{m}   \right)   } 
\label{eq8}
\end{equation}
where, $m=1, 2, ..., M$ is an index identifying the pads  sharing the Cherenkov ring. The minimization of the above $\chi^{2}$ provides an estimation of the MIP arrival time as:
\begin{equation}
\hat {T}_{comb.}=\frac{\sum\limits_{m=1,M}\dfrac{\left(    T_{f-corr.}^{m} -\tau\left( Q_{e}^{m}\right)  \right)^{2} }{     \sigma^{2}\left( Q_{e}^{m}   \right)   } }{\sum\limits_{m=1,M}\dfrac{1 }{     \sigma^{2}\left( Q_{e}^{m}   \right)   }}
\label{eq9}
\end{equation}
whilst the estimation uncertainty is evaluated by error propagation.
\\

For a demonstration, Eq. (\ref{eq8}) is applied to estimate the arrival time of MIPs passing within 2 mm from the common corner of pads No. 4, 7 and 8. In this case, the photoelectrons are shared almost equally between the  three anode segments and the induced  signals are of similar amplitude. The top row of Fig. \ref{fig:pad_contr} shows the SAT distributions of each of the  pads separately, corrected for shifts due to the $Q_{e}$ dependence (i.e. $T_{f-corr.}^{m} -\tau\left( Q_{e}^{m}\right) $). In this case, the spread (RMS) of a single pad distribution, i.e. the resolution in timing the MIP arrival using a single-pad measurement, is about 68 ps.  Moreover, the observed single-pad time resolution is  in a good agreement with  the expected resolution, which is expressed by the calibration function $\sigma\left( Q^{m}_{e}   \right)$, as it is demonstrated by the pull (i.e. $\dfrac{\left(    T_{f-corr.}^{m} -\tau\left( Q_{e}^{m}\right)  \right) }{     \sigma\left( Q_{e}^{m}   \right)   }$) distributions shown at the bottom row of Fig. \ref{fig:pad_contr}. 
\begin{figure}
\centering
\begin{minipage}{.48\textwidth}
\includegraphics[width=1.\textwidth]{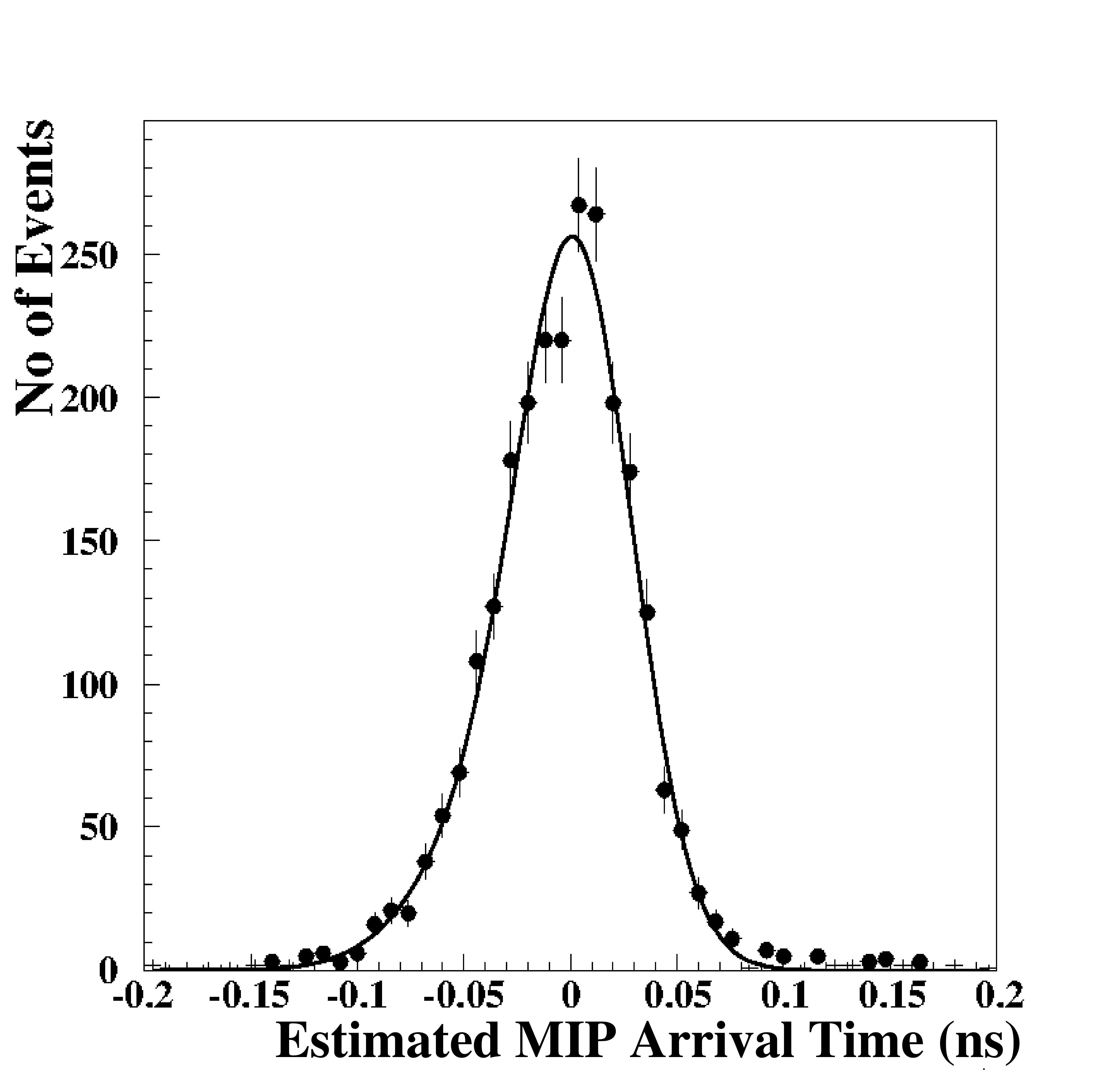}
\end{minipage}
\centering
\begin{minipage}{.48\textwidth}
\includegraphics[width=1.\textwidth]{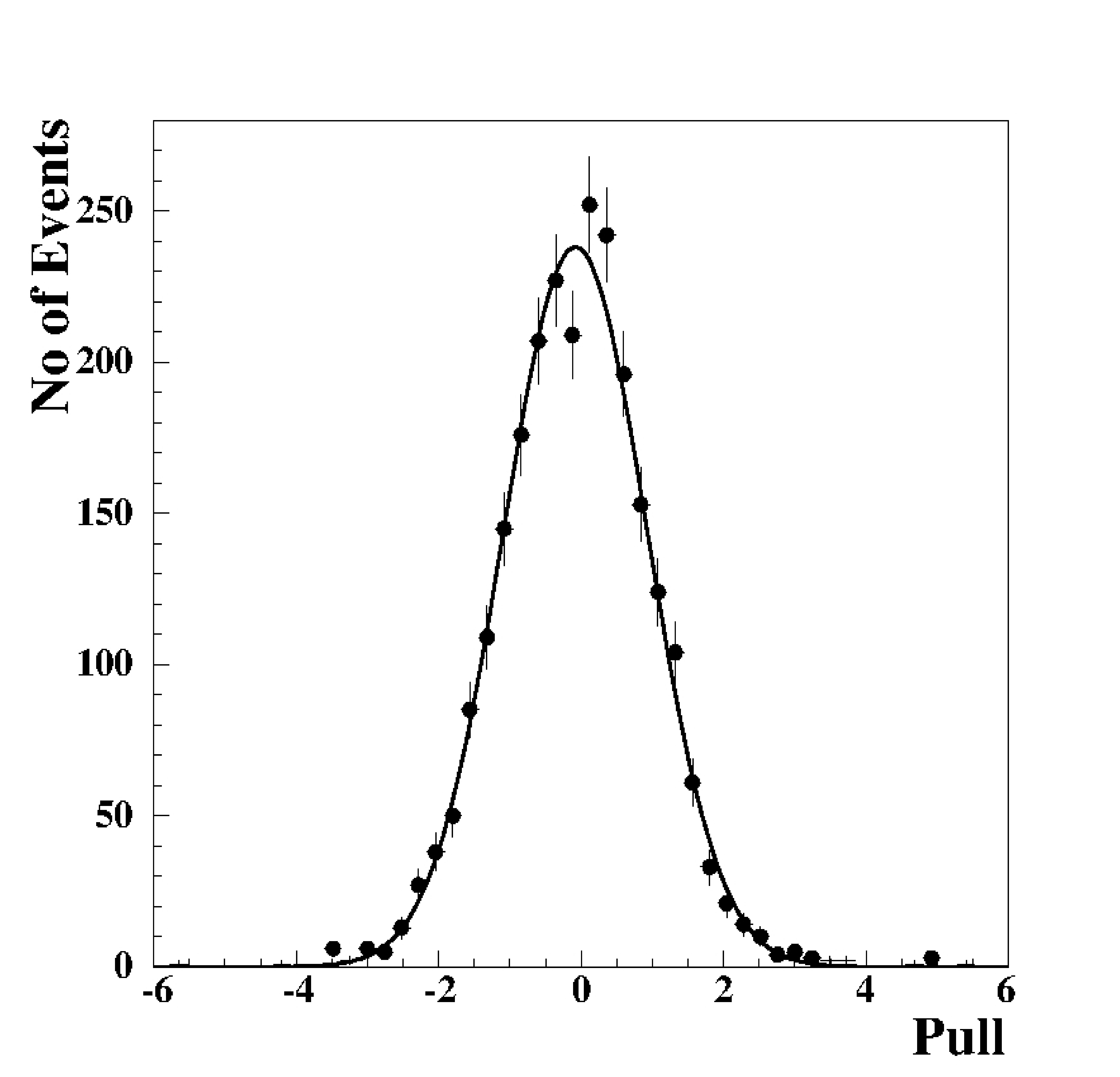}
\end{minipage}
\caption{ (left) Distribution of the arrival time of MIPs, passing within 2 mm of a common pad corner (pads No. 4, 7 and 8), estimated by Eq. (\ref{eq9}) combining  the individual single-pad measurements and their expected errors.  The solid line represents a fit to the data points by a sum of two Gaussian functions corresponding to an RMS of 32.2 $\pm$ 0.5 ps.   (right) Pull distribution of estimated arrival times by Eq.(\ref{eq9}). The solid line represents a Gaussian fit to the data points, consistent with  mean and  $\sigma$ values equal to 0 and 1 respectively.}
\label{fig:allpad_comb}
\end{figure}

Then, the MIP arrival time is also estimated by combining the individual single-pad measurements and their expected errors using   Eq. (\ref{eq9}). The distribution of the above estimations  is shown in Fig. \ref{fig:allpad_comb} (left plot) with a mean value that is  consistent with zero, demonstrating the no-bias of the timing procedure. The accuracy in estimating the MIP arrival time is expressed by the RMS of the distribution, which is   32.2 $\pm$ 0.5 ps.  Moreover, by evaluating the statistical error of each  estimation by Eq. (\ref{eq9}), we form the pull of the respective  arrival time estimation from its true value.  The emerging  pull distribution, shown  in the right plot of Fig. \ref{fig:allpad_comb}, follows a standard normal shape, signifying that the estimation error is  consistently evaluated  on an event by event basis. 
\\

\begin{figure}
\centering
\begin{minipage}{.48\textwidth}
\includegraphics[width=1.\textwidth]{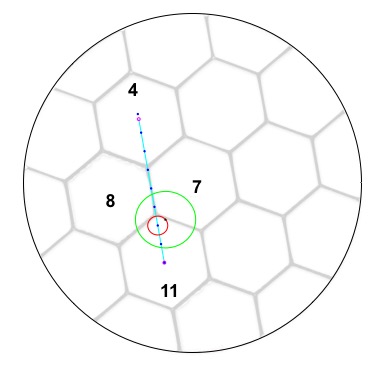}
\end{minipage}
\centering
\begin{minipage}{.48\textwidth}
\includegraphics[width=1.\textwidth]{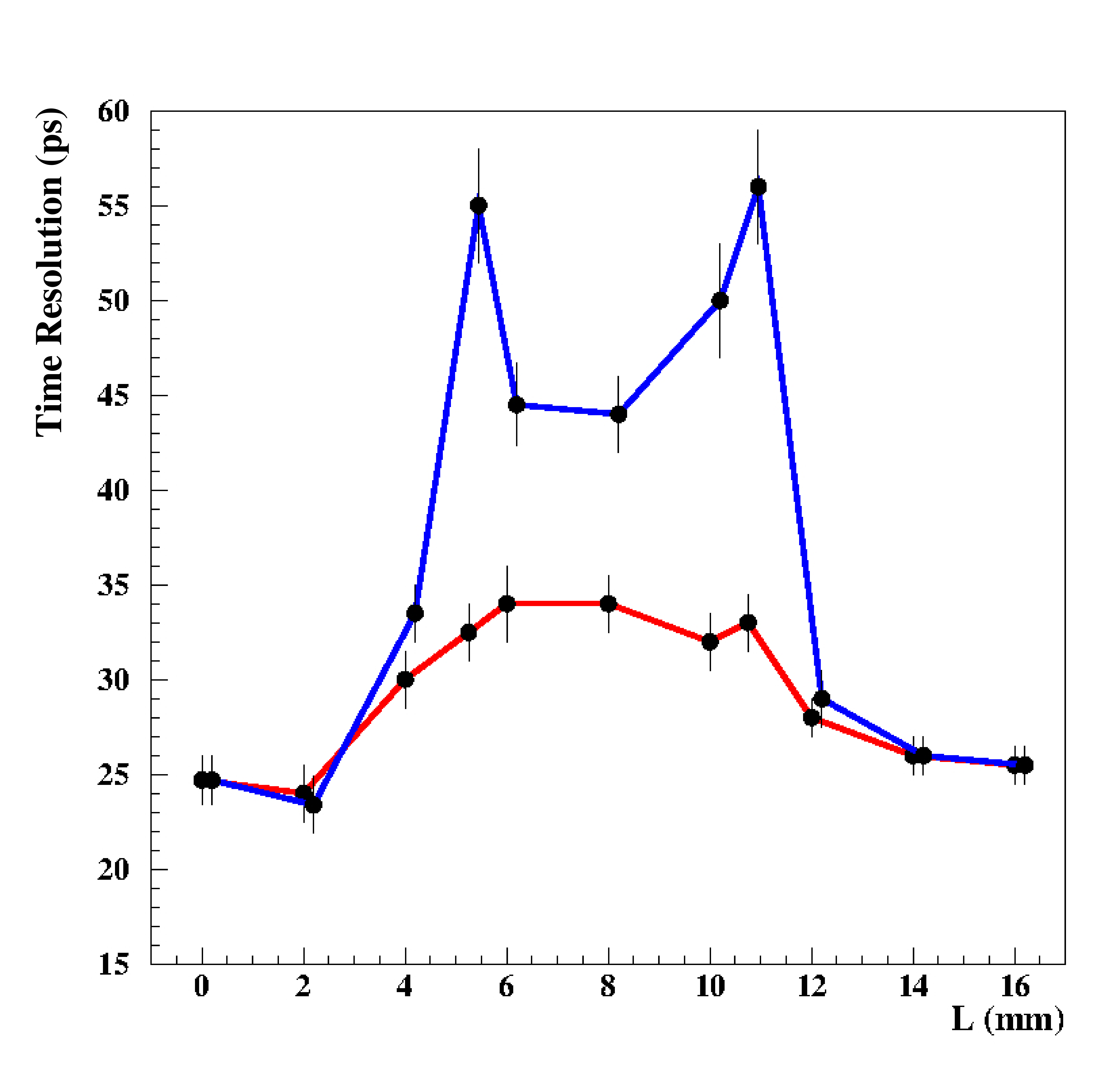}
\end{minipage}
\caption{ The resolution of the multi-pad PICOSEC-Micromegas prototype in timing the arrival  of MIPs that impact the detector plane along the line connecting the centres of pad No. 11 and pad No. 4. In the left plot, the red circle, 1 mm in radius, displays the area used to select MIPs  associated  with the sampling-point shown at its centre. The green circle, 3 mm in radius, illustrates the Cherenkov disc of a track passing through a point on the periphery of the red circle. In the right plot, the dots connected by red line segments represent the resolution in timing the arrival of MIPs passing around a sampling-point at distance L from the centre of pad No. 11. The dots connected with blue line segments denote the best single-pad timing performance  among the active anode pads. The points  at L = 5.25 mm and L = 10.75 mm display the respective time resolution for tracks passing around the common corners of three pads.}
\label{fig:allpad_line}
\end{figure}

For completeness, we  monitor the accuracy in estimating the MIP arrival time as its track impacts the detector along the line connecting the centres of pad No. 11 and pad No. 4. The left plot in Fig. \ref{fig:allpad_line} shows the sampling-points (blue marks) used to select MIPs passing around them within 1 mm distance (e.g. the red circle represents such a region). We label these sets of tracks by the distance L of the respective sampling-point to the centre of pad No. 11. In the right plot of Fig. \ref{fig:allpad_line}, the dots connected with  red line segments represent the resolution in estimating the MIP arrival time by combining information from the active pads using Eq. (\ref{eq9}) versus the distance L. The points connected with blue line segments illustrate the best single-pad timing performance (resolution) among the active anode segments. Depending on how the  photoelectrons  are shared between the neighboring segments, the time resolution is by 5 - 8  ps ($\sim$4 sigma) worse than in the case that all the Cherenkov photons are measured by a single pad. This  difference, however small, is possibly due to the limitations of  the empirical flatness corrections, detailed in Section \ref{spad}, to fully correct and equalize the timing performance of the  prototype PICOSEC-Micromegas pads. 
Nevertheless, the worsening of the time resolution maybe also due to other effects such as: a distortion of the amplification field in the vicinity of the  gaps between pads resulting to inefficiencies and extra timing errors or effects (e.g. cross talk between adjacent anode segments) which cause correlation between neighboring  pad measurements. This type of possible effects will be investigated in the near future with a new, improved flatness, multi-pad PICOSEC-Micromegas prototype currently under construction.

\section{Concluding Remarks} \label{concl}

We developed a multi-pad PICOSEC-Micromegas prototype consisting of 19 hexagonal pads and we evaluated its performance with data collected in a muon test beam over 4 instrumented pads. In the present paper, we studied in detail the effect of the drift gap thickness non-uniformity on the time resolution and devised empirical flatness corrections to obtain a uniform timing response over the whole detector coverage. We measured a time resolution of $25.8 \pm 0.6$ ps when the muons impact the detector close to a pad center. However, the timing error rises up to $32.2 \pm 0.5$  ps when the track approaches the pad corners and the Cherenkov ring is shared among three pads. This increase is possibly due to remaining systematics, i.e. to the limitation of the applied corrections to fully mitigate the systematic effects caused by the drift field non-uniformity. Nevertheless, there may be other effects contributing to the worsening of the time resolution when the Cherenkov photons initiate avalanches close to the pad edges (e.g. field distortions, cross-talk, etc). Such effects, introducing inefficiencies or/and correlations between adjacent pads, will be investigated in the near future with a new improved detector. \\

The Micromegas PCB of the current prototype suffers from initial deformations (see Fig. 9), most likely caused by the production process and mesh stretching, which can be additionally pronounced by tightening the PCB to the detector housing. In a new prototype, currently under construction, special care is taken to ensure precise flattening and polishing (e.g. by using more rigid and thicker PCB that combines a ceramic core for the rigidity and thin FR4 outer layers). Such an improved detector will have a drift thickness uniformity of better than $10\,\textrm{\selectlanguage{greek}m\selectlanguage{english}m}$ across the area in front of the Cherenkov radiator and flatness corrections will not be needed.

\section*{Acknowledgments}
We acknowledge the support of the RD51 collaboration in the framework of RD51 common projects, the support of Photek Ltd. for lending us the MCP-PMT during the beam tests and the financial support of the Cross-Disciplinary Program on Instrumentation and Detection of CEA, the French Alternative Energies and Atomic Energy Commission. M. Gallinaro acknowledges the support from the Funda\c c\~ao para a Ci\^encia e a Tecnologia (FCT), Portugal. S. White acknowledges partial support through the US CMS program under DOE contract No. DE-AC02-07CH11359.

\bibliography{multi_pico_v6_18Jan_rev}

\newpage
\appendix

%\end{linenumbers}
\end{document}